\documentclass[11pt,a4paper]{article}

\usepackage[a4paper,top=0.88in,bottom=0.92in,left=0.96in,right=0.96in,headheight=15pt,headsep=19pt]{geometry}
\usepackage[T1]{fontenc}
\usepackage{newtxtext,newtxmath}
\usepackage{microtype}
\usepackage{amsmath,mathtools,bm}

\usepackage{amsthm}
\usepackage{graphicx}
\usepackage{titlesec}
\usepackage{fancyhdr}
\usepackage{booktabs,array,tabularx,multirow}
\usepackage{enumitem}
\usepackage{algorithm}
\usepackage[section]{placeins}
\usepackage{algpseudocode}
\usepackage{caption,subcaption}
\usepackage{tikz}
\usetikzlibrary{arrows.meta,positioning,calc,fit,decorations.pathreplacing}
\usepackage[dvipsnames]{xcolor}
\usepackage{siunitx}
\usepackage[colorlinks=true,linkcolor=NavyBlue,citecolor=NavyBlue,urlcolor=NavyBlue,hypertexnames=false]{hyperref}
\usepackage[nameinlink,noabbrev]{cleveref}
\makeatletter
\providecommand{\theHALG@line}{\thealgorithm.\arabic{ALG@line}}
\makeatother
\usepackage{xurl}
\hypersetup{
	pdftitle={An Efficient Sequential Algorithm for
		Structured Quantum State Preparation},
	pdfauthor={Alok Shukla},
	pdfsubject={Geometric and Fourier state preparation on finite integer intervals},
	pdfkeywords={quantum state preparation, norm distribution, geometric states, Fourier states, finite intervals, quantum circuits}
}

\titleformat{\section}{\normalfont\Large\bfseries}{\thesection}{0.72em}{}
\titleformat{\subsection}{\normalfont\large\bfseries}{\thesubsection}{0.68em}{}
\titleformat{\subsubsection}{\normalfont\normalsize\bfseries}{\thesubsubsection}{0.62em}{}
\titlespacing*{\section}{0pt}{2.15ex plus .5ex minus .2ex}{0.85ex}
\titlespacing*{\subsection}{0pt}{1.65ex plus .4ex minus .2ex}{0.6ex}
\titlespacing*{\subsubsection}{0pt}{1.25ex plus .3ex minus .15ex}{0.4ex}

\setlist[enumerate]{leftmargin=2em,itemsep=3pt,topsep=4pt,parsep=0pt}
\setlist[itemize]{leftmargin=1.6em,itemsep=2pt,topsep=4pt,parsep=0pt}
\fancypagestyle{plain}{\fancyhf{}\fancyfoot[C]{\small\thepage}}

\newcommand{\C}{\mathbb C}
\newcommand{\R}{\mathbb R}

\newcommand{\ket}[1]{\lvert #1\rangle}
\newcommand{\bra}[1]{\langle #1\rvert}

\newcommand{\diag}{\operatorname{diag}}

\newcommand{\vtwo}{v_2}

\newcommand{\Ry}{R_y}

\newcommand{\atanTwo}{\operatorname{atan2}}
\newcommand{\e}{\mathrm e}
\newcommand{\xor}{\mathbin{\oplus}}

\newcommand{\cgate}[2]{\operatorname{C}_{#1}(#2)}
\newcommand{\keywords}[1]{\par\vspace{0.35em}\noindent\small\textbf{Keywords---} #1\par}

\numberwithin{equation}{section}
\newtheoremstyle{journalplain}{7pt}{7pt}{\itshape}{}{\bfseries}{.}{0.55em}{}
\newtheoremstyle{journaldefinition}{7pt}{7pt}{\normalfont}{}{\bfseries}{.}{0.55em}{}
\newtheoremstyle{journalremark}{6pt}{6pt}{\normalfont}{}{\itshape}{.}{0.55em}{}
\theoremstyle{journalplain}
\newtheorem{theorem}{Theorem}[section]
\newtheorem{lemma}[theorem]{Lemma}
\newtheorem{proposition}[theorem]{Proposition}

\theoremstyle{journaldefinition}
\newtheorem{definition}[theorem]{Definition}
\newtheorem{example}[theorem]{Example}
\theoremstyle{journalremark}
\newtheorem{remark}[theorem]{Remark}

\crefname{theorem}{theorem}{theorems}
\Crefname{theorem}{Theorem}{Theorems}
\crefname{lemma}{lemma}{lemmas}
\Crefname{lemma}{Lemma}{Lemmas}
\crefname{proposition}{proposition}{propositions}
\Crefname{proposition}{Proposition}{Propositions}
\crefname{corollary}{corollary}{corollaries}
\Crefname{corollary}{Corollary}{Corollaries}
\crefname{definition}{definition}{definitions}
\Crefname{definition}{Definition}{Definitions}

\begin{document}
	\title{An Efficient Sequential Algorithm for Structured Quantum State Preparation}
	\author{Alok Shukla\\
		\small School of Arts and Sciences, Ahmedabad University, Ahmedabad, Gujarat, India\\
		\small \texttt{alok.shukla@ahduni.edu.in}}
	\date{}
	\maketitle

\begin{abstract}
	\noindent
	We develop an exact sequential algorithm for quantum state preparation based on two-dimensional norm-preserving representations of residual amplitude profiles. It uses a single memory qubit drawn from the register itself, yielding exact, ancilla-free circuits. Writing $N=2^m$, we prepare shifted geometric states with computational-basis amplitudes proportional to $r^{x-k}$ for $k\leq x<N$, where $0\leq k<N$ is the starting label and $r\in\mathbb C\setminus\{0\}$ is the common ratio, and real single-frequency Fourier states with amplitudes proportional to $\sin(2\pi\nu x/N+\phi)$ for $0\leq x<N$, where $\nu\in\{0,\ldots,N-1\}$ is the discrete frequency and $\phi\in\mathbb R$ is the phase offset, provided the profile is nonzero. Both constructions extend to arbitrary finite integer intervals with nonzero restricted norm.
	For full-register states, we prove optimal adjacent two-qubit interaction counts on a fixed linear nearest-neighbour ordering. These counts are at most linear in $m$. Each completed local two-qubit gate admits an exact synthesis using at most two CNOT gates and single-qubit gates, so the corresponding full-register circuits have CNOT-count upper bounds of $O(m)$. For finite intervals, a common dyadic partition into $h\geq2$ blocks admits a branch-and-block selector using at most $3h-3$ one-qubit or singly controlled one-qubit gates. The resulting finite-window circuits are exact and ancilla-free at the logical-circuit level. Numerical statevector simulations verify the full-register circuits and the logical finite-window state-preparation constructions.
\end{abstract}


\section{Introduction}\label{sec:intro}

Quantum computing has developed into a promising computational paradigm, leading to algorithms for problems that can be difficult for classical methods. Over the past several decades, a substantial collection of quantum algorithms has been established, with provable advantages for particular structured computational tasks. Representative examples include the Deutsch--Jozsa algorithm \cite{deutsch1992rapid}, the Bernstein--Vazirani algorithm and its probabilistic generalization \cite{bernstein1993quantum,shukla2023generalizationBV}, Simon's algorithm \cite{simon1997power}, and quantum search \cite{grover1996fast,shukla2025efficientGrover}. Other fundamental algorithmic tools include the quantum Fourier transform and quantum phase estimation \cite{nielsen2002quantum,kitaev1995quantum,dobvsivcek2007arbitrary,shukla2025towardsQPE}, together with quantum amplitude estimation \cite{brassard2002quantum,giurgica2022low,shukla2025modularQAE}. Shor's algorithms for integer factorization and discrete logarithms provide another important example of quantum speedup \cite{shor1999polynomial,shukla2025modularShor}.

Beyond these foundational developments, quantum computing is being investigated as a computational tool for a growing range of scientific and engineering problems. Examples include trajectory optimization \cite{shukla2019trajectory}, the solution of large-scale linear systems \cite{harrow2009quantum}, and linear and nonlinear differential equations \cite{childs2020quantum,berry2014high,shukla2023hybriddiffeq}. Quantum approaches have also been developed for digital image and signal processing, as well as spectral analysis \cite{shukla2023hybridimageprocessing,shukla2023quantumDSP,rohida2024hybridPolar,rohida2025quantumEdgeDetect}. Further areas of investigation include quantum optimal control \cite{sandesara2024quantumOptimalcontrol}, quantum machine learning \cite{biamonte2017quantum}, and numerical integration \cite{shimada2020quantum,abrams1999fast,shukla2025efficientPartialSum}. Many of these applications involve processing structured classical data, such as discretized functions, images, signals, coefficient arrays, or training data, within a quantum computational framework.

Realizing the potential of these algorithms requires attention not only to the main quantum computation but also to the preparation of its input state and the extraction of its output through measurement. An efficient quantum routine does not, by itself, guarantee an end-to-end quantum advantage: the advantage can be substantially reduced or even lost if preparing the required input state, or obtaining the desired classical information from the output, incurs a prohibitive cost. This issue is particularly relevant in application-oriented settings such as differential equation solvers, quantum image and signal processing, optimal control, and quantum machine learning, where structured classical data may need to be encoded repeatedly or at different stages of a computation. Efficient state preparation is therefore an essential component of quantum algorithm design, and its resource requirements must be included in any meaningful assessment of computational advantage.

The preparation cost can be especially significant when the target is an arbitrary quantum state. An $m$-qubit register has Hilbert-space dimension $N=2^m$, and a general pure state in this space has $2N-2$ independent real parameters. Accordingly, generic exact state-synthesis procedures require $O(N)$ elementary gates in the worst case \cite{mottonen2004transformation,ShendeBullockMarkov2006,PleschBrukner2011}. This is exponential in the number of qubits, even though the register itself has only $m$ qubits. The central question is whether useful structured families of states can be prepared much more efficiently, with circuit complexity governed by the register width rather than by the dimension of the Hilbert space. For many structured quantum states, however, the preparation cost can be substantially smaller. For example, the algorithm in \cite{shukla2024efficient} prepares a uniform superposition over \(M\) basis states with gate complexity and circuit depth \(O(\log_2 M)\), without requiring ancillary qubits or multi-controlled gates.

The principal contribution of this work is the development of an exact sequential framework for quantum state preparation based on the distribution of amplitude norms. The proposed approach exploits the low-dimensional structure of the residual amplitude profiles of the target quantum states. For the state families considered here, the residual states corresponding to different prefixes (see \Cref{def:binary-cuts}) admit a representation in terms of a two-dimensional coherent memory. Importantly, this memory is encoded directly in the register, eliminating the need for additional ancilla qubits. At each stage, positive branch operators govern the distribution of squared norms between the two possible extensions of the current prefix, while the associated polar factors preserve the coherent evolution of the remaining suffix. By an appropriate common change of memory basis, one branch factor can be absorbed, reducing the branch-dependent coherent correction to a single relative unitary. This leads to a reusable local circuit in which norm redistribution and coherent memory evolution are implemented together.

The above approach enables efficient quantum state preparation
for two families of structured amplitudes: shifted geometric
profiles and real single-frequency Fourier profiles. To specify
these families, let $N=2^m$. For $0\leq k<N$ and
$r\in\mathbb C\setminus\{0\}$, the normalized shifted geometric
state is
\[
\ket{\Gamma_{m,k}(r)}
=
\frac{1}{\sqrt{Z_{m,k}(r)}}
\sum_{x=k}^{N-1}r^{x-k}\ket{x},
\qquad
Z_{m,k}(r)=\sum_{j=0}^{N-k-1}|r|^{2j}.
\]
For a discrete frequency $\nu\in\{0,\ldots,N-1\}$ and
phase offset $\phi\in\mathbb R$, the real single-frequency
Fourier state is
\[
\ket{F_{m,\nu,\phi}}
=
\frac{1}{\sqrt{Z_{m,\nu,\phi}}}
\sum_{x=0}^{N-1}
\sin\left(\frac{2\pi\nu x}{N}+\phi\right)\ket{x},
\qquad
Z_{m,\nu,\phi}
=
\sum_{x=0}^{N-1}
\sin^2\left(\frac{2\pi\nu x}{N}+\phi\right),
\]
provided the unnormalized vector is nonzero. The latter states
are expressed in the computational basis: their amplitudes
are samples of a real sinusoid, rather than coefficients of
a state written in the Fourier basis.

We first consider these states on full binary registers and
then extend the constructions to finite integer intervals.
Although the amplitudes in both families have compact
mathematical descriptions, this compactness does not
automatically translate into an efficient quantum circuit.
For a shifted geometric profile, the offset determines how
the support is distributed relative to the binary structure
of the register. For a sinusoidal profile, the frequency and
phase generally introduce dependencies among several binary
digits. Therefore, the corresponding states do not
generally factor into one-qubit states. A direct sequential
construction can divide the squared norm between the two
possible values of each successive qubit, but the required
division may depend on the entire sequence of values already
prepared. In general, this leads to a different controlled
operation for many such sequences, and the number of
operations can grow with the number of computational-basis
states, $N=2^m$. The objective is to exploit the structure
shared by these amplitude profiles so that their preparation
requires only a small number of coherent operations.

The local updates must act coherently on every possible state of the memory qubit. It is not enough to prescribe fixed scalar probabilities for the two branches, since the branch weights depend on the incoming memory state and the relative amplitudes must remain coherent. We instead represent the two branches by matrices $A_0$ and $A_1$ satisfying
\[
A_0A_0^\dagger+A_1A_1^\dagger=I_2.
\]
For an incoming memory state represented by a normalized row vector $\lambda$, the squared norm assigned to branch $b$ is $\lambda A_bA_b^\dagger\lambda^\dagger$. Thus, the positive matrices $A_bA_b^\dagger$ determine the norm distribution for every incoming memory state, while the corresponding polar factors describe the coherent transformations of the memory. A suitable common choice of memory basis absorbs one branch factor and leaves only a relative unitary for the other. This representation separates the distribution of norm between the branches from the coherent transport of the memory, without losing the phase information required for exact state preparation.

For the full-register constructions, the resulting interaction counts are linear in the register width. A shifted geometric state on an $m$-qubit register requires at most $m-1$ adjacent two-qubit interactions: one interaction is used across each cut where the state does not factor, and none is needed across a product cut. This count is optimal for the prescribed linear nearest-neighbour ordering, as follows from Schmidt-rank lower bounds. For a non-product real single-frequency Fourier state, writing the frequency as $\nu=2^s u$, with $u$ odd, isolates an active core of $d=m-s-1$ qubits after the product factors are removed. The construction uses exactly $d-1$ adjacent interactions, one across each internal cut of this core. Since every such cut is non-product, the Schmidt-rank lower bound gives the matching optimality result. Thus, both constructions have interaction complexity $O(m)$, in contrast with the $O(2^m)$ scaling of generic exact state preparation.

The local gate parameters are generated by short scalar and $2\times2$ matrix recurrences. Moreover, each real two-qubit update admits an exact constructive decomposition using at most two CNOT gates together with one-qubit gates. Applying these decompositions gives linear CNOT upper bounds for the full-register circuits. The optimality statements in this work concern the number of adjacent two-qubit interactions; the two-CNOT constructions establish efficient synthesis upper bounds. For finite intervals, we use a common dyadic partition to organize the local preparation, together with a branch-and-block selector requiring at most $3h-3$ one-qubit or singly controlled one-qubit gates when the partition contains $h\geq 2$ blocks.
These resource bounds can be put in context by comparing our
constructions with the corresponding implementations in
\emph{PyEncode}, an open-source library for structured quantum
state preparation \cite{Suresh2026PyEncode}.
PyEncode provides exact circuit constructions for several classes of structured amplitude profiles, including geometric and Fourier states. For the geometric profile \(f_i=c r^i\) without a shift, the state is a product state and can be prepared using \(O(m)\) single-qubit gates and no entangling operations. For a general nonzero starting index \(k\), however, the profile is supported on the tail interval \([k,2^m)\). The current PyEncode implementation handles this case through a dyadic decomposition and lists \(O(m^2)\) gate complexity. In contrast, our construction prepares shifted geometric states using exactly
\[
E_m(k)=\max\left\{0,\left\lfloor\log_2(2^m-k-1)\right\rfloor-v_2(k)\right\}
\]
adjacent two-qubit interactions, for \(1\leq k<2^m-1\). This count is at most \(m-1\), is optimal for the prescribed linear nearest-neighbour ordering, and each local interaction admits an exact synthesis using at most two CNOT gates together with single-qubit gates. Thus, for general nonuniform shifted geometric profiles, our construction improves the documented quadratic gate complexity to a linear bound, while also providing an exact interaction count.

For Fourier profiles, PyEncode prepares a sampled sinusoidal mode by loading its nonzero discrete Fourier coefficients and applying an inverse quantum Fourier transform, resulting in \(O(m^2)\) gate complexity. Our construction instead exploits the binary structure of a single Fourier frequency. Writing a nonexceptional frequency as \(\nu=2^s u\), with \(u\) odd, reduces the state to an active register of \(d=m-s-1\) qubits. We show that exactly \(d-1\) adjacent two-qubit interactions are both necessary and sufficient, with an exact synthesis using at most two CNOT gates per interaction. Therefore, our single-mode Fourier construction has linear gate complexity. These comparisons concern the two specific amplitude families studied here; PyEncode remains a more general framework supporting a wider range of structured states, including multi-mode Fourier profiles.

The remainder of the paper is organized as follows.
\Cref{sec:scalar} develops the norm-preserving representation
and explains how the branch effects and relative memory
updates determine the local norm distribution.
\Cref{sec:local-circuit} derives the corresponding
two-qubit implementation. The full-register constructions
for shifted geometric and real single-mode Fourier states,
together with their optimal adjacent-interaction bounds,
are presented in \Cref{sec:geometric,sec:fourier}.
\Cref{sec:dyadic} extends these constructions to arbitrary
finite intervals through a common dyadic partition.
\Cref{sec:synthesis} develops the standard-gate synthesis,
establishes the gatewise CNOT criteria, and reports
representative synthesis and numerical verification
results. Finally, \Cref{sec:conclusion} summarizes the
main contributions, their complexity implications, and
directions for further work. The appendices provide
additional details on norm-preserving coordinates,
geometric and Fourier formulas, and the finite-interval
block-selection construction.

\section{Binary norm splitting and norm-preserving updates}\label{sec:scalar}
	
	We consider an $n$-qubit register, with $N=2^n$, and use the usual most-significant-bit-first binary representation. The terms prefix, suffix, and cut will be used throughout the paper, so we fix their meaning at the outset.
	
	\begin{definition}[Prefixes, suffixes, and cuts]\label{def:binary-cuts}
		Let $\mathcal S\subseteq\{0,1\}^n$ be the set of allowed basis labels, and let
		$	|\psi\rangle=\sum_{x\in\mathcal S}f(x)|x\rangle$
		be a target state, with its amplitude function understood to be zero outside $\mathcal S$. For $0\leq j\leq n$, write a binary string $x=x_1\cdots x_n$ uniquely as $x=pz$, where $p=x_1\cdots x_j$ consists of the first $j$ bits and $z=x_{j+1}\cdots x_n$ consists of the remaining $n-j$ bits. We call $p$ the \emph{prefix} and $z$ the corresponding \emph{suffix} at cut $j$; $pz$ denotes their concatenation. The cut at position $j$ separates the first $j$ qubits from the remaining $n-j$ qubits. For a fixed prefix $p\in\{0,1\}^j$, its set of allowed suffix completions is
		$\mathcal C(p)=\{z\in\{0,1\}^{n-j}:pz\in\mathcal S\}$.
		The vector
		$	f_p^{(j)}=\bigl(f(pz)\bigr)_{z\in\{0,1\}^{n-j}}$,
		with zero entries for disallowed completions, is the \emph{suffix-amplitude vector} below $p$. Stacking these vectors over all prefixes gives the cut coefficient matrix $F^{(j)}\in\mathbb C^{2^j\times 2^{n-j}}$, with entries $F^{(j)}_{p,z}=f(pz)$. Its rank is the Schmidt rank of $|\psi\rangle$ across the cut. Accordingly, an interior cut ($1\leq j<n$) is called \emph{product} if $\operatorname{rank}F^{(j)}=1$, and \emph{non-product} if $\operatorname{rank}F^{(j)}>1$. Equivalently, a product cut is one across which the target state factors as a tensor product of states on the two groups of qubits. For $b\in\{0,1\}$ and $|p|<n$, we write $p_b:=pb$ for the child prefix obtained by appending the bit $b$.
	\end{definition}
	
	Thus, at a fixed cut, varying the prefix selects different rows of the same coefficient matrix, while the suffix coordinate records the unprepared part of the basis label. This viewpoint is useful both for the norm calculations below and for the adjacent-interaction lower bounds: a non-product cut has Schmidt rank greater than one and therefore carries entanglement between the two sides of the cut. Throughout, amplitude vectors are written as row vectors and matrices act from left to right. We write $A^\dagger$ and $A^T$ for the conjugate transpose and transpose, respectively.
	
	\subsection{Binary norm splitting}
	
	Let $f$ be a normalized target amplitude profile on $n$-bit strings. For a prefix $p$ of length $j<n$, define its remaining squared norm (the total weight of all allowed suffix completions) by
	$W(p)=\sum_{z\in\mathcal C(p)}|f(pz)|^2$.
	Every allowed suffix completion begins with exactly one of the two possible next bits. With the child-prefix notation from Definition~\ref{def:binary-cuts}, this gives $W(p)=W(p_0)+W(p_1)$. Whenever $W(p)>0$, the angle
	\begin{equation}\label{eq:scalar-angle}
		\theta_p=2\arccos\sqrt{\frac{W(p_0)}{W(p)}}
	\end{equation}
	implements the corresponding norm split:
	\[
	\Ry(\theta_p)\ket0=
	\sqrt{\frac{W(p_0)}{W(p)}}\ket0+
	\sqrt{\frac{W(p_1)}{W(p)}}\ket1.
	\]
	Thus a binary tree of prefix-controlled $R_y$ rotations prepares any prescribed nonnegative amplitude profile. For a general complex profile, these rotations determine the magnitudes; the phase information must be supplied separately.
	
	This scalar construction applies to arbitrary target magnitudes, but its rotations generally depend on the complete prefix. In the absence of additional structure, as many as $2^n-1$ prefix-conditioned rotations may be required. The aim here is to replace these separate scalar splits by a small collection of local matrix updates that works for every coherent state of a low-dimensional memory. The two-dimensional case is the one needed for the geometric and Fourier families considered in this paper; it is a structural property of those families, not a representation available for every amplitude vector.
	
	\subsection{A norm-preserving representation}
	
	We encode the amplitude profile using a two-dimensional memory. Let $\lambda\in\mathbb C^{1\times2}$ be a unit row vector, let $A_j^{(b)}\in\mathbb C^{2\times2}$ for $1\leq j<n$ and $b\in\{0,1\}$, and let $g_0,g_1\in\mathbb C^{2\times1}$ be terminal columns. For $\mathbf b=(b_1,\ldots,b_n)\in\{0,1\}^n$, define
	\begin{equation}\label{eq:balanced-amplitude}
		a(\mathbf b)=\lambda A_1^{(b_1)}A_2^{(b_2)}\cdots A_{n-1}^{(b_{n-1})}g_{b_n}.
	\end{equation}
	We call this a \emph{norm-preserving update representation} when
	\begin{align}
		\lambda\lambda^\dagger&=1,\label{eq:balanced-boundary}\\
		A_j^{(0)}A_j^{(0)\dagger}+A_j^{(1)}A_j^{(1)\dagger}&=I_2,
		\qquad 1\leq j<n,\label{eq:balanced-update}\\
		g_0g_0^\dagger+g_1g_1^\dagger&=I_2.\label{eq:balanced-terminal}
	\end{align}
	The matrices $A_j^{(b)}$ update the memory coordinates when the next output bit is $b$. The terminal columns specify the final output bit. 
	Let $G=[g_0\;g_1]$. Then  the terminal condition \eqref{eq:balanced-terminal} implies $GG^\dagger=I_2$.
	
	\begin{proposition}[Normalization by local norm conservation]\label{prop:balanced-normalization}
		If \eqref{eq:balanced-boundary}--\eqref{eq:balanced-terminal} hold, then the amplitudes in \eqref{eq:balanced-amplitude} are normalized:
		\[
		\sum_{\mathbf b\in\{0,1\}^n}|a(\mathbf b)|^2=1.
		\]
	\end{proposition}
%	\begin{proof}
%		For a fixed prefix, first sum over the final bit. The identity
%		$g_0g_0^\dagger+g_1g_1^\dagger=I_2$ removes the terminal columns and preserves the squared norm of the incoming memory row. Summing over the preceding bit then uses
%		$A_j^{(0)}A_j^{(0)\dagger}+A_j^{(1)}A_j^{(1)\dagger}=I_2$.
%		Iterating this contraction leaves $\lambda\lambda^\dagger=1$.
%	\end{proof}
%
%
%\begin{proof}
%	For each fixed choice of $b_1,\ldots,b_{n-1}$, write
%	$v_{n-1}=\lambda A_1^{(b_1)}\cdots A_{n-1}^{(b_{n-1})}$.
%	Using the terminal identity
%	$g_0g_0^\dagger+g_1g_1^\dagger=I_2$, we obtain
%	\[
%	\sum_{\mathbf b\in\{0,1\}^n}|a(\mathbf b)|^2
%	=
%	\sum_{b_1,\ldots,b_{n-1}\in\{0,1\}}
%	v_{n-1}v_{n-1}^\dagger.
%	\]
%	Here and below, each $v_j$ depends on the bits
%	$b_1,\ldots,b_j$.
%	
%	We now perform the remaining sums successively, starting
%	with $b_{n-1}$. For fixed $b_1,\ldots,b_{j-1}$, put
%	$v_{j-1}=\lambda A_1^{(b_1)}\cdots A_{j-1}^{(b_{j-1})}$.
%	Then the update identity gives
%	\[
%	\sum_{b_j\in\{0,1\}}
%	v_{j-1}A_j^{(b_j)}A_j^{(b_j)\dagger}v_{j-1}^\dagger
%	=v_{j-1}v_{j-1}^\dagger.
%	\]
%	Thus, summing over $b_j$ eliminates the corresponding
%	update matrices. Repeating this contraction for
%	$j=n-1,\ldots,1$ leaves
%	$\lambda\lambda^\dagger=1$, by the boundary condition.
%	Hence the amplitudes are normalized.
%\end{proof}
%

\begin{proof}
	For $j=0,\ldots,n-1$, write
	$v_j=\lambda A_1^{(b_1)}\cdots A_j^{(b_j)}$,
	with $v_0=\lambda$. We first sum over the last bit $b_n$.
	Since $|a(\mathbf b)|^2=v_{n-1}g_{b_n}g_{b_n}^\dagger
	v_{n-1}^\dagger$, the terminal identity gives
	\[
	\sum_{\mathbf b\in\{0,1\}^n}|a(\mathbf b)|^2
	=\sum_{b_1,\ldots,b_{n-1}}v_{n-1}v_{n-1}^\dagger.
	\]
	Here the sum over the remaining $n-1$ bits is retained.
	We next sum over $b_{n-1}$. Writing
	$v_{n-1}=v_{n-2}A_{n-1}^{(b_{n-1})}$ and using
	$A_{n-1}^{(0)}A_{n-1}^{(0)\dagger}
	+A_{n-1}^{(1)}A_{n-1}^{(1)\dagger}=I_2$, we obtain
	\[
	\sum_{\mathbf b\in\{0,1\}^n}|a(\mathbf b)|^2
	=\sum_{b_1,\ldots,b_{n-2}}v_{n-2}v_{n-2}^\dagger.
	\]
	The same argument applies to each preceding bit. More
	precisely, whenever the sums over $b_n,\ldots,b_{j+1}$
	have been performed, the remaining expression is
	$\sum_{b_1,\ldots,b_j}v_jv_j^\dagger$. Summing over $b_j$
	and applying the corresponding update identity reduces it
	to $\sum_{b_1,\ldots,b_{j-1}}v_{j-1}v_{j-1}^\dagger$.
	Repeating this contraction down to $j=1$ leaves
	$v_0v_0^\dagger=\lambda\lambda^\dagger=1$, by the
	boundary condition. Hence the amplitudes are normalized.
\end{proof}

The same contraction argument gives a useful identity for
any fixed initial string $b_1\cdots b_j$, which we call a
prefix of length $j$. For $0\leq j<n$, keep
$b_1,\ldots,b_j$ fixed and sum over all possible choices of
the remaining bits $b_{j+1},\ldots,b_n$. The successive
contractions give
$\sum_{b_{j+1},\ldots,b_n\in\{0,1\}}
|a(b_1,\ldots,b_n)|^2=v_jv_j^\dagger$.
Thus, the total squared norm of all amplitudes extending a
fixed prefix is determined entirely by the corresponding
two-component memory vector $v_j$. This identity will also
be useful in the finite-interval construction.
	
	\subsection{Branch effects and the relative memory update}
	
	Fix one step and write its two branch maps as $A_0,A_1$. The norm-preserving condition gives
	\begin{equation}\label{eq:effects-main}
		E_0=A_0A_0^\dagger,\qquad
		E_1=A_1A_1^\dagger=I_2-E_0.
	\end{equation}
	These matrices are positive semidefinite and sum to the identity. We call them the \emph{branch effects}: for a normalized incoming memory row $\alpha$, the squared norm assigned to branch $b$ is
	$p_b(\alpha)=\alpha E_b\alpha^\dagger$.
	The term ``effect'' is used here only for this positive-operator description of branch norms; no measurement is performed. In particular, the scalar norm split at a particular prefix is recovered by evaluating the two quadratic forms on its current memory state.
	
	Since $E_1=I_2-E_0$, the two effects have a common orthonormal eigenbasis. Choose a unitary $Q$ such that
	$Q^\dagger E_0Q=\Lambda=\operatorname{diag}(\lambda_0,\lambda_1)$,
	$0\leq\lambda_a\leq1$.
	In this memory basis, the controlled rotation associated with basis state $\ket a$ allocates the two branch norms according to
	\begin{equation}\label{eq:controlled-angles}
		\theta_a=2\arccos\sqrt{\lambda_a},\qquad
		\Ry(\theta_a)\ket0=
		\sqrt{\lambda_a}\ket0+\sqrt{1-\lambda_a}\ket1,
		\quad a=0,1.
	\end{equation}
	For a coherent memory state $\alpha_0\ket0+\alpha_1\ket1$, the probability of output zero under this controlled rotation is
	$|\alpha_0|^2\lambda_0+|\alpha_1|^2\lambda_1$, and the probability of output one is its complement. The relative phase between the memory components is retained in the joint state. The controlled rotation therefore performs the norm allocation, but does not by itself specify the complete coherent memory update.
	
	That remaining information is contained in the polar factors of the branch maps. For a square matrix $C$, write its left polar decomposition as
	$C=DW$ and $D=(CC^\dagger)^{1/2}$.
	If $C$ is invertible, $W$ is uniquely determined and unitary. If $C$ is singular, the polar partial isometry can be extended to a unitary on the memory space; the extension is not fixed on the null space of $C$. This freedom will be relevant for the real geometric construction.
	
	A common change $T$ of outgoing memory basis right-multiplies both branch maps by the same unitary. It leaves the positive factors $D_b$ unchanged and replaces each polar unitary $W_b$ by $W_bT$. Therefore, the relative product
	$\mathcal R=W_1W_0^\dagger$
	is independent of this common basis choice. The following proposition records the resulting reduction.
	
	\begin{proposition}[Reduction to a single relative memory update]\label{prop:relative-memory}
		Let $C_0,C_1$ be square branch maps on the same finite-dimensional memory space and suppose
		$C_0C_0^\dagger+C_1C_1^\dagger=I$. Write their polar decompositions as
		$C_b=D_bW_b$ and $D_b=(C_bC_b^\dagger)^{1/2}$
		where, in the singular case, each $W_b$ is a unitary extension of the polar partial isometry. Under a common change $T$ of outgoing memory basis, the branch maps become $C_bT=D_b(W_bT)$ and
		$(W_1T)(W_0T)^\dagger=W_1W_0^\dagger$.
		In particular, choosing $T=W_0^\dagger$ gives
		\begin{equation}\label{eq:relative-basis-form}
			C_0'=D_0,\qquad
			C_1'=D_1\mathcal R,
			\qquad \mathcal R=W_1W_0^\dagger.
		\end{equation}
		Thus the positive factors determine the branch norms, while a single relative unitary contains the remaining branch-dependent memory transformation.
	\end{proposition}
	\begin{proof}
		Right multiplication by a common unitary $T$ leaves each positive factor $D_b$ unchanged and replaces $W_b$ by $W_bT$. The relative product is therefore invariant. Taking $T=W_0^\dagger$ makes the zero-branch polar unitary the identity and leaves $W_1W_0^\dagger$ on the one-branch map, which gives \eqref{eq:relative-basis-form}.
	\end{proof}
	
	Finally, let $G=[g_0\;g_1]$. The terminal condition \eqref{eq:balanced-terminal} says $GG^\dagger=I_2$, so $G$ is unitary. With the row-amplitude convention used here, the last output bit is emitted by the one-qubit gate $G^T$. The terminal step therefore requires no additional two-qubit interaction. In the next section, we combine the controlled norm split with the relative memory update to obtain an explicit two-qubit implementation of each local step.

	\begin{example}[Constructing a norm-preserving representation]
		\label{ex:two-qubit-completion}
		
		Consider the normalized two-qubit state
		\[
		\ket{\psi}=
		\frac{\sqrt3}{2}\ket{00}
		+\frac12\ket{11}.
		\]
		We illustrate how its amplitudes determine part of the branch matrices and how the remaining entries can be chosen to satisfy the norm-preserving condition.
		Across the cut separating the first qubit from the second, the coefficient matrix is
		\[
		F^{(1)}
		=
		\begin{pmatrix}
			\sqrt3/2&0\\
			0&1/2
		\end{pmatrix}.
		\]
		Its rank is two, so the state is not a product state. The two suffix-amplitude vectors are $f_0=(\sqrt3/2,0)$ and $f_1=(0,1/2)$. Their normalized versions are the computational basis vectors. We therefore choose the initial memory vector and terminal columns as
		$\lambda=(1,0)$, $g_0=e_0$, and $g_1=e_1$, where $e_0=(1,0)^T$ and $e_1=(0,1)^T$. In particular, the terminal matrix is $G=I_2$.
		The amplitude representation
		$a(b_1b_2)=\lambda A_1^{(b_1)}g_{b_2}$
		now determines the first row of each branch matrix. Indeed,
		\[
		\lambda A_1^{(0)}
		=\left(\frac{\sqrt3}{2},0\right),
		\qquad
		\lambda A_1^{(1)}
		=\left(0,\frac12\right).
		\]
		Thus, at this stage, the matrices have the form
		\[
		A_1^{(0)}=
		\begin{pmatrix}
			\sqrt3/2&0\\ x_0&x_1
		\end{pmatrix},
		\qquad
		A_1^{(1)}=
		\begin{pmatrix}
			0&1/2\\ y_0&y_1
		\end{pmatrix},
		\]
		where the second rows remain to be chosen.
		To impose norm preservation, form the horizontal concatenation $B_1=[\,A_1^{(0)}\;A_1^{(1)}\,]$. The condition in \eqref{eq:balanced-update} is equivalent to $B_1B_1^\dagger=I_2$, so the two rows of $B_1$ must be orthonormal. Its first row is already fixed and has unit norm:
		\[
		r_1=\left(\frac{\sqrt3}{2},0,0,\frac12\right).
		\]
		We may therefore choose any unit vector orthogonal to $r_1$ as the second row. One convenient choice is
		$r_2=(0,1/2,\sqrt3/2,0)$.
		Separating the resulting four entries into two $2\times2$ blocks gives
		\begin{equation}
				A_1^{(0)}
				=
				\begin{pmatrix}
					\sqrt3/2&0\\
					0&1/2
				\end{pmatrix},
				\qquad
				A_1^{(1)}
				=
				\begin{pmatrix}
					0&1/2\\
					\sqrt3/2&0
				\end{pmatrix}.
			\label{eq:two-qubit-completion}
		\end{equation}
		By construction, these matrices satisfy
		$A_1^{(0)}A_1^{(0)\dagger}
		+A_1^{(1)}A_1^{(1)\dagger}=I_2$.
		Moreover, contraction with $\lambda$, $g_0$, and $g_1$ reproduces the four target amplitudes exactly.
		We next examine the branch effects. They are
		\[
		E_0=
		\begin{pmatrix}3/4&0\\0&1/4\end{pmatrix},
		\qquad
		E_1=
		\begin{pmatrix}1/4&0\\0&3/4\end{pmatrix}.
		\]
		The computational memory basis already diagonalizes both effects. The corresponding positive polar factors are
		$D_0=\operatorname{diag}(\sqrt3/2,1/2)$ and
		$D_1=\operatorname{diag}(1/2,\sqrt3/2)$.
		The polar decompositions are $A_1^{(0)}=D_0I_2$ and
		$A_1^{(1)}=D_1X$, where
		$X=\begin{psmallmatrix}0&1\\1&0\end{psmallmatrix}$.
		Therefore, the relative memory update is $\mathcal R=X$.
		
		The eigenvalues of $E_0$ are $3/4$ and $1/4$. Hence the two controlled-rotation angles from \eqref{eq:controlled-angles} are $\theta_0=\pi/3$ and $\theta_1=2\pi/3$. Notice that the branch norm distribution depends on the incoming memory state: for a normalized memory row $\alpha=(\alpha_0,\alpha_1)$, the probability weight assigned to branch zero is $3|\alpha_0|^2/4+|\alpha_1|^2/4$.
		
		The completion is not unique. For example, choosing instead the second row
		$r_2'=(0,\sqrt3/2,1/2,0)$ gives
		\[
		\widetilde A_1^{(0)}=\frac{\sqrt3}{2}I_2,
		\qquad
		\widetilde A_1^{(1)}=\frac12X.
		\]
		This pair also reproduces the target state and satisfies norm preservation, but its effects are $(3/4)I_2$ and $(1/4)I_2$. Thus the first branch norm is now independent of the incoming memory state. The two choices describe the same target amplitudes but lead to different local implementations.
		
		This example illustrates the central construction principle: the target amplitudes determine the action of the branch matrices on the initially occupied memory direction, while the norm-preserving condition constrains how that action is extended to the entire memory space. The freedom in this completion can subsequently be exploited to simplify the branch effects and relative memory updates.
		
	\end{example}

	\section{Local two-qubit implementation}\label{sec:local-circuit}

	\subsection{A circuit for one norm-preserving update}
	
	Fix one step and suppress its position index. By Section~\ref{sec:scalar}, in an incoming memory basis that diagonalizes the branch effect $E_0$, and after a common choice of outgoing memory coordinates, the branch maps take the form $\widetilde A_0=D_0=\sqrt{\Lambda}$ and $\widetilde A_1=D_1\mathcal R$, where $D_1=\sqrt{I_2-\Lambda}$ and $\Lambda=\operatorname{diag}(\lambda_0,\lambda_1)$. The corresponding rotation angles are $\theta_a=2\arccos\sqrt{\lambda_a}$ for $a=0,1$, as in \eqref{eq:controlled-angles}.
	
	Use these angles to define the two-angle multiplexed rotation $U_{\mathrm{rot}}=\sum_{a=0}^1|a\rangle\!\langle a|_M\otimes R_y(\theta_a)$, with the current memory $M$ as selector and the blank register qubit $B$ as target. On a computational-basis memory state, it acts as $U_{\mathrm{rot}}|a\rangle_M|0\rangle_B=|a\rangle_M(\sqrt{\lambda_a}|0\rangle_B+\sqrt{1-\lambda_a}|1\rangle_B)$. The operation therefore allocates the two branch norms coherently, without measuring or dephasing the memory. It does not, however, implement the remaining memory update in $\widetilde A_1$. Since amplitudes are represented by row vectors, that update is implemented on memory kets by $\mathcal R^T$.
	
	\begin{proposition}[A two-qubit realization of one update]\label{prop:local-gate}
		Suppose the branch maps have the form just described. The basis-adjusted local isometry $|a\rangle_M|0\rangle_B\mapsto\sum_{b,\gamma=0}^1(\widetilde A_b)_{a\gamma}|b\rangle_{\mathrm{out}}|\gamma\rangle_{\mathrm{mem}}$ is implemented by the following operations, applied in order: (i) apply $U_{\mathrm{rot}}$ to the current memory and the blank register qubit; (ii) swap the two wires, so that the newly generated bit occupies the current memory wire and the old memory state moves to the next wire; and (iii) apply $\mathcal R^T$ to the new memory, controlled on the emitted bit being one. No correction is applied when the emitted bit is zero. The resulting circuit is unitary on the full two-qubit space and agrees with the stated isometry whenever the second input qubit is $|0\rangle$.
	\end{proposition}
	
	\begin{proof}
		After the multiplexed rotation, the coefficient of $|a\rangle_M|b\rangle_B$ is $(D_b)_{aa}$. The SWAP places $b$ on the current memory wire, which becomes the output wire, and moves the old memory label $a$ to the next wire. For $b=0$, no further operation is applied, so the resulting coefficient is $(D_0)_{aa}\delta_{a\gamma}=(\widetilde A_0)_{a\gamma}$. For $b=1$, the controlled application of $\mathcal R^T$ gives $(D_1)_{aa}(\mathcal R^T)_{\gamma a}=(D_1)_{aa}\mathcal R_{a\gamma}=(D_1\mathcal R)_{a\gamma}=(\widetilde A_1)_{a\gamma}$. Each operation is unitary, so the construction specifies a unitary on the full two-qubit space, not merely an isometry on the subspace with blank input $|0\rangle_B$.
	\end{proof}
	
	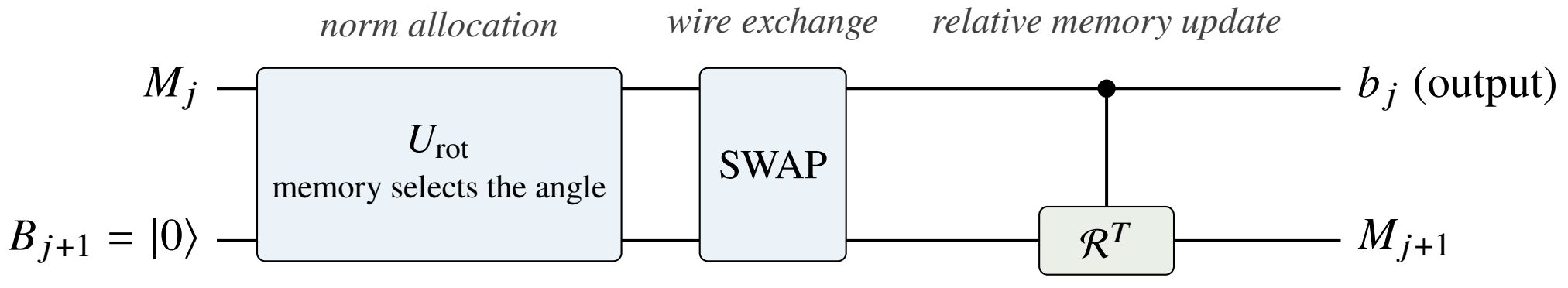
\begin{figure}[tbp]
		\centering
		\begin{tikzpicture}[x=1cm,y=1cm,>=Latex,
			wire/.style={draw=black,line width=0.75pt},
			op/.style={draw,rounded corners=2pt,fill=MidnightBlue!5,
				minimum height=1.65cm,minimum width=2.15cm,
				align=center,font=\small},
			swapop/.style={draw,rounded corners=2pt,fill=MidnightBlue!5,
				minimum height=1.65cm,minimum width=1.25cm,
				align=center,font=\small},
			stage/.style={font=\footnotesize\itshape,text=black!75},
			target/.style={draw,rounded corners=2pt,fill=OliveGreen!8,
				minimum width=1.15cm,minimum height=0.58cm,
				align=center,font=\small}]
			\draw[wire] (0.45,0.65)--(10.05,0.65);
			\draw[wire] (0.45,-0.65)--(10.05,-0.65);
			\node[left] at (0.45,0.65) {$M_j$};
			\node[left] at (0.45,-0.65) {$B_{j+1}=|0\rangle$};
			\node[stage] at (2.35,1.2) {norm allocation};
			\node[op] at (2.35,0) {$U_{\mathrm{rot}}$\\[-1pt]
				{\scriptsize memory selects the angle}};
			\node[stage] at (5.2,1.2) {wire exchange};
			\node[swapop] at (5.2,0) {SWAP};
			\node[stage] at (8.05,1.2) {relative memory update};
			\fill (8.05,0.65) circle (2.2pt);
			\draw[wire] (8.05,0.65)--(8.05,-0.36);
			\node[target] at (8.05,-0.65) {$\mathcal R^T$};
			\node[right,align=left] at (10.05,0.65) {$b_j$ (output)};
			\node[right,align=left] at (10.05,-0.65) {$M_{j+1}$};
		\end{tikzpicture}
		\caption{One local update. The upper wire initially carries the current memory $M_j$, while the lower wire is the blank register qubit $B_{j+1}$. The multiplexed rotation allocates the branch norms according to the memory value. The SWAP emits the bit $b_j$ on the upper wire and moves the old memory state to the lower wire. The final controlled gate applies $\mathcal R^T$ to that memory only when $b_j=1$, producing the updated memory $M_{j+1}$.}
		\label{fig:local-step}
	\end{figure}

\subsection{Sequential composition and the alternating output--memory formulation}
\label{subsec:sequential-memory}

The local circuit of Proposition~\ref{prop:local-gate} has a useful interpretation when applied successively along the register. At each step, a two-dimensional memory produces one output qubit and transfers its remaining amplitude information to the next register qubit. Thus the memory moves from left to right through the register, while the qubits already passed by the memory retain the output bits that have been generated.

We describe this process first in terms of local isometries and then in terms of the physical two-qubit circuit. This distinction clarifies the connection between the norm-preserving matrix representation and its circuit implementation.

\subsubsection{The local isometry}

Fix a step $j$, and let $M_j$ denote the incoming two-dimensional memory, $P_j$ the output qubit produced at this step, and $M_{j+1}$ the updated memory. The two branch matrices define the linear map
\begin{equation}
	\mathcal V_j
	=
	\ket{0}_{P_j}\otimes(A_j^{(0)})^T
	+
	\ket{1}_{P_j}\otimes(A_j^{(1)})^T,
	\label{eq:sequential-isometry}
\end{equation}
from $\mathcal H_{M_j}$ to $\mathcal H_{P_j}\otimes\mathcal H_{M_{j+1}}$. The transposes appear because the amplitude representation in \eqref{eq:balanced-amplitude} uses row vectors, whereas memory states in the circuit are represented by columns.
For an incoming memory row vector $\alpha$, the corresponding memory ket has coefficient column $\alpha^T$. The action of the local map is therefore
\[
\mathcal V_j\ket{\alpha^T}
=
\ket{0}\otimes(A_j^{(0)})^T\alpha^T
+
\ket{1}\otimes(A_j^{(1)})^T\alpha^T.
\]
The two terms describe the two possible output bits and their associated updated memory states. Importantly, both terms remain in coherent superposition; no measurement of the output bit or the memory is performed.
The norm-preserving identity \eqref{eq:balanced-update} guarantees that $\mathcal V_j$ is an isometry. Indeed, orthogonality of the two output states gives
\[
\mathcal V_j^\dagger\mathcal V_j
=
\left(
A_j^{(0)}A_j^{(0)\dagger}
+
A_j^{(1)}A_j^{(1)\dagger}
\right)^T
=I_2.
\]
Thus each local update preserves the norm of every incoming memory state, not merely that of the particular state encountered along one computational branch.

\subsubsection{Successive updates and the alternating process}

We now apply the local isometries successively. Initially, the memory $M_1$ is prepared in the state whose computational-basis amplitudes are the entries of $\lambda$. We denote this state by $\ket{\lambda}_{M_1}$, with coefficient column $\lambda^T$. All other register qubits are initially in $\ket{0}$.
After the first update, the state is
\[
\ket{\Psi_1}=\mathcal V_1\ket{\lambda}_{M_1}
=
\sum_{b_1\in\{0,1\}}
\ket{b_1}_{P_1}\otimes
\ket{\bigl(\lambda A_1^{(b_1)}\bigr)^T}_{M_2}.
\]
The first output bit has been produced, while the memory now occupies the next register position.

At the second step, $\mathcal V_2$ acts only on the memory, leaving the first output qubit unchanged. Thus $\ket{\Psi_2}=(I_{P_1}\otimes\mathcal V_2)\ket{\Psi_1}$, and explicitly
\[
\ket{\Psi_2}
=
\sum_{b_1,b_2\in\{0,1\}}
\ket{b_1b_2}_{P_1P_2}
\otimes
\ket{\bigl(\lambda A_1^{(b_1)}A_2^{(b_2)}\bigr)^T}_{M_3}.
\]
Here and below, a row vector inside the ket notation denotes the corresponding memory state with its entries interpreted as computational-basis amplitudes.

The same argument applies at every subsequent position. To make the structure explicit, for a prefix $p=b_1\cdots b_j$ define the row vector
$v_p=\lambda A_1^{(b_1)}\cdots A_j^{(b_j)}$.
An induction on $j$ then gives
\begin{equation}
		\ket{\Psi_j}
		=
		\sum_{b_1,\ldots,b_j\in\{0,1\}}
		\ket{b_1\cdots b_j}_{P_1\cdots P_j}
		\otimes
		\ket{v_{b_1\cdots b_j}^{\,T}}_{M_{j+1}}.
	\label{eq:sequential-prefix-state}
\end{equation}
This formula describes the alternating output--memory process. The output prefix is stored in the qubits already passed by the memory, whereas the current memory state retains the information needed to generate the remaining suffix amplitudes. In particular, the memory dimension remains two at every stage, regardless of the number of qubits already emitted.
Equivalently, the first $n-1$ updates can be written as the operator composition
\begin{equation}
	\begin{split}
		\ket{\Psi_{n-1}}
		={}&
		\left(I_{P_1\cdots P_{n-2}}\otimes\mathcal V_{n-1}\right)
		\cdots 
		\quad\cdots
		\left(I_{P_1}\otimes\mathcal V_2\right)
		\mathcal V_1\ket{\lambda}_{M_1}.
	\end{split}
	\label{eq:sequential-composition}
\end{equation}
The identity factors are essential: each new isometry acts on the current memory and the next output qubit, leaving all previously generated output bits unchanged.

\subsubsection{The terminal operation}

After the first $n-1$ updates, the first $n-1$ output bits have been generated, but the final memory is still a two-dimensional quantum state. The last output bit is obtained by converting this memory directly into the computational basis of the final register qubit.
Recall that $G=[g_0\;g_1]$. The terminal condition \eqref{eq:balanced-terminal} gives $GG^\dagger=I_2$, so $G$ is unitary. Therefore, $G^T$ is also unitary and can be applied as a single-qubit gate.

For a fixed output prefix $p=b_1\cdots b_{n-1}$, the corresponding memory row vector is $v_p=\lambda A_1^{(b_1)}\cdots A_{n-1}^{(b_{n-1})}$. Applying $G^T$ to its coefficient column produces the two-component vector $(v_pg_0,v_pg_1)^T$. These components are precisely the amplitudes associated with the two possible values of the final bit. Therefore, the complete state is
\begin{equation*}
	\ket{\psi}
	=
	\left(I_{P_1\cdots P_{n-1}}\otimes G^T\right)
	\ket{\Psi_{n-1}} = 
	\sum_{\mathbf b\in\{0,1\}^n}
	\left[
	\lambda A_1^{(b_1)}\cdots
	A_{n-1}^{(b_{n-1})}g_{b_n}
	\right]\ket{b_1\cdots b_n}.
\end{equation*}
Thus the sequential construction recovers exactly the amplitude representation \eqref{eq:balanced-amplitude}. The terminal operation does not require another two-qubit interaction: it simply converts the final memory into the last output qubit.

\subsubsection{Physical realization: the moving-memory circuit}

The operator description above has a direct interpretation in terms of the physical register wires. At step $j$, the current memory occupies the physical wire $P_j$, while the next wire $P_{j+1}$ is blank. The two-qubit circuit of Proposition~\ref{prop:local-gate} acts on this pair. It emits the bit $b_j$ on the first wire and transfers the updated memory to the second wire.

More precisely, in the compatible memory bases selected for the local construction, the two-qubit unitary $U_j$ satisfies
\[
U_j\left(\ket{a}_{P_j}\ket{0}_{P_{j+1}}\right)
=
\sum_{b,\gamma=0}^1
\bigl(\widetilde A_j^{(b)}\bigr)_{a\gamma}
\ket{b}_{P_j}\ket{\gamma}_{P_{j+1}},
\]
where $\widetilde A_j^{(b)}$ denotes the branch matrix in the incoming and outgoing memory bases used by Proposition~\ref{prop:local-gate}. The matrix coefficient is exactly the amplitude associated with emitting bit $b$ and updating the memory from $a$ to $\gamma$.
The physical sequence is therefore
\[
\begin{array}{ccl}
	P_1:\ \text{memory}
	&\longrightarrow&
	P_1:\ \text{output bit }b_1,\\
	P_2:\ \text{blank}
	&\longrightarrow&
	P_2:\ \text{memory},
\end{array}
\]
followed by the same operation on the pair $(P_2,P_3)$, and then on $(P_3,P_4)$, and so forth. The role of the physical wires alternates: a wire first receives the updated memory and subsequently becomes an output wire when the next local operation is applied.
In terms of the physical two-qubit unitaries, the complete preparation circuit is
\[
\begin{split}
	\ket{\psi}
	={}&
	(G^T)_{P_n}
	U_{n-1}^{(P_{n-1},P_n)}
	\cdots U_2^{(P_2,P_3)}
	U_1^{(P_1,P_2)} \times
	\left(
	\ket{\lambda}_{P_1}
	\otimes\ket{0}_{P_2}\otimes\cdots\otimes
	\ket{0}_{P_n}
	\right).
\end{split}
\label{eq:sequential-physical-circuit}
\]
The rightmost local unitary acts first. Each $U_j$ is the two-qubit realization described in Proposition~\ref{prop:local-gate}, with the required memory-basis changes included in its local implementation. No additional memory register or ancillary qubit is introduced: one of the original register qubits carries the memory until the next output bit is generated.

\subsubsection{The role of branch effects and relative memory updates}

The preceding discussion also clarifies why both the branch effects and the relative memory unitary are needed in the local construction.
At each step, the effects $E_j^{(0)}$ and $E_j^{(1)}$ determine how the squared norm of an incoming memory state is distributed between the two output branches. In their common eigenbasis, the positive polar factors $D_{j,0}$ and $D_{j,1}$ are implemented by the memory-controlled rotation of Proposition~\ref{prop:local-gate}. This operation performs the required norm allocation, but does not by itself reproduce the complete branch matrices.

The remaining coherent transformation is specified by the relative unitary $\mathcal R_j$. After the SWAP has placed the emitted bit on the current output wire and moved the old memory to the next wire, a controlled application of $\mathcal R_j^T$ supplies the required memory update on the branch where the emitted bit is one. No such correction is needed on the zero branch.
Thus each local circuit has three conceptually distinct operations: a controlled rotation to allocate the branch norms, a SWAP to exchange the roles of the two physical wires, and a controlled relative memory update to complete the coherent transformation. Their combination realizes the full local isometry, rather than merely reproducing its branch probabilities.

\subsubsection{Compatibility of memory bases}

The local diagonalizations and polar decompositions must be expressed in compatible memory coordinates at successive cuts. This is possible because the amplitude representation is invariant under a change of basis in an intermediate memory space.
Indeed, for a unitary $T_j$ acting at the cut between steps $j$ and $j+1$, replace the two adjacent updates by $A_j^{(b)}T_j$ and $T_j^\dagger A_{j+1}^{(b)}$, respectively. Their product is unchanged:
\[
\bigl(A_j^{(b_j)}T_j\bigr)
\bigl(T_j^\dagger A_{j+1}^{(b_{j+1})}\bigr)
=
A_j^{(b_j)}A_{j+1}^{(b_{j+1})}.
\]
Hence every target amplitude in \eqref{eq:balanced-amplitude} remains unchanged.
This freedom permits the outgoing memory basis at one step to be chosen as the incoming memory basis at the next. The initial vector $\lambda$ and terminal columns $g_0,g_1$ are expressed in the corresponding endpoint bases. The resulting local circuits therefore form a compatible sequence acting on a single moving memory qubit.

The alternating formulation reveals the central mechanism of the construction: each update converts the current memory into an output bit and a new memory, while retaining all information required for the remaining suffix. The branch effects specify the norm distribution, and the relative unitary supplies the remaining coherent memory transformation. Together, these give an exact sequential realization of the norm-preserving amplitude representation.

\begin{theorem}[Exact preparation from norm-preserving local steps]
	\label{thm:balanced-steps}
	Let $\lambda$, $A_j^{(b)}$, and $g_b$ satisfy the
	norm-preserving conditions
	\eqref{eq:balanced-boundary}--\eqref{eq:balanced-terminal},
	with all updates expressed in compatible memory coordinates.
	Starting from $\ket{0}^{\otimes n}$, prepare the first
	register qubit in the state specified by $\lambda$, apply
	the local circuit of Proposition~\ref{prop:local-gate}
	successively for $j=1,\ldots,n-1$, and finish by applying
	$G^T$, where $G=[g_0\;g_1]$, to the remaining memory.
	The resulting state has amplitude
	$\lambda A_1^{(b_1)}\cdots A_{n-1}^{(b_{n-1})}g_{b_n}$
	on $\ket{b_1\cdots b_n}$. Thus the construction prepares
	exactly the normalized amplitude profile
	\eqref{eq:balanced-amplitude} using one moving memory
	qubit and no additional ancillas.
\end{theorem}

\begin{proof}
	By \eqref{eq:sequential-prefix-state}, after the first
	$n-1$ local updates, the memory associated with the
	output prefix $p=b_1\cdots b_{n-1}$ has coefficient
	column
	$\left(\lambda A_1^{(b_1)}\cdots
	A_{n-1}^{(b_{n-1})}\right)^T$.
	The terminal gate $G^T$ converts this memory into the
	last output qubit. Its component corresponding to
	$b_n$ is therefore
	$\lambda A_1^{(b_1)}\cdots A_{n-1}^{(b_{n-1})}g_{b_n}$,
	which is precisely the amplitude in
	\eqref{eq:balanced-amplitude}. The norm-preserving
	conditions guarantee normalization, and the sequential
	circuit uses only the original register qubits.
\end{proof}

	Algorithm~\ref{alg:balanced-step} summarizes the matrix factorization and gate construction for one pair of branch updates.
	
		\begin{algorithm}[H]
			\caption{Constructing one local norm-distribution circuit}
			\label{alg:balanced-step}
			\begin{algorithmic}[1]
				\Require Norm-preserving updates $A_0,A_1$ and reference incoming/outgoing memory bases $Q_{\rm in},Q_{\rm out}$
				\Require The incoming memory is expressed in the $Q_{\rm in}$ coordinates
				
				\State Form $\widehat A_b=Q_{\rm in}^\dagger A_bQ_{\rm out}$ for $b=0,1$
				
				\State Compute $E_0=\widehat A_0\widehat A_0^\dagger$ and choose $Q$ such that $Q^\dagger E_0Q=\Lambda$
				
				\State Set $Q_{\rm in}^{\rm new}=Q_{\rm in}Q$ and $C_b=Q^\dagger\widehat A_b$, with
				$D_0=\sqrt{\Lambda}$ and $D_1=\sqrt{I_2-\Lambda}$
				
				\State Take left polar factorizations $C_b=D_bW_b$, extending $W_b$ to a unitary when $C_b$ is singular
				
				\State Set $\mathcal R=W_1W_0^\dagger$ and choose the common outgoing change $T=W_0^\dagger$
				
				\State Replace $C_b$ by $C_bT$, so that $C_0T=D_0$ and $C_1T=D_1\mathcal R$
				
				\State Set $Q_{\rm out}^{\rm new}=Q_{\rm out}T$
				
				\State For $a=0,1$, set $\theta_a=2\arccos\sqrt{\Lambda_{aa}}$
				
				\State Apply $Q^T$ to the incoming memory, unless this change has been absorbed into the preceding step's outgoing basis
				
				\State Apply $U_{\mathrm{rot}}$, swap the wires, and apply $\mathcal R^T$ to the new memory controlled on output one
				
				\Ensure The local circuit realizes the branch maps
				$C_bT=Q_{\rm in}^{\rm new\dagger}A_bQ_{\rm out}^{\rm new}$.
				Pass $Q_{\rm out}^{\rm new}$ as the reference incoming basis to the next step.
			\end{algorithmic}
		\end{algorithm}
	
The diagonalizing unitary $Q$ changes the incoming memory
basis from $Q_{\rm in}$ to $Q_{\rm in}^{\rm new}
=Q_{\rm in}Q$. Since amplitudes are represented by row
vectors whereas memory states are represented by columns,
this change is implemented on the memory ket by $Q^T$.
It may be applied immediately before $U_{\mathrm{rot}}$,
or absorbed into the preceding step by choosing its
outgoing basis to coincide with $Q_{\rm in}^{\rm new}$.
The common right change $T$ updates the outgoing memory
basis to $Q_{\rm out}^{\rm new}=Q_{\rm out}T$, which is
then used as the reference incoming basis at the next
step. Thus the successive local updates remain compatible
and require no additional memory qubits. In the real
examples below, the basis changes and polar factors are
planar rotations or reflections.

\subsection{Rank-one branches and reflection completions}
	
	If one branch map has rank one, one diagonal entry of its positive polar factor is zero. The corresponding row of the polar unitary does not affect the branch map and can be chosen to complete that unitary. In a real two-dimensional memory, this freedom can be used to choose $\det\mathcal R=-1$. The relative update then belongs to $O(2)$ and is a reflection. Every such reflection has the form $R(\phi)ZR(-\phi)$, where $Z=\operatorname{diag}(1,-1)$ and $R(\phi)=\begin{psmallmatrix}\cos\phi&-\sin\phi\\\sin\phi&\cos\phi\end{psmallmatrix}$ is a planar rotation. It is therefore symmetric. Hence, the controlled memory-ket operation $\mathcal R^T=\mathcal R$ is locally equivalent to controlled-$Z$ and hence to CNOT.
	
	For later use, define $c=\sqrt{\zeta}$ and $s=\sqrt{1-\zeta}$, where $\zeta\in(0,1)$, and set
	\begin{equation}\label{eq:reflections}
		J_{\zeta}=\begin{pmatrix}c&s\\s&-c\end{pmatrix},\qquad
		K_{\zeta}=\begin{pmatrix}-s&c\\c&s\end{pmatrix}.
	\end{equation}
	Both are orthogonal reflections. If $\eta=\arccos\sqrt{\zeta}$, then $J_{\zeta}=R(\eta/2)ZR(-\eta/2)$ and $K_{\zeta}=R(\pi/4+\eta/2)ZR(-\pi/4-\eta/2)$. In the geometric comparison transitions, the nonzero polar row fixes one of these reflections exactly; the other row is free. The four choices are listed in \cref{tab:geometric-reflections}.

	\begin{remark}[A SWAP-free construction on a star topology]
		The moving-memory construction can also be implemented without
		SWAP gates by keeping the memory on a fixed register qubit.
		Specifically, place the initial memory state $\ket{\lambda}$ on
		$P_n$ and initialize $P_1,\ldots,P_{n-1}$ to $\ket{0}$.
		At step $j$, apply the local norm-distribution rotation to
		the pair $(P_n,P_j)$, with $P_n$ as the memory qubit and
		$P_j$ as the output qubit, followed by the relative memory
		update $\mathcal R_j^T$ controlled on the output qubit being
		one. The resulting local operation has the action
		\[
		\ket{a}_{P_n}\ket{0}_{P_j}
		\longmapsto
		\sum_{b,\gamma=0}^1
		(\widetilde A_j^{(b)})_{a\gamma}
		\ket{\gamma}_{P_n}\ket{b}_{P_j}.
		\]
		Thus the output bit is written directly onto $P_j$, while
		the updated memory remains on $P_n$. Applying these operations
		successively for $j=1,\ldots,n-1$, and then applying $G^T$
		to $P_n$, prepares exactly the state
		\eqref{eq:balanced-amplitude}. This construction uses no
		SWAP gates and requires only $n-1$ two-qubit interactions,
		provided the hardware supports direct interactions between
		the fixed memory qubit and every other register qubit.
		It is therefore naturally suited to a star connectivity.
		On a linear nearest-neighbour architecture, routing the
		interactions to the fixed memory qubit may require additional
		SWAP gates, so the moving-memory construction remains
		preferable in that setting.
	\end{remark}

	\begin{remark}[Conceptual decomposition versus joint synthesis]
		Proposition~\ref{prop:local-gate} gives a transparent unitary extension of the required local isometry: a multiplexed norm split, a wire exchange, and one branch-dependent memory operation. This displayed extension need not itself have determinant one. For resource accounting, Section~\ref{sec:synthesis} changes its action only on the unused input subspace to obtain an equivalent $SO(4)$ completion, without changing its action when the blank output qubit is in $|0\rangle$. That completion admits a two-CNOT synthesis. Thus the three pieces in the figure explain the construction, while the jointly synthesized $SO(4)$ completion provides the native-gate benchmark.
	\end{remark}

	\subsection{A complete three-qubit construction and circuit}
	\label{subsec:three-qubit-example}
	
	We illustrate the complete construction using a three-qubit state. The example demonstrates how successive pairs of norm-preserving matrices are obtained from the target amplitudes, how their branch effects determine the local rotations, and how the resulting updates are combined into an exact quantum circuit.
	
	Consider the normalized state
	\begin{equation}
		\ket{\psi}
		=
		\frac{1}{\sqrt2}\ket{000}
		+\frac12\ket{011}
		+\frac14\ket{101}
		+\frac{\sqrt3}{4}\ket{110}.
		\label{eq:three-qubit-target}
	\end{equation}
	We seek a representation of the form
	$a(b_1b_2b_3)=\lambda A_1^{(b_1)}A_2^{(b_2)}g_{b_3}$,
	with a two-dimensional memory.
	
	\paragraph{The first cut and construction of the first update.}
	
	Separate the first qubit from the remaining two. In the suffix ordering $00,01,10,11$, the two suffix-amplitude vectors are
	$f_0=(1/\sqrt2,0,0,1/2)$ and
	$f_1=(0,1/4,\sqrt3/4,0)$.
	Their squared norms are $W(0)=3/4$ and $W(1)=1/4$. The corresponding normalized vectors are
	\[
	u_0=\left(\sqrt{\frac23},0,0,\frac1{\sqrt3}\right),
	\qquad
	u_1=\left(0,\frac12,\frac{\sqrt3}{2},0\right).
	\]
	These vectors are orthonormal, so the first cut has Schmidt rank two. We choose the memory basis so that its two basis directions represent $u_0$ and $u_1$, respectively.
	
	Take $\lambda=(1,0)$ and $g_0=e_0$, $g_1=e_1$. Matching the target amplitudes then fixes the first rows of the two branch matrices:
	\[
	\lambda A_1^{(0)}
	=\left(\frac{\sqrt3}{2},0\right),
	\qquad
	\lambda A_1^{(1)}
	=\left(0,\frac12\right).
	\]
	As in Example~\ref{ex:two-qubit-completion}, we combine the two matrices horizontally. The first row of
	$B_1=[\,A_1^{(0)}\;A_1^{(1)}\,]$
	is $r_1=(\sqrt3/2,0,0,1/2)$.
	We now choose the orthogonal unit row
	$r_2=(0,\sqrt3/2,1/2,0)$.
	This is the alternative completion discussed in Example~\ref{ex:two-qubit-completion}. It gives
	\begin{equation}
			A_1^{(0)}=\frac{\sqrt3}{2}I_2,
			\qquad
			A_1^{(1)}=\frac12X.
		\label{eq:three-qubit-A1}
	\end{equation}
	The first norm-preserving identity follows immediately:
	$A_1^{(0)}A_1^{(0)\dagger}=(3/4)I_2$ and
	$A_1^{(1)}A_1^{(1)\dagger}=(1/4)I_2$.
	
	\paragraph{Construction of the second update.}
	
	We next determine the updates associated with the second bit. Expressing the two normalized suffix vectors in terms of the last two qubits gives
	\[
	u_0=\sqrt{\frac23}\ket{00}
	+\frac1{\sqrt3}\ket{11},
	\qquad
	u_1=\frac12\ket{01}
	+\frac{\sqrt3}{2}\ket{10}.
	\]
	The rows of the second pair of matrices correspond to the incoming memory states $\ket0,\ket1$, while their columns correspond to the outgoing memory states, identified with the final output bit through $g_0$ and $g_1$.
	For incoming memory state $\ket0$, the second bit is zero with final bit zero and amplitude $\sqrt{2/3}$, whereas the second bit is one with final bit one and amplitude $1/\sqrt3$. This determines the first rows of $A_2^{(0)}$ and $A_2^{(1)}$ as $(\sqrt{2/3},0)$ and $(0,1/\sqrt3)$, respectively.
	
	For incoming memory state $\ket1$, the corresponding possibilities are second bit zero with final bit one and amplitude $1/2$, and second bit one with final bit zero and amplitude $\sqrt3/2$. Thus the second rows are $(0,1/2)$ and $(\sqrt3/2,0)$. We obtain
	\begin{equation}
			A_2^{(0)}
			=
			\begin{pmatrix}
				\sqrt{2/3}&0\\
				0&1/2
			\end{pmatrix},
			\qquad
			A_2^{(1)}
			=
			\begin{pmatrix}
				0&1/\sqrt3\\
				\sqrt3/2&0
			\end{pmatrix}.
		\label{eq:three-qubit-A2}
	\end{equation}
	The norm-preserving identity can be verified directly. Indeed,
	$A_2^{(0)}A_2^{(0)\dagger}=\operatorname{diag}(2/3,1/4)$ and
	$A_2^{(1)}A_2^{(1)\dagger}=\operatorname{diag}(1/3,3/4)$, so their sum is $I_2$. Together with the first-step identity and the terminal condition $g_0g_0^\dagger+g_1g_1^\dagger=I_2$, all the required conditions are satisfied.
	
	\paragraph{Verification of the amplitudes.}
	
	The four nonzero amplitudes obtained from the matrix representation are
	\[
	\begin{aligned}
		a(000)&=\frac{\sqrt3}{2}\sqrt{\frac23}
		=\frac1{\sqrt2},&
		a(011)&=\frac{\sqrt3}{2}\frac1{\sqrt3}
		=\frac12,\\
		a(101)&=\frac12\frac12
		=\frac14,&
		a(110)&=\frac12\frac{\sqrt3}{2}
		=\frac{\sqrt3}{4}.
	\end{aligned}
	\]
	All other amplitudes vanish. Thus the representation reproduces exactly the target state \eqref{eq:three-qubit-target}.
	
	\paragraph{Branch effects and relative memory updates.}
	
	We now determine the local circuit parameters. At the first step, the branch effects are $E_1^{(0)}=(3/4)I_2$ and $E_1^{(1)}=(1/4)I_2$. Both are already diagonal in the computational memory basis. The two rotation angles are therefore equal: $\theta_{1,0}=\theta_{1,1}=\pi/3$. The positive polar factors are $D_{1,0}=(\sqrt3/2)I_2$ and $D_{1,1}=(1/2)I_2$. Since $A_1^{(0)}=D_{1,0}$ and $A_1^{(1)}=D_{1,1}X$, the relative memory update is $\mathcal R_1=X$.
	At the second step, the branch effects are
	\[
	E_2^{(0)}
	=
	\begin{pmatrix}2/3&0\\0&1/4\end{pmatrix},
	\qquad
	E_2^{(1)}
	=
	\begin{pmatrix}1/3&0\\0&3/4\end{pmatrix}.
	\]
	Again, the computational memory basis diagonalizes both effects. The two controlled-rotation angles are
	\[
	\theta_{2,0}=2\arccos\sqrt{\frac23},
	\qquad
	\theta_{2,1}=\frac{2\pi}{3}.
	\]
	The positive polar factors are
	$D_{2,0}=\operatorname{diag}(\sqrt{2/3},1/2)$ and
	$D_{2,1}=\operatorname{diag}(1/\sqrt3,\sqrt3/2)$.
	Since $A_2^{(0)}=D_{2,0}$ and $A_2^{(1)}=D_{2,1}X$, the relative memory update is again $\mathcal R_2=X$.
	Thus the two steps have the same relative memory operation, although their norm distributions differ. The first uses an unconditional rotation, while the second requires a genuinely memory-controlled rotation.
	
	\paragraph{Physical circuit construction.}
	
	We now implement the two updates on the physical register qubits $P_1,P_2,P_3$. Initially, $P_1$ carries the memory state $\ket0$, while $P_2$ and $P_3$ are blank.
	
	For the first update, apply $R_y(\pi/3)$ to $P_2$. Since the two rotation angles are equal, no control is required. Next, swap $P_1$ and $P_2$, and apply a CNOT with control $P_1$ and target $P_2$. This CNOT implements the relative memory update $\mathcal R_1^T=X$ when the emitted first bit is one. The resulting state is
	\[
	\left(
	\frac{\sqrt3}{2}\ket{00}
	+\frac12\ket{11}
	\right)_{P_1P_2}
	\otimes\ket0_{P_3}.
	\]
	The first output bit is now stored in $P_1$, and $P_2$ carries the updated memory.
	For the second update, the memory qubit $P_2$ selects the rotation applied to the blank qubit $P_3$. The multiplexed rotation is
	\[
	U_{\mathrm{rot},2}
	=
	\ket0\!\bra0_{P_2}\otimes R_y(\theta_{2,0})_{P_3}
	+
	\ket1\!\bra1_{P_2}\otimes R_y(\theta_{2,1})_{P_3}.
	\]
	After applying this rotation, the state is
	\[
	\frac1{\sqrt2}\ket{000}
	+\frac12\ket{001}
	+\frac14\ket{110}
	+\frac{\sqrt3}{4}\ket{111}.
	\]
	We next swap $P_2$ and $P_3$. The state becomes
	\[
	\frac1{\sqrt2}\ket{000}
	+\frac12\ket{010}
	+\frac14\ket{101}
	+\frac{\sqrt3}{4}\ket{111}.
	\]
	Finally, apply a CNOT with control $P_2$ and target $P_3$. This realizes $\mathcal R_2^T=X$ on the branch where the second output bit is one, giving
	\begin{equation}
		\boxed{
			\ket{\psi}
			=
			\frac1{\sqrt2}\ket{000}
			+\frac12\ket{011}
			+\frac14\ket{101}
			+\frac{\sqrt3}{4}\ket{110}.
		}
		\label{eq:three-qubit-circuit-result}
	\end{equation}
	This is precisely the target state \eqref{eq:three-qubit-target}. Since $G=I_2$, the terminal memory is already expressed in the computational basis of the third output qubit, and no further gate is necessary.
	
\begin{figure}[tbp]
		\centering
		\resizebox{0.98\textwidth}{!}{%
			\begin{tikzpicture}[
				x=1cm,y=1cm,>=Latex,
				wire/.style={draw=black,line width=0.75pt},
				gate/.style={
					draw,rounded corners=2pt,fill=white,
					minimum height=0.65cm,minimum width=1.65cm,
					align=center,font=\small
				},
				swapmark/.style={font=\large},
				target/.style={
					circle,draw=black,fill=white,
					minimum size=5mm,inner sep=0pt,font=\small
				}
				]
				% Register wires
				\draw[wire] (0.4,2)--(14.5,2);
				\draw[wire] (0.4,1)--(14.5,1);
				\draw[wire] (0.4,0)--(14.5,0);
				
				% Input labels
				\node[left] at (0.4,2) {$P_1:\ket0$ (memory)};
				\node[left] at (0.4,1) {$P_2:\ket0$ (blank)};
				\node[left] at (0.4,0) {$P_3:\ket0$ (blank)};
				
				% First update
				\node[font=\footnotesize\itshape] at (3.8,2.7)
				{First local update};
				
				\node[gate,minimum width=1.5cm] at (2.5,1)
				{$R_y(\pi/3)$};
				
				% First SWAP
				\draw[wire] (4.8,1)--(4.8,2);
				\node[swapmark] at (4.8,2) {$\times$};
				\node[swapmark] at (4.8,1) {$\times$};
				
				% First CNOT
				\fill (6.3,2) circle (2pt);
				\draw[wire] (6.3,2)--(6.3,1);
				\node[target] at (6.3,1) {$\oplus$};
				
				% Second update
				\node[font=\footnotesize\itshape] at (10.3,2.7)
				{Second local update};
				
				% Controlled multiplexed rotation
				\fill (8.2,1) circle (2pt);
				\draw[wire] (8.2,1)--(8.2,0.38);
				\node[gate,minimum width=2.35cm] at (8.2,0)
				{$U_{\mathrm{rot},2}$};
				
				% Second SWAP
				\draw[wire] (11.1,0)--(11.1,1);
				\node[swapmark] at (11.1,1) {$\times$};
				\node[swapmark] at (11.1,0) {$\times$};
				
				% Second CNOT
				\fill (12.7,1) circle (2pt);
				\draw[wire] (12.7,1)--(12.7,0);
				\node[target] at (12.7,0) {$\oplus$};
				
				% Output labels
				\node[right] at (14.5,2) {$b_1$};
				\node[right] at (14.5,1) {$b_2$};
				\node[right] at (14.5,0) {$b_3$};
				
			\end{tikzpicture}%
		}
		\caption{Exact three-qubit preparation using two successive norm-preserving updates. The first update uses an unconditional rotation, followed by a SWAP and a CNOT. The second uses a memory-controlled two-angle rotation, followed by another SWAP and CNOT. The final memory already represents the third output bit because the terminal matrix is the identity.}
		\label{fig:three-qubit-circuit}
	\end{figure}
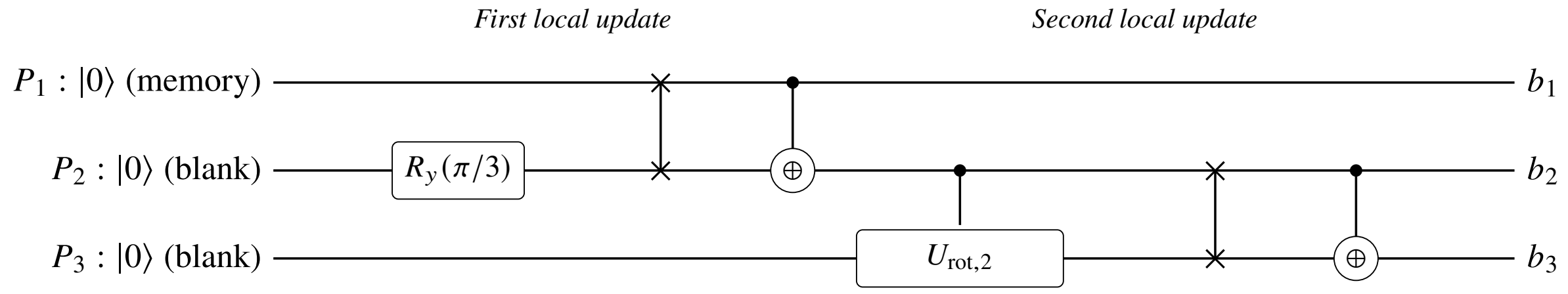

	The complete circuit is shown in \Cref{fig:three-qubit-circuit}. It consists of two successive local updates. Each update comprises a norm-distribution operation, a SWAP, and a controlled relative memory update. The first norm-distribution operation is an ordinary one-qubit rotation, whereas the second is a memory-controlled rotation. The example therefore illustrates both forms of local norm allocation within a single exact three-qubit preparation.
	
	This example illustrates how the choice of orthonormal completion influences the local circuit. The first completion produces branch effects proportional to the identity, eliminating the need for a memory-controlled rotation at the first step. The second update has nontrivial diagonal effects and therefore requires two different rotation angles. In both cases, the relative memory operation is $X$. The example makes explicit the connection between amplitude factorization, norm-preserving matrix updates, and their sequential quantum-circuit realization.

	\section{Shifted geometric states}\label{sec:geometric}

\subsection{Target state and removal of product factors}

Let $N=2^m$, where $m$ is the number of qubits, let
$0\leq k<N$, and let $r\in\mathbb C\setminus\{0\}$.
The integer $k$ is the first allowed basis label, and $r$
is the ratio between consecutive amplitudes. We consider
the normalized shifted geometric state
\begin{equation}\label{eq:geom-target}
	\ket{\Gamma_{m,k}(r)}
	=
	\frac{1}{\sqrt{Z_{m,k}(r)}}
	\sum_{x=k}^{N-1}r^{x-k}\ket{x},
	\qquad
	Z_{m,k}(r)=\sum_{j=0}^{N-k-1}|r|^{2j}.
\end{equation}
Write $r=\sigma e^{\mathrm{i}\varphi}$, with $\sigma>0$.
The phase can be restored after preparing the positive-amplitude
state by applying the product of one-qubit gates
\[
D_\varphi=
\bigotimes_{j=1}^{m}
\operatorname{diag}
\left(1,e^{\mathrm{i}2^{m-j}\varphi}\right).
\]
Indeed, writing the computational-basis label as
$x=\sum_{j=1}^{m}2^{m-j}b_j$, we have
$D_\varphi\ket{x}=e^{\mathrm{i}\varphi x}\ket{x}$.
It follows that
$D_\varphi\ket{\Gamma_{m,k}(\sigma)}
=e^{\mathrm{i}\varphi k}\ket{\Gamma_{m,k}(r)}$.
The factor $e^{\mathrm{i}\varphi k}$ is global and has no
physical effect. It is therefore sufficient to construct
the state for $\sigma>0$.

The unshifted case $k=0$ is a product state. Writing
$x=\sum_{j=0}^{m-1}2^j b_j$ gives
$\sigma^x=\prod_{j=0}^{m-1}\sigma^{2^j b_j}$ and hence
\begin{equation}\label{eq:geom-product}
	\ket{\Gamma_{m,0}(\sigma)}
	=
	\bigotimes_{j=m-1}^{0}
	\frac{\ket{0}+\sigma^{2^j}\ket{1}}
	{\sqrt{1+\sigma^{2^{j+1}}}}.
\end{equation}
Here the tensor factors are ordered from the most significant
qubit to the least significant qubit. For numerical stability,
when $\sigma^{2^j}>1$, the corresponding unnormalized factor
may instead be written as
$\sigma^{-2^j}\ket{0}+\ket{1}$, with its normalization
adjusted accordingly.

Now suppose $k>0$. Write $\nu=v_2(k)$, so that
$k=2^\nu a$ with $a$ odd, and set
$\ell=m-\nu$ and $M=2^\ell$. Decompose each supported
index as $x=2^\nu u+s$, where $0\leq s<2^\nu$.
Since $k=2^\nu a$, the condition $x\geq k$ is equivalent
to $u\geq a$. Moreover,
\[
\sigma^{x-k}=\sigma^s R^{u-a},
\qquad R=\sigma^{2^\nu}.
\]
Thus the $\nu$ least significant qubits separate into
one-qubit product factors, while the remaining $\ell$
qubits carry the shifted geometric state
$\ket{\Gamma_{\ell,a}(R)}$.
The normalized low-order factors are
\begin{equation}\label{eq:low-factors}
	\ket{\eta_j(\sigma)}
	=
	\begin{cases}
		\dfrac{\ket{0}+\sigma^{2^j}\ket{1}}
		{\sqrt{1+\sigma^{2^{j+1}}}},
		&\sigma\leq1,\\[8pt]
		\dfrac{\sigma^{-2^j}\ket{0}+\ket{1}}
		{\sqrt{1+\sigma^{-2^{j+1}}}},
		&\sigma>1,
	\end{cases}
	\qquad 0\leq j<\nu.
\end{equation}
The second expression is obtained by rescaling the
unnormalized factor by $\sigma^{-2^j}$ and is useful when
$\sigma>1$. In the original register ordering, these factors
occupy the $\nu$ least significant qubits, in the order
$\ket{\eta_{\nu-1}(\sigma)}\otimes\cdots
\otimes\ket{\eta_0(\sigma)}$.

Let $L=M-a$ be the number of supported values in the
reduced register. If $L=1$, then $a=M-1$, and the reduced
register is simply $\ket{1}^{\otimes\ell}$. The entire state
is therefore a product state. Assume henceforth that $L>1$.
Since $M$ is even and $a$ is odd, $L$ is odd. Define
$Q=\lfloor\log_2(L-1)\rfloor$, $t=Q+1$,
$c=2^t-L$, and $m_{\rm fix}=\ell-t$.
Since $2^{t-1}<L<2^t$, the integer $c$ is odd and satisfies
$1\leq c<2^{t-1}$. Also, $t\leq\ell$, so $m_{\rm fix}\geq0$.
Using $M=2^\ell$, we obtain
$a=M-L=(2^{m_{\rm fix}}-1)2^t+c$.
Therefore, the $m_{\rm fix}$ most significant bits of
the reduced register are fixed to one, and the remaining
$t$ qubits form the active core
\begin{equation}\label{eq:geom-core-state}
	\ket{\Gamma_{t,c}(R)}
	=
	\frac{1}{\sqrt{Z_{t,c}(R)}}
	\sum_{x=c}^{2^t-1}R^{x-c}\ket{x}.
\end{equation}
In particular, the full state factors as
$\ket{1}^{\otimes m_{\rm fix}}\otimes
\ket{\Gamma_{t,c}(R)}\otimes
\ket{\eta_{\nu-1}(\sigma)}\otimes\cdots
\otimes\ket{\eta_0(\sigma)}$,
with empty tensor factors omitted. Thus all non-product
structure is confined to the $t$-qubit active core.

It remains to reduce the active-core problem to a ratio
no larger than one. Put
$h=2^\nu\log\sigma$ and $\tau=e^{-|h|}\in(0,1]$.
If $h\leq0$, then $R=e^h=\tau$, and the active core is
a decreasing upper tail supported on
$\{c,\ldots,2^t-1\}$. If $h>0$, then $R=e^h=\tau^{-1}$.
In this case, reverse the active-core index by setting
$y=2^t-1-x$. Since $L=2^t-c$, we have
$x-c=L-1-y$, and therefore
$R^{x-c}=\tau^{-(L-1)}\tau^y$.
The first factor is independent of $y$ and hence contributes
only an overall normalization. It is therefore sufficient
to prepare the decreasing initial-segment state
proportional to $\sum_{y=0}^{L-1}\tau^y\ket{y}$ and then
apply $X^{\otimes t}$.
Thus the active-core preparation reduces to one of two
forms: a decreasing upper tail with threshold $c$, or a
decreasing initial segment of length $L$. In both cases,
the ratio satisfies $0<\tau\leq1$.

	\subsection{Binary comparison and the scalar norm recurrence}
	
	We now explain how the support of a geometric state can be
	handled through a sequence of binary comparisons. Recall
	that we need to prepare either an upper-tail state, supported
	on $c\leq y<2^t$, or an initial-segment state, supported on
	$0\leq y<L$. The number $c$ is the inclusive lower threshold
	in the first case, while $L$ is the exclusive upper endpoint
	in the second.
	
	The main idea is to read the binary digits of the basis label
	from most significant to least significant and compare them
	with the corresponding digits of the threshold. At any stage,
	the comparison has three possible outcomes. The bits read so
	far may agree with the threshold, in which case the comparison
	is still undecided; they may already place the label inside
	the allowed region, in which case all remaining choices are
	allowed; or they may place it outside the allowed region, in
	which case no continuation is supported. The last possibility
	has zero amplitude and need not be represented. Thus only
	two comparison states are needed: an undecided state and an
	already-accepted state. These are the two states carried by
	the memory qubit.
	
	We next determine how the squared norm is divided between
	the allowed and excluded parts at each comparison step.
	The geometric form of the amplitudes makes this division
	particularly simple: it can be described by a single real
	parameter. This scalar description will determine the
	rotation angles of the circuit, while the corresponding
	coherent transformations of the memory will be considered
	in the next subsection.
	
	For $q\geq0$ and $0\leq d\leq2^q$, define
	$T_q(d)=\sum_{y=0}^{d-1}\tau^{2y}$ and
	$S_q(d)=\sum_{y=d}^{2^q-1}\tau^{2y}$.
	Thus $T_q(d)$ is the squared norm of the initial segment
	$0\leq y<d$, while $S_q(d)$ is the squared norm of the
	upper tail $d\leq y<2^q$. Empty sums are zero. For
	$\tau\ne1$, the corresponding finite-series expressions are
	$T_q(d)=(1-\tau^{2d})/(1-\tau^2)$ and
	$S_q(d)=\tau^{2d}(1-\tau^{2(2^q-d)})/(1-\tau^2)$.
	When $\tau=1$, they reduce to $d$ and $2^q-d$,
	respectively. We also write
	$G_q=T_q(2^q)=S_q(0)$ for the squared norm of the
	complete $q$-qubit geometric profile.
	
	Consider a comparison involving $q\geq1$ bits, with residual
	threshold $d$. Write $d=\kappa 2^{q-1}+d_0$, where
	$\kappa\in\{0,1\}$ is the most significant bit of the
	threshold and $0\leq d_0<2^{q-1}$. The current output bit
	is compared with $\kappa$, and $d_0$ is the threshold that
	remains if the two bits agree. The two halves of the register
	are weighted differently: choosing the current output bit
	to be one multiplies every amplitude in the corresponding
	half by $\gamma=\tau^{2^{q-1}}$. Its squared-norm contribution
	is therefore multiplied by $\gamma^2$.
	
	Let $G=G_{q-1}$. We denote by $S$ the squared norm in the
	allowed part of the lower half and by $T$ the squared norm
	in its excluded part. For an upper-tail comparison these
	are $S=S_{q-1}(d_0)$ and $T=T_{q-1}(d_0)$; for an
	initial-segment comparison their roles are interchanged,
	so that $S=T_{q-1}(d_0)$ and $T=S_{q-1}(d_0)$.
	In either case, $S+T=G$. Let $S_{\rm par}$ and
	$T_{\rm par}$ denote the corresponding allowed and excluded
	squared norms before the current threshold bit is processed.
	The normalized parameter is $\zeta=S/(S+T)$.
	Splitting the two halves according to the current threshold
	bit gives the following recurrence.
	
	\begin{proposition}[Scalar threshold recurrence]
		\label{prop:geom-scalar}
		Let $\zeta=S/(S+T)$ and $\gamma\in(0,1]$. The normalized
		allowed squared norm at the parent stage is
		\begin{equation}\label{eq:geom-p-recurrence}
			\zeta_{\rm par}=
			\begin{cases}
				\dfrac{\zeta+\gamma^2}{1+\gamma^2},
				&\text{upper tail},\quad \kappa=0,\\[5pt]
				\dfrac{\gamma^2\zeta}{1+\gamma^2},
				&\text{upper tail},\quad \kappa=1,\\[5pt]
				\dfrac{\zeta}{1+\gamma^2},
				&\text{initial segment},\quad \kappa=0,\\[5pt]
				\dfrac{1+\gamma^2\zeta}{1+\gamma^2},
				&\text{initial segment},\quad \kappa=1.
			\end{cases}
		\end{equation}
		In particular, the norm distribution at each comparison
		step is determined entirely by the current scalar $\zeta$
		and the geometric factor $\gamma$.
	\end{proposition}
	
	\begin{proof}
		The parent squared-norm contributions can be obtained by
		adding the contributions from the two halves. For an
		upper-tail comparison with $\kappa=0$, the entire upper
		half is allowed, whereas only the upper tail of the lower
		half is allowed. Hence
		$(S_{\rm par},T_{\rm par})=(S+\gamma^2G,T)$.
		For $\kappa=1$, the entire lower half is excluded, while
		only the upper tail of the upper half is allowed. Thus
		$(S_{\rm par},T_{\rm par})=(\gamma^2S,G+\gamma^2T)$.
		
		For an initial-segment comparison with $\kappa=0$, only
		the initial segment of the lower half is allowed, giving
		$(S_{\rm par},T_{\rm par})=(S,T+\gamma^2G)$.
		When $\kappa=1$, the entire lower half is allowed, together
		with the initial segment of the upper half, so that
		$(S_{\rm par},T_{\rm par})=(G+\gamma^2S,\gamma^2T)$.
		
		In all four cases,
		$S_{\rm par}+T_{\rm par}=(1+\gamma^2)G$.
		Dividing $S_{\rm par}$ by this total and using
		$\zeta=S/G$ gives \eqref{eq:geom-p-recurrence}.
	\end{proof}
	
	The recurrence provides the scalar norm information needed
	at each stage of the comparison. Unlike the general
	norm-preserving matrix construction of Section~\ref{sec:scalar},
	the geometric case requires only one real parameter to
	determine the local norm split. The next subsection uses
	this parameter to construct the coherent memory updates
	and the corresponding controlled rotation angles.

\subsection{Explicit local gates}

The scalar recurrence determines how the squared norm is
divided between the two output branches. A quantum circuit
must also preserve the coherence of the memory and update
it according to the branch that is produced. We now construct
the corresponding local two-qubit operations.

Consider one nonterminal comparison step. Let $S$ and $T$
be the allowed and excluded squared norms in the lower half,
respectively, and let $S'$ and $T'$ be the corresponding
parent squared norms. As before, put $G=S+T$ and
$\zeta=S/G$. For convenience, write
$\alpha=\sqrt S$, $\beta=\sqrt T$,
$\alpha'=\sqrt{S'}$, and $\beta'=\sqrt{T'}$.

In suitable orthonormal memory coordinates, the
norm-preserving branch maps for the four comparison cases
are given below. These are the branch maps after normalizing
the comparison states at the parent and child stages.

\begin{center}
	\small
	\begin{tabularx}{\textwidth}{@{}c c X X@{}}
		\toprule
		Support form & $\kappa$ & $C_0$ & $C_1$\\
		\midrule
		Upper tail & $0$
		& $\operatorname{diag}(\alpha/\alpha',
		\beta/\beta')$
		& $\begin{psmallmatrix}
			\gamma\alpha/\alpha'&\gamma\beta/\alpha'\\
			0&0
		\end{psmallmatrix}$\\[4pt]
		Upper tail & $1$
		& $\begin{psmallmatrix}
			0&0\\
			\alpha/\beta'&\beta/\beta'
		\end{psmallmatrix}$
		& $\operatorname{diag}(1,\gamma\beta/\beta')$\\[4pt]
		Initial segment & $0$
		& $\operatorname{diag}(1,\beta/\beta')$
		& $\begin{psmallmatrix}
			0&0\\
			\gamma\alpha/\beta'&\gamma\beta/\beta'
		\end{psmallmatrix}$\\[4pt]
		Initial segment & $1$
		& $\begin{psmallmatrix}
			\alpha/\alpha'&\beta/\alpha'\\
			0&0
		\end{psmallmatrix}$
		& $\operatorname{diag}(\gamma\alpha/\alpha',1)$\\
		\bottomrule
	\end{tabularx}
\end{center}

Each pair satisfies the norm-preserving identity
$C_0C_0^T+C_1C_1^T=I_2$. Thus the two branch maps together
define an isometry from the incoming memory space to the
output-bit and updated-memory spaces.
We first determine the rotation that distributes the norm
between the two branches. The zero-branch effect is
$E_0=C_0C_0^T$. Its eigenvalues, in the memory coordinates
used above, are

\begin{equation}\label{eq:geom-lambdas}
	\begin{array}{c|c}
		\text{Case}&(\lambda_0,\lambda_1)\\ \hline
		\text{Upper tail},\ \kappa=0 &(S/S',1)\\
		\text{Upper tail},\ \kappa=1 &(0,G/T')\\
		\text{Initial segment},\ \kappa=0 &(1,T/T')\\
		\text{Initial segment},\ \kappa=1 &(G/S',0).
	\end{array}
\end{equation}

The corresponding two rotation angles are
$\theta_j=2\arccos\sqrt{\lambda_j}$, for $j=0,1$.
In every case, one eigenvalue is zero or one. The
corresponding memory basis state therefore selects a
deterministic output branch, with rotation angle either
$\pi$ or zero. The other memory basis state carries the
nontrivial norm split. Its rotation angle is

\begin{equation}\label{eq:geom-active-angle}
	\theta_{\rm act}=
	\begin{cases}
		2\arctan(\gamma/\sqrt{\zeta}),
		&\text{upper tail},\ \kappa=0,\\
		2\arctan(\gamma\sqrt{1-\zeta}),
		&\text{upper tail},\ \kappa=1,\\
		2\arctan(\gamma/\sqrt{1-\zeta}),
		&\text{initial segment},\ \kappa=0,\\
		2\arctan(\gamma\sqrt{\zeta}),
		&\text{initial segment},\ \kappa=1.
	\end{cases}
\end{equation}

It remains to specify the coherent transformation of the
memory. For this purpose, consider the left polar
decompositions $C_b=D_bW_b$, where
$D_b=(C_bC_b^T)^{1/2}$ and $W_b$ is an orthogonal
extension of the polar partial isometry when $C_b$ is
singular. Since one branch map has rank one in each of the
four cases, its polar factor is not uniquely determined
on the null space. We use this freedom to choose a
particularly simple relative memory update.

For $0<\zeta<1$, put
$\eta=\arccos\sqrt{\zeta}$, so that
$\cos\eta=\sqrt{\zeta}$ and
$\sin\eta=\sqrt{1-\zeta}$. Recall the two reflections
$J_\zeta$ and $K_\zeta$ defined in
\eqref{eq:reflections}. The following table gives
convenient orthogonal extensions of the polar factors.

\begin{table}[t]
	\centering
	\small
	\caption{Reflection-based polar completions for the four
		geometric comparison transitions. For a rank-one branch,
		the unused orthogonal direction is chosen so that the
		relative update is a reflection.}
	\label{tab:geometric-reflections}
	\begin{tabular}{@{}l c c c@{}}
		\toprule
		Case & $W_0$ & $W_1$ & $\mathcal R=W_1W_0^T$\\
		\midrule
		Upper tail, $\kappa=0$
		& $I_2$ & $J_\zeta$ & $J_\zeta$\\
		Upper tail, $\kappa=1$
		& $K_\zeta$ & $I_2$ & $K_\zeta$\\
		Initial segment, $\kappa=0$
		& $I_2$ & $K_\zeta$ & $K_\zeta$\\
		Initial segment, $\kappa=1$
		& $J_\zeta$ & $I_2$ & $J_\zeta$\\
		\bottomrule
	\end{tabular}
\end{table}
For the chosen polar-unitary extensions, the relative
update is $\mathcal R=W_1W_0^T$. It is invariant under
a common change of outgoing memory basis. In the common
basis in which $W_0=I_2$, the branch maps have the form
$C_0'=D_0$ and $C_1'=D_1\mathcal R$. Since the amplitudes
are represented by row vectors, the corresponding
operation on memory kets is $\mathcal R^T$, controlled
on the output bit being one.

For a rank-one branch map, only the action of its polar
unitary on the support of the positive polar factor is
fixed; the action on the orthogonal direction is free.
The table uses this freedom to make $\mathcal R$ one of
the two reflections. Both $J_\zeta$ and $K_\zeta$ are
symmetric orthogonal reflections. Each is a one-qubit
conjugate of $Z$, so its controlled implementation is
a controlled-$Z$ conjugated by one-qubit memory rotations.
The choice of reflection changes the polar completion,
but not the branch effects, their eigenvalues, or the
controlled-rotation angles in
\eqref{eq:geom-lambdas}--\eqref{eq:geom-active-angle}.

For the two-qubit updates considered here, the residual
threshold is odd. At every nonterminal step, at least
two comparison bits remain, and the threshold in the
lower half is strictly between zero and the size of that
half. Therefore $0<\zeta<1$, as required for the
reflection formulas. The final remaining bit is handled
by the terminal one-qubit gate.

\subsection{Direct preparation and optimal interaction count}

The comparison circuit processes the bits of the
$t$-qubit register from most significant to least
significant, beginning with the memory in the
undecided-comparison state $\ket0$. At each nonterminal
step, the current threshold bit $\kappa$ selects the
appropriate case in \cref{tab:geometric-reflections}.
If $q$ bits remain to be processed, including the current
bit, then $\gamma=\tau^{2^{q-1}}$. The parent norm
parameter is obtained from the child parameter using
\eqref{eq:geom-p-recurrence}.

Because the recurrence expresses a parent parameter in
terms of the corresponding child parameter, all scalar
parameters can be computed from the terminal comparison
back toward the root before the quantum circuit is
applied. The local operations are then applied in the
forward, most-significant-bit-first order. Each operation
uses the two-angle memory-controlled rotation and the
corresponding relative memory reflection, with the wire
exchange in the explicit local-gate factorization.
After the $t-1$ two-qubit updates, a terminal one-qubit
gate completes the final output bit. In the
initial-segment case, applying $X^{\otimes t}$ reverses
the index, as described above.

For interaction counting, each local update is regarded
as one complete real two-qubit gate acting across the
corresponding adjacent pair of register qubits. The
controlled rotation, wire exchange, and controlled
reflection describe a useful factorization of this gate;
they are not counted as three separate interactions.
As discussed in Section~\ref{sec:synthesis}, the complete
local isometry can be extended to an equivalent
$SO(4)$ gate and jointly synthesized.

\begin{theorem}[Shifted-geometric preparation]
	\label{thm:geom-main}
	Let $N=2^m$, $0\leq k<N$, and $r\neq0$. The state
	$\ket{\Gamma_{m,k}(r)}$ can be prepared without additional
	ancillas using exactly
	\begin{equation}\label{eq:geom-interactions}
		E_m(k)=
		\begin{cases}
			0,& k=0\text{ or }k=N-1,\\[2pt]
			\max\!\left\{
			0,\left\lfloor\log_2(N-k-1)\right\rfloor-v_2(k)
			\right\},
			&1\leq k<N-1,
		\end{cases}
	\end{equation}
	real adjacent two-qubit interactions on the prescribed
	linear ordering. The states for $k=0$ and $k=N-1$ are
	product states. More generally, for an interior offset,
	if $N-k$ is a power of two, the support is an aligned
	dyadic block and the state is also a product state.
	For every remaining offset, the construction uses one
	adjacent interaction across each non-product cut and
	none across a product cut.
\end{theorem}

\begin{proof}
	The cases $k=0$ and $k=N-1$ are product states. Now
	suppose $1\leq k<N-1$, and write
	$\nu=v_2(k)$ and $k=2^\nu a$, with $a$ odd.
	The low-order product reduction leaves a register of
	$\ell=m-\nu$ qubits, with $M=2^\ell$ and
	$L=M-a=(N-k)/2^\nu$ supported values. Since $L$ is odd,
	there are two cases.
	
	If $L=1$, then $a=M-1$ and the reduced register is
	the basis state $\ket{M-1}=\ket1^{\otimes\ell}$.
	Together with the independently prepared low-order
	qubits, the entire state is a product state. In this
	case $N-k=2^\nu$, and the expression in
	\eqref{eq:geom-interactions} is zero.
	
	Suppose henceforth that $L>1$. Set
	$Q=\lfloor\log_2(L-1)\rfloor$ and $t=Q+1$.
	Since $L$ is odd and $2^Q\leq L-1<2^{Q+1}$, we have
	$2^{\nu+Q}\leq N-k-1<2^{\nu+Q+1}$. Therefore,
	$Q=\lfloor\log_2(N-k-1)\rfloor-\nu$.
	The reduction leaves a $t$-qubit comparison register.
	The comparison construction uses one adjacent two-qubit
	interaction at each of its $t-1=Q$ internal boundaries.
	All other qubits are fixed or independently prepared
	one-qubit product factors. The phase-restoration layer
	$D_\varphi$ also consists entirely of one-qubit gates.
	This proves the required upper bound.
	
	For the lower bound, consider the cut separating the
	$q$ least significant qubits from the remaining register,
	and put $B=2^q$. The state is a product across this cut
	if $B$ divides $k$, since the support then begins at a
	block boundary, or if $k\geq N-B$, since the support
	then lies entirely in the final block.
	
	Suppose neither condition holds. Write $k=pB+s$,
	where $1\leq s<B$. Since $k<N-B$, the next block
	$p+1$ is also present. Consider the two rows of the
	cut coefficient matrix corresponding to the blocks
	$p$ and $p+1$, and the two columns corresponding to
	the lower-bit labels $s-1$ and $s$. The resulting
	$2\times2$ submatrix is
	$\frac{1}{\sqrt{Z_{m,k}(r)}}
	\begin{pmatrix}
		0&1\\
		r^{B-1}&r^B
	\end{pmatrix}$.
	Its determinant is
	$-r^{B-1}/Z_{m,k}(r)\neq0$. Hence the cut is
	non-product.
	It follows that the non-product cuts are precisely
	those satisfying
	$v_2(k)<q\leq\lfloor\log_2(N-k-1)\rfloor$.
	Their number is exactly $E_m(k)$. A nearest-neighbour
	two-qubit gate crosses only one such cut, so at least
	one interaction is necessary across each non-product
	cut. The construction uses exactly one across every
	such cut, establishing the matching lower bound.
\end{proof}

\begin{algorithm}[t]
	\caption{Shifted-geometric preparation by scalar norm distribution}
	\label{alg:geom-steps}
	\begin{algorithmic}[1]
		\Require Width $m$, offset $k$, nonzero ratio $r$
		\State Set $N=2^m$
		\If{$k=0$}
		\State Prepare the product state \eqref{eq:geom-product}
		\State Apply the phase layer $D_\varphi$ and stop
		\EndIf
		\If{$k=N-1$}
		\State Prepare $\ket1^{\otimes m}$ and stop
		\EndIf
		\State Set $\nu=v_2(k)$ and remove the $\nu$ low-order
		product factors
		\State Determine the reduced support length $L$ and
		the comparison-register parameters $t,c$
		\If{$L=1$}
		\State Prepare the remaining fixed qubits in
		$\ket1^{\otimes(m-\nu)}$
		\State Prepare the low-order one-qubit product factors
		\State Apply the phase layer $D_\varphi$ and stop
		\EndIf
		\State Set $\tau=\exp(-2^\nu|\log|r||)$
		\State Choose the upper-tail or initial-segment form
		according to the sign of $\log|r|$
		\State Compute the scalar norm parameters recursively
		from the terminal comparison toward the root
		\For{each of the first $t-1$ comparison steps,
			in most-significant-bit-first order}
		\State Set $\gamma=\tau^{2^{q-1}}$, where $q$
		bits remain, including the current bit
		\State Obtain the parent scalar from
		\eqref{eq:geom-p-recurrence}
		\State Select the polar completion from
		\cref{tab:geometric-reflections}
		\State Apply the corresponding local two-qubit update
		\EndFor
		\State Apply the terminal one-qubit gate
		\If{initial-segment form}
		\State Apply $X^{\otimes t}$ to reverse the index
		\EndIf
		\State Restore the fixed and low-order product factors
		\State Apply the phase layer $D_\varphi$
	\end{algorithmic}
\end{algorithm}

\begin{example}[A four-qubit shifted geometric state with offset five]
	\label{ex:geom-16}
	
	We illustrate the construction with $N=16$, $k=5$, and
	$r=1/2$. The target is the normalized state
	\[
	\ket{\Gamma_{4,5}(1/2)}
	=
	\frac{1}{\sqrt{Z}}
	\sum_{x=5}^{15}2^{-(x-5)}\ket{x},
	\qquad
	Z=\sum_{j=0}^{10}4^{-j}
	=\frac43(1-4^{-11}).
	\]
	Thus the first nonzero amplitude occurs at $\ket5$,
	and each successive amplitude is half the preceding one.
	
	We first examine the binary structure of the support.
	The threshold is $k=5=(0101)_2$, so the allowed labels
	are precisely the four-bit strings that represent integers
	at least $0101$. The comparison proceeds from left to
	right, starting with the most significant bit:
	\begin{itemize}
		\item If the first bit is $1$, the label is already
		greater than the threshold, and every choice of the
		remaining bits is allowed.
		\item If the first bit is $0$, the comparison continues.
		If the second bit is then $0$, the label is below the
		threshold and is rejected. If it is $1$, the comparison
		continues.
		\item With the first two bits equal to $01$, a third bit
		equal to $1$ places the label inside the allowed region.
		If it is $0$, the comparison remains undecided.
		\item In the final case, the first three bits are $010$.
		The last bit must then be $1$ for the label to be
		accepted.
	\end{itemize}
	For example, $1000$ is accepted immediately, $0011$ is
	rejected after the second bit, and $0101$ is accepted only
	after the final bit is read. This illustrates why the
	comparison requires only two memory states: one for a
	comparison that is still undecided and one for a comparison
	that has already been accepted. Rejected strings have no
	allowed continuation and carry zero amplitude.
	
	We now determine the parameters of the construction.
	Since $k=5$ is odd, $\nu=v_2(k)=0$, so no low-order
	product factors need to be removed. We have
	$L=N-k=11$, $Q=\lfloor\log_2(10)\rfloor=3$,
	$t=Q+1=4$, and $c=2^t-L=5$.
	There are no fixed leading qubits either. Thus the entire
	four-qubit register is the comparison register, and its
	support is the upper tail beginning at $c=5$.
	Since $r=1/2$ is positive and less than one, no index
	reversal or phase-restoration gate is required. In the
	notation of the construction, $\tau=1/2$, and the
	squared-amplitude ratio between successive labels is
	$\rho=\tau^2=1/4$.
	
	\paragraph{Computing the scalar norm parameters.}
	
	The threshold $0101$ is processed from most significant
	to least significant bit. The scalar parameters are
	calculated in the reverse direction, starting with the
	last comparison bit. We show the calculation explicitly.
	
	With one bit remaining, the residual threshold is $1$.
	The allowed label is $1$, whose squared amplitude is
	$\tau^2=1/4$, whereas the excluded label is $0$, whose
	squared amplitude is $1$. Therefore,
	$S_1(1)=1/4$ and $T_1(1)=1$, giving
	$\zeta= S/(S+T)=1/5$.
	
	With two bits remaining, the residual threshold is $01$,
	again corresponding to $d=1$. The allowed squared norm is
	$S_2(1)=1/4+1/16+1/64=21/64$, while
	$T_2(1)=1$. Hence the normalized allowed squared norm is
	$\zeta=21/85$.
	
	With three bits remaining, the residual threshold is
	$101$, corresponding to $d=5$. In this case,
	$S_3(5)=1/4^5+1/4^6+1/4^7=21/16384$, and
	$T_3(5)=1+1/4+1/16+1/64+1/256=341/256$.
	Therefore,
	$\zeta=21/21845$.
	
	These calculations give the child norm parameters needed
	by the three local two-qubit updates. The successive
	parent parameters follow from
	\eqref{eq:geom-p-recurrence}. In particular, the
	bottom-up recursion is
	\[
	\frac15
	\longmapsto
	\frac{21}{85}
	\longmapsto
	\frac{21}{21845}
	\longmapsto
	\frac{1398101}{1431655765}.
	\]
	The final value is the allowed squared-norm fraction
	for the complete four-bit comparison problem.
	
	\paragraph{The three local updates.}
	
	We now apply the gates in the forward direction, starting
	with the most significant bit. At each step, $q$ denotes
	the number of bits still to be processed, including the
	current bit, and
	$\gamma=\tau^{2^{q-1}}$. The threshold bits are
	$0,1,0$, respectively, for the three two-qubit updates.
	The fourth bit is handled by the terminal one-qubit gate.
	
	The required scalar parameters, rotation angles, and
	relative memory updates are summarized below. The angles
	$\theta_0,\theta_1$ refer to the two computational memory
	basis states in the diagonal basis of the zero-branch
	effect.
	
	\begin{center}
		\small
		\begin{tabular}{@{}c c c c c c@{}}
			\toprule
			$q$ & $\kappa$ & $\zeta$ & $\gamma$
			& $(\theta_0,\theta_1)$ & $\mathcal R$\\
			\midrule
			$4$ & $0$ & $\dfrac{21}{21845}$
			& $\dfrac1{256}$
			& $(0.2507,0)$ & $J_{21/21845}$\\[5pt]
			$3$ & $1$ & $\dfrac{21}{85}$
			& $\dfrac1{16}$
			& $(\pi,0.1084)$ & $K_{21/85}$\\[5pt]
			$2$ & $0$ & $\dfrac15$
			& $\dfrac14$
			& $(1.0195,0)$ & $J_{1/5}$\\
			\bottomrule
		\end{tabular}
	\end{center}
	
	The angles are in radians. They are obtained directly
	from \eqref{eq:geom-active-angle}. More explicitly, the
	three nontrivial angles are
	\[
	\begin{aligned}
		\theta_{\rm act}^{(4)}
		=2\arctan\left(
		\frac{1}{256}\sqrt{\frac{21845}{21}}\right),\quad
		\theta_{\rm act}^{(3)} 
		=2\arctan\left(\frac{1}{2\sqrt{85}}\right), \quad
		\theta_{\rm act}^{(2)} 
		=2\arctan\left(\frac{\sqrt5}{4}\right).
	\end{aligned}
	\]
	The other angle at each step is either zero or $\pi$,
	as shown in the table. Thus the first and third updates
	use a nontrivial rotation on one memory basis state and
	no rotation on the other. In the second update, one
	memory basis state selects the deterministic branch
	with angle $\pi$, while the other undergoes the
	nontrivial rotation.
	
	The relative memory transformations are obtained from
	\Cref{tab:geometric-reflections}. For the first and third
	updates they are $J_\zeta$, and for the second update
	they are $K_\zeta$. These reflections complete the
	coherent memory transformation after the norm-distribution
	rotation. In the explicit local factorization, the
	corresponding reflection is applied to the memory,
	controlled on the emitted output bit being one.
	
	\paragraph{The resulting circuit and interaction count.}
	
	The first update generates the most significant output
	bit, the second generates the next bit, and the third
	generates the third bit. The memory is passed successively
	through the four register qubits. After these three
	two-qubit updates, the remaining memory is converted
	into the final output bit by the terminal one-qubit gate.
	The circuit therefore uses interactions on the three
	adjacent pairs $(P_1,P_2)$, $(P_2,P_3)$, and $(P_3,P_4)$.
	All three corresponding cuts are non-product. Indeed,
	$v_2(5)=0$ and $\lfloor\log_2(16-5-1)\rfloor=3$,
	so the optimal interaction count is
	$E_4(5)=3$. The construction attains this count.
	Finally, since $r=1/2$ is real and positive, the
	phase-restoration layer is the identity. The resulting
	state is exactly $\ket{\Gamma_{4,5}(1/2)}$.
\end{example}

	\section{Single-mode Fourier states}\label{sec:fourier}
	
	\subsection{Target state and product-state reduction}
	
	Let $N=2^m$, let $\nu\in\{0,\ldots,N-1\}$ be a discrete frequency, and let $\phi\in\R$ be a phase offset. We consider the real sinusoidal profile sampled on the computational-basis labels, namely
	\begin{equation}\label{eq:fourier-target}
		\ket{F_{m,\nu,\phi}}=\frac1{\sqrt{Z_{m,\nu,\phi}}}
		\sum_{i=0}^{N-1}\sin\left(\frac{2\pi\nu i}{N}+\phi\right)\ket i,
		\qquad Z_{m,\nu,\phi}=\sum_{i=0}^{N-1}\sin^2\left(\frac{2\pi\nu i}{N}+\phi\right),
	\end{equation}
	whenever the unnormalised vector is nonzero. These are computational-basis states whose amplitudes follow a sampled sine wave; they are not states expressed in the Fourier basis. The cases $\nu=0$ and $\nu=N/2$ are product-state cases:
	when $\sin\phi\ne0$ they give, up to a global sign,
	$\ket+^{\otimes m}$ and
	$\ket+^{\otimes(m-1)}\otimes\ket-$, respectively.
	When $\sin\phi=0$, the target vector vanishes.
	For every other frequency, write $\nu=2^s u$ with $u$ odd and put $d=m-s-1$. Thus $d\geq1$. The phase depends only on the lowest $d+1$ bits of the label. Within those bits, changing the most significant bit changes the sign of every amplitude, because $u$ is odd. The remaining $s$ high-order bits are therefore uniform product factors, while this sign change produces one $\ket-$ factor. In the bit ordering used below, this gives
	\begin{equation}\label{eq:fourier-factor}
		\ket{F_{m,\nu,\phi}}=\ket+^{\otimes s}\otimes\ket-\otimes\ket{\Phi_{d,u,\phi}},
	\end{equation}
	where the active $d$-qubit state is
	\begin{equation}\label{eq:fourier-core}
		\ket{\Phi_{d,u,\phi}}=\frac1{\sqrt{2^{d-1}}}
		\sum_{x=0}^{2^d-1}\sin\left(\phi+\frac{2\pi u x}{2^{d+1}}\right)\ket x.
	\end{equation}
	The normalization follows by pairing the terms indexed by $x$ and $x+2^{d-1}$: their phases differ by $\pi u/2$, so their squared sine values sum to one. Therefore, $Z_{m,\nu,\phi}=2^{m-1}$ for every nontrivial frequency, independently of $\phi$.
	
	\subsection{Binary updates and the scalar norm formula}
	
	Write the active-core bits in the order $x=x_1\cdots x_d$, from most significant to least significant, and define $\theta_j=2\pi u/2^{j+1}$ for $1\leq j\leq d$. The contribution of bit $x_j$ to the phase is $x_j\theta_j$. With
	$R(\theta)=\begin{psmallmatrix}\cos\theta&-\sin\theta\\\sin\theta&\cos\theta\end{psmallmatrix}$,
	the identity $(\sin\alpha,\cos\alpha)R(\theta)=(\sin(\alpha+\theta),\cos(\alpha+\theta))$ shows that the amplitude profile is generated by the initial row vector $\lambda=(\sin\phi,\cos\phi)$, the branch maps $B_j^{(0)}=I$ and $B_j^{(1)}=R(\theta_j)$, and the terminal column $\rho=(1,0)^T$. In particular, the product selected by the bits of $x$ has terminal amplitude $\sin(\phi+\sum_jx_j\theta_j)$, which is exactly the coefficient in \eqref{eq:fourier-core}.
	
	The special feature of this family is that the remaining-norm data can be computed from one complex scalar. For $0\leq j\leq d$, let $m_j=2^{d-j}$ and define
	\begin{equation}\label{eq:fourier-P}
		P_j=\prod_{\ell=j+1}^{d}(1+\e^{2i\theta_\ell}),\qquad P_d=1.
	\end{equation}
Thus $P_{j-1}=P_j(1+\e^{2i\theta_j})$. To see why this product is useful, consider the remaining bits $x_{j+1},\ldots,x_d$. For a fixed choice $z=(z_{j+1},\ldots,z_d)$, write $\beta_z=\sum_{\ell=j+1}^d z_\ell\theta_\ell$. The amplitude associated with these remaining bits, for an incoming row vector $v$, is $vR(\beta_z)e_0$, where $e_0=(1,0)^T$. Summing the squared amplitudes over all possible choices of the remaining bits gives $vH_jv^T$, where
\[
H_j=\sum_{z\in\{0,1\}^{d-j}}
R(\beta_z)e_0e_0^TR(\beta_z)^T.
\]
Thus $H_j$ is the matrix that records the remaining squared norm for every incoming memory vector.
The number of terms in this sum is $m_j=2^{d-j}$. Moreover,
\[
\sum_z \e^{2i\beta_z}
=\prod_{\ell=j+1}^d(1+\e^{2i\theta_\ell})
=P_j.
\]
Using the definition of $R(\beta_z)$, we can therefore evaluate the matrix explicitly:
\[
H_j=\frac12
\begin{pmatrix}
	m_j+\operatorname{Re}P_j&\operatorname{Im}P_j\\
	\operatorname{Im}P_j&m_j-\operatorname{Re}P_j
\end{pmatrix}.
\]
For $j<d$, this matrix is positive definite. We choose $F_j$ to be its positive square root, $F_j=H_j^{1/2}$. The standard square-root formula for a positive $2\times2$ matrix gives a closed expression involving only $P_j$ and $m_j$. In particular, setting $\Delta_j=\sqrt{m_j^2-|P_j|^2}$, we obtain
	\begin{equation}\label{eq:fourier-F}
		F_j=\frac1{2\sqrt{m_j+\Delta_j}}
		\begin{pmatrix}
			m_j+\operatorname{Re}P_j+\Delta_j&\operatorname{Im}P_j\\
			\operatorname{Im}P_j&m_j-\operatorname{Re}P_j+\Delta_j
		\end{pmatrix}.
	\end{equation}
	This formula uses only $P_j$ and the known integer $m_j$. Since $u$ is odd, $1+\e^{2i\theta_1}=0$, hence $P_0=0$ and $F_0=2^{(d-1)/2}I_2$.
	For $1\leq j<d$, the two norm-preserving branch maps are
	\begin{equation}\label{eq:fourier-norm-preservations}
		C_j^{(0)}=F_{j-1}^{-1}F_j,\qquad C_j^{(1)}=F_{j-1}^{-1}R(\theta_j)F_j.
	\end{equation}
	They obey $C_j^{(0)}C_j^{(0)T}+C_j^{(1)}C_j^{(1)T}=I_2$. The eigenvalues of the first branch effect $E_j^{(0)}=C_j^{(0)}C_j^{(0)T}$ are $\lambda_\pm=(1\pm\xi_j)/2$, where
	\begin{equation}\label{eq:fourier-eigenvalues}
		\xi_j=\frac{|P_j|\,|\sin\theta_j|}{\sqrt{m_j^2-|P_j|^2\cos^2\theta_j}}.
	\end{equation}
	In the eigenbasis of $E_j^{(0)}$, the two branch probabilities are therefore implemented by the angles
	\begin{equation}\label{eq:fourier-angles}
		\alpha_{j,+}=2\arccos\sqrt{\frac{1+\xi_j}{2}},\qquad
		\alpha_{j,-}=2\arccos\sqrt{\frac{1-\xi_j}{2}},
		\qquad \alpha_{j,+}+\alpha_{j,-}=\pi.
	\end{equation}
	These are the two memory-conditioned $R_y$ angles in the local circuit of Section~\ref{sec:local-circuit}; the notation $\alpha_{j,\pm}$ distinguishes them from the phase increments $\theta_j$.
	
	It remains to determine the relative memory update. Each $F_j$ is positive definite, and each $R(\theta_j)$ has determinant one, so both $C_j^{(0)}$ and $C_j^{(1)}$ are nonsingular with positive determinant. Their orthogonal polar factors are consequently unique planar rotations. For a real $2\times2$ matrix $M$ with positive determinant, write the angle of its orthogonal polar factor as
	\[
	\chi(M)=\atanTwo(M_{21}-M_{12},M_{11}+M_{22}).
	\]
	If $W_j^{(b)}=R(\chi(C_j^{(b)}))$, the relative row-amplitude update is
	\begin{equation}\label{eq:fourier-relative-angle}
		\mathcal R_j=W_j^{(1)}(W_j^{(0)})^T=R(\delta_j),\qquad \delta_j=\chi(C_j^{(1)})-\chi(C_j^{(0)}).
	\end{equation}
	This relative rotation is invariant under a common change
	of the outgoing-memory basis, as established in
	Proposition~\ref{prop:relative-memory}. Choosing the common
	outgoing-memory basis so that $W_j^{(0)}=I$ leaves the
	memory-ket operation $\mathcal R_j^T=R(-\delta_j)$,
	as in the local circuit of Section~\ref{sec:local-circuit}. Because both branch maps have full rank, their polar factors have no null-space freedom; the relative angle is therefore fixed by the maps themselves. In this two-dimensional setting it is obtained directly from the displayed polar-angle formula, without a general two-qubit-unitary completion.
	
	The final active bit is treated by a terminal one-qubit gate rather than another two-qubit interaction. With $e_0=(1,0)^T$, define
	\begin{equation}\label{eq:fourier-terminal-pair}
		g_0=F_{d-1}^{-1}e_0,\qquad g_1=F_{d-1}^{-1}R(\theta_d)e_0,\qquad G=[g_0\;g_1].
	\end{equation} The defining identity for $F_{d-1}$ gives $g_0g_0^T+g_1g_1^T=I_2$, so $G$ is orthogonal; the terminal gate is $G^T$. The initial memory is $(\sin\phi,\cos\phi)^T$, since $F_0=2^{(d-1)/2}I_2$ and the active core has squared norm $2^{d-1}$. When $d=1$, there are no two-qubit steps and the same terminal prescription gives the required one-qubit state.
	
	\begin{example}[Three active Fourier qubits]\label{ex:fourier-d3}
		For $d=3$ and $u=1$, the phase increments are $\theta_1=\pi/2$, $\theta_2=\pi/4$, and $\theta_3=\pi/8$. The suffix products are $P_3=1$, $P_2=1+\e^{i\pi/4}$, $P_1=(1+i)(1+\e^{i\pi/4})=1+i(1+\sqrt2)$, and $P_0=0$. Substitution into \eqref{eq:fourier-eigenvalues} gives representative gate parameters:
		\begin{center}\small
			\begin{tabular}{@{}c c c c@{}}
				\toprule
				Step & $(\lambda_+,\lambda_-)$ & $(\alpha_{j,+},\alpha_{j,-})$ & $\delta_j$\\
				\midrule
				1 & $(0.82664074,0.17335926)$ & $(0.85888576,2.28270689)$ & $\pi/2$\\
				2 & $(0.93142810,0.06857190)$ & $(0.52990279,2.61168986)$ & $1.20530657$\\
				\bottomrule
			\end{tabular}
		\end{center}
		The listed $\delta_j$ are the row-amplitude relative angles; the corresponding operations on memory kets are $R(-\delta_j)$. The phase offset $\phi$ changes only the initial memory state, not the local entangling gates. The active state therefore uses two adjacent two-qubit interactions.
	\end{example}
	
	\subsection{A two-qubit illustration of the norm split}
	\label{sec:fourier-small-example}
	
For $d=2$, $u=1$, and the illustrative choice $\phi=0$,
the active amplitudes are $\sin(\pi x/4)$ for
$x=0,1,2,3$. The unnormalised vector is $(0,1/\sqrt2,1,1/\sqrt2)$ and its squared norm is $2$. Splitting at the most significant bit gives branch weights $1/2$ and $3/2$, so the first qubit is prepared by $R_y(2\pi/3)$. The lower branch has normalized remaining state $\ket1$, while the upper branch has normalized state $\sqrt{2/3}\ket0+\ket1/\sqrt3$. A negative-control $X$ on the lower branch and a positive-control $R_y(2\arctan(1/\sqrt2))$ on the upper branch therefore prepare
	\[
	\frac12\ket{01}+\frac1{\sqrt2}\ket{10}+\frac12\ket{11}.
	\]
	This is the ordinary scalar-tree realization of the same norm split. In the full construction, the two conditional states are instead represented coherently in the memory and updated by the local gate of Proposition~\ref{prop:local-gate}; the pair of angles in \eqref{eq:fourier-angles} gives a reusable description of each step.
	
	\subsection{Optimality of the Fourier interaction count}

\begin{proposition}[Non-product cuts of the active Fourier state]
	\label{prop:fourier-cuts}
	For $d\geq2$, the state $\ket{\Phi_{d,u,\phi}}$ is
	non-product across every cut between consecutive active qubits.
\end{proposition}
\begin{proof}
	Fix an internal cut after the first $j$ active bits,
	where $1\leq j\leq d-1$. Write a binary label as $x=pz$,
	where $p$ consists of the first $j$ bits and $z$ consists
	of the remaining $d-j$ bits. The phase contribution of
	these two parts is
	\[
	\sum_{\ell=1}^{j}p_\ell\theta_\ell
	+\sum_{\ell=j+1}^{d}z_\ell\theta_\ell.
	\]
	
	Choose two prefixes
	$p^{(0)}=(0,\ldots,0)$ and
	$p^{(1)}=(1,0,\ldots,0)$, whose phase contributions are
	$0$ and $\theta_1$, respectively. Similarly, choose two
	suffixes $z^{(0)}=(0,\ldots,0)$ and
	$z^{(1)}=(1,0,\ldots,0)$, whose phase contributions are
	$0$ and $\theta_{j+1}$, respectively.
	The corresponding $2\times2$ minor of the cut coefficient
	matrix is
	\[
	M_j=\frac{1}{\sqrt{2^{d-1}}}
	\begin{pmatrix}
		\sin\phi&
		\sin(\phi+\theta_{j+1})\\
		\sin(\phi+\theta_1)&
		\sin(\phi+\theta_1+\theta_{j+1})
	\end{pmatrix}.
	\]
	Its determinant is
	\[
	\begin{aligned}
		\det M_j
		&=\frac{1}{2^{d-1}}\big[
		\sin\phi\sin(\phi+\theta_1+\theta_{j+1})
	-
		\sin(\phi+\theta_{j+1})\sin(\phi+\theta_1)
		\big]\\
		&=-\frac{\sin\theta_1\sin\theta_{j+1}}{2^{d-1}}.
	\end{aligned}
	\]
	Since $u$ is odd, $\sin\theta_1\ne0$. Moreover,
	$\theta_{j+1}=\pi u/2^{j+1}$, so
	$\sin\theta_{j+1}\ne0$ for $j\geq1$.
	Thus $\det M_j\ne0$, and the cut coefficient matrix
	has rank at least two. By Definition~2.1, the state is
	non-product across this cut.
	Since the cut was arbitrary, the proposition follows.
\end{proof}

\begin{theorem}[Optimal Fourier interaction count]
	\label{thm:fourier-optimal}
	For a nontrivial frequency $\nu=2^su$ with $u$ odd and
	$d=m-s-1$, any ancilla-free nearest-neighbour circuit on the
	prescribed qubit order that prepares
	$\ket{F_{m,\nu,\phi}}$ from a product input requires at least
	$d-1$ two-qubit gates. The construction above uses exactly
	$d-1$ adjacent interactions and is therefore optimal in
	adjacent-interaction count.
\end{theorem}
\begin{proof}
	If $d=1$, the active Fourier block consists of a single qubit.
	The construction then requires no two-qubit interactions,
	and the claimed lower bound is zero. We may therefore assume
	$d\geq2$.
	
	By Proposition~\ref{prop:fourier-cuts}, the active Fourier
	state is non-product across each of its $d-1$ internal cuts.
	The product factors in \eqref{eq:fourier-factor} do not change
	the Schmidt rank across these cuts, so the same conclusion
	holds for the full $m$-qubit state.
	
	A circuit starting from a product input must contain at least
	one two-qubit gate crossing each such cut. Indeed, if no gate
	crosses a given cut, every operation in the circuit acts
	entirely on one side or the other, and the circuit unitary
	factors across that bipartition. It therefore cannot produce
	a state that is non-product across the cut.
	On the prescribed linear nearest-neighbour ordering, a
	two-qubit gate acts on adjacent qubits and crosses only one
	cut. Hence at least $d-1$ two-qubit gates are necessary.
	The construction uses exactly one interaction across each
	internal boundary of the active block, for a total of
	$d-1$ interactions. The lower bound is therefore attained.
\end{proof}

	\begin{algorithm}[t]
		\caption{Full-register Fourier circuit parameters}
		\label{alg:fourier-steps}
		\begin{algorithmic}[1]
			\Require Register width $m$, frequency $\nu$, phase $\phi$
			\If{$\nu=0$ or $\nu=2^{m-1}$}
			\State Prepare the corresponding product state and stop (provided it is nonzero)
			\EndIf
			\State Write $\nu=2^s u$ with $u$ odd; set $d=m-s-1$
			\State Compute $\theta_j=2\pi u/2^{j+1}$ for $1\leq j\leq d$
%			\For{$j=d,d-1,\ldots,0$}
%			\State Compute $P_j$ from \eqref{eq:fourier-P} and form $F_j$ from \eqref{eq:fourier-F}
%			\EndFor
			\State Compute $P_d=1$ and then
			$P_{j-1}=P_j(1+\e^{2i\theta_j})$ for $j=d,\ldots,1$
			\State Set $F_0=2^{(d-1)/2}I_2$ and form
			$F_1,\ldots,F_{d-1}$ from \eqref{eq:fourier-F}
			\For{$j=1,\ldots,d-1$}
			\State Form $C_j^{(0)},C_j^{(1)}$ from \eqref{eq:fourier-norm-preservations}
			\State Diagonalize $E_j^{(0)}=C_j^{(0)}C_j^{(0)T}$ and compute $\alpha_{j,\pm}$ from \eqref{eq:fourier-angles}
			\State Compute $\delta_j$ from the polar angles of $C_j^{(0)}$ and $C_j^{(1)}$
			\EndFor
			\State Prepare the initial memory $(\sin\phi,\cos\phi)^T$ and apply the $d-1$ local steps
			\State Apply the terminal one-qubit gate; prepend $\ket+^{\otimes s}\otimes\ket-$
		\end{algorithmic}
	\end{algorithm}

\section{Finite windows and the common dyadic partition}\label{sec:dyadic}

The preceding sections prepare a full shifted geometric tail or a full-register Fourier mode. We now restrict these profiles to an arbitrary nonempty interval $I=[a,a+M)\subseteq[0,2^n)$, where $[a,a+M)=\{a,a+1,\ldots,a+M-1\}$. The main difficulty is that the two sides of the interval may begin at different binary alignments. A common dyadic partition resolves this mismatch by giving the two sides the same possible block positions and lengths, while recording separately which blocks are occupied on each side. The circuit first prepares the branch and block labels from their squared norms and then prepares the appropriate profile inside the selected block. Here a \emph{branch} means the lower or upper side of the interval split, not the next-bit branch $b\in\{0,1\}$ in the norm-preserving updates.

If $M=1$, the interval consists of the single label $a$. The geometric target is then $|a\rangle$. The Fourier target is $|a\rangle$ up to a global sign, provided its amplitude at $a$ is nonzero; if that amplitude is zero, there is no normalized target state. We therefore assume $M>1$ in the common-partition construction below.

\subsection{Common partition and coordinate maps}

Put $b=a+M-1$ and define
\begin{equation}\label{eq:dyadic-midpoint}
	k=\lfloor\log_2(a\xor b)\rfloor,\qquad
	C=2^{k+1}\left\lfloor\frac{a}{2^{k+1}}\right\rfloor,\qquad
	D=C+2^k.
\end{equation}
Then $[C,C+2^{k+1})$ is the smallest aligned dyadic block containing both endpoints, and $D$ is its midpoint. The interval splits into the lower and upper branches of lengths $L=D-a$ and $R=b+1-D$, respectively. In particular, $1\leq L,R\leq 2^k$.

A branch of length $2^k$ is a complete dyadic half and is represented by a single block. To define the common marker positions, set $L^\circ=0$ if $L=2^k$ and $L^\circ=L$ otherwise; similarly, set $R^\circ=0$ if $R=2^k$ and $R^\circ=R$ otherwise.
Let $S=L^\circ\lor R^\circ$, where $\lor$ denotes bitwise OR, and write $S=\sum_{j=0}^{h-1}2^{\lambda_j}$ with $\lambda_0>\lambda_1>\cdots>\lambda_{h-1}\geq0$.
When $S=0$, we take $h=0$. For $h\geq1$, define $u_0=0$ and
$u_j=\sum_{i<j}2^{\lambda_i}$. The common partial-block positions are the consecutive dyadic intervals
$B_j=[u_j,u_j+2^{\lambda_j})$. A partial branch of length $m\in\{L,R\}$ occupies block position $j$ precisely when bit $\lambda_j$ of $m$ is one. Thus the two partial branches use the same ordered list of possible block lengths, although some positions may be absent on either side. A complete-half branch is not represented by these marker positions; it is represented by its single block of length $2^k$.

For a partial branch of length $m<2^k$, define
\begin{equation}\label{eq:sigma-m}
	\sigma_m=(S\xor m)\mathbin{\&}\bigl(2^k-2^{\vtwo(m)+1}\bigr).
\end{equation}
The mask in \eqref{eq:sigma-m} sets all bits below $\vtwo(m)+1$ to zero. For a complete-half branch set its corresponding mask equal to zero. Thus $\sigma_L=0$ if $L=2^k$ and $\sigma_R=0$ if $R=2^k$. Put $\tau_0=(2^k-1)\xor\sigma_L$ and $\tau_1=\sigma_R$. For an integer $q$, write $q_t$ for its bit in position $t$, with $t=0$ the least significant bit. For a partial block of exponent $e=\lambda_j$, define $p_{m,j}=2^e\lfloor(u_j\xor\sigma_m)/2^e\rfloor$. Its physical aligned bases are $X_{0,j}=C+2^k-2^e-p_{L,j}$ and $X_{1,j}=C+2^k+p_{R,j}$.
For a complete-half branch, use one block, denoted by $j=*$, with exponent $e=k$, offset parameter $p=0$, and base $X_{0,*}=C$ on the lower side or $X_{1,*}=D$ on the upper side. This convention gives the local masks
$\mu_{s,j}=\tau_s\bmod 2^e$ for both partial and complete blocks. In particular, a complete lower half has $\mu_{0,*}=2^k-1$, while a complete upper half has $\mu_{1,*}=0$.

For every occupied block $(s,j)$ of exponent $e$, the physical index corresponding to its local coordinate $z$ is
\begin{equation}\label{eq:dyadic-coordinate}
	x=X_{s,j}+(z\xor\mu_{s,j}).
\end{equation}
Here $z$ is the local coordinate immediately before the final coordinate mask; if a geometric lower-block reflection is used, it has already been applied to $z$. The bases $X_{s,j}$ are divisible by $2^e$, and the occupied physical blocks are disjoint and partition the original interval. A complete lower half is mapped by $z\mapsto 2^k-1-z$, whereas a complete upper half uses the unmodified local coordinate.

For a geometric lower block, the reflection $X^{\otimes e}$ and the final mask $X^{\mu_{0,j}}$ combine into the effective mask $X^{\eta_{0,j}}$, where $\eta_{0,j}=(2^e-1)\xor\mu_{0,j}=\sigma_L\bmod 2^e$. On an upper geometric block the effective mask is $X^{\eta_{1,j}}$ with $\eta_{1,j}=\mu_{1,j}=\sigma_R\bmod 2^e$. Fourier blocks require no profile reflection, so their effective mask is simply $X^{\mu_{s,j}}$. In particular, both complete-half cases have zero effective geometric mask after the lower reflection and coordinate mask are combined.

\subsection{Block restriction in norm-preserving coordinates}

Consider a normalized two-branch profile of the form
\[
f(x)=\ell^\dagger A_{n-1}^{(x_{n-1})}\cdots A_1^{(x_1)}g_{x_0},
\]
with the norm-preserving conditions of Section~\ref{sec:scalar}. On an occupied block $(s,j)$ of exponent $e$, the high bits of its aligned base $X_{s,j}$ are fixed. Contracting those fixed bits into the terminal vector gives the block boundary vector
\begin{equation}\label{eq:block-boundary-balanced}
	\ell_{s,j}=A_e^{((X_{s,j})_e)\dagger}A_{e+1}^{((X_{s,j})_{e+1})\dagger}\cdots
	A_{n-1}^{((X_{s,j})_{n-1})\dagger}\ell,
\end{equation}
where $(X_{s,j})_t$ denotes bit $t$ of the aligned base and the order is inherited from the profile factorization; equivalently,
$\ell_{s,j}^\dagger=\ell^\dagger A_{n-1}^{((X_{s,j})_{n-1})}\cdots A_e^{((X_{s,j})_e)}$. The local coordinate mask changes the remaining updates to
$B_{s,t}^{(b)}=A_t^{(b\xor(\tau_s)_t)}$ for $1\leq t<e$, and changes the terminal pair to $g_{s,b}=g_{b\xor(\tau_s)_0}$. Interchanging the two branches preserves the norm-preservation identities.

Repeated use of these identities shows that the total squared amplitude on the block is just the squared Euclidean norm of its boundary vector, namely $W_{s,j}=\|\ell_{s,j}\|^2$.
For $W_{s,j}>0$, the initial memory state of the local block circuit is
$\alpha_{s,j}=\ell_{s,j}^\dagger/\sqrt{W_{s,j}}$. The within-block transitions are the masked norm-preserving updates, followed by the masked terminal pair. Thus the block boundary affects only a scalar weight and an initial memory state; it does not change the local transition gates. A singleton block ($e=0$) has no internal transition and is handled directly by its scalar amplitude.

\begin{lemma}[Block-weight identity]\label{lem:balanced-block}
	For every occupied block of exponent $e\geq1$, the squared norm of the target restricted to that block equals $\|\ell_{s,j}\|^2$. In particular, if the block has nonzero weight, the local initial memory state $\ell_{s,j}^\dagger/\sqrt{W_{s,j}}$ is normalized.
\end{lemma}
\begin{proof}
	Sum the squared amplitudes over the $e$ local bits. The terminal sum replaces $\sum_b g_{s,b}g_{s,b}^\dagger$ by $I_2$, and each preceding local sum replaces $\sum_b B_{s,t}^{(b)}B_{s,t}^{(b)\dagger}$ by $I_2$. The remaining expression is $\ell_{s,j}^\dagger\ell_{s,j}$.
\end{proof}

Therefore, the outer selector needs only the scalar block weights $W_{s,j}$; it does not require a separate matrix of remaining norms for each block.

\subsection{Weighted block selection}

Let $W_s$ be the sum of the occupied block weights on branch $s$, including the single block weight if that branch is a complete half, and put $Z=W_0+W_1$. The branch qubit $s$ records whether the lower or upper side of the interval is selected. It is prepared with $R_y(\eta)|0\rangle$, where $\eta=2\arccos\sqrt{W_0/Z}$, so that the two branch probabilities are $W_0/Z$ and $W_1/Z$. If one branch has zero weight, the branch rotation is interpreted in the corresponding deterministic limit.

The marker positions are the qubits at $\lambda_0,\ldots,\lambda_{h-2}$; there are $h-1$ such positions when $h\geq1$. For a partial branch $s$, put zero in the block-weight list at every absent common position and define the residual mass $T_{s,j}=\sum_{\ell\geq j}W_{s,\ell}$. The angle that selects block $j$ from the still-unassigned branch mass is
\begin{equation}\label{eq:weighted-block-angle}
	\theta_{s,j}=2\arccos\sqrt{\frac{W_{s,j}}{T_{s,j}}},
	\qquad 0\leq j\leq h-2,
\end{equation}
whenever $T_{s,j}>0$. If $T_{s,j}>0$ and $W_{s,j}=0$, this angle is $\pi$; if $W_{s,j}=T_{s,j}$, it is zero. If $T_{s,j}=0$, the angle is immaterial because that residual branch has no support. For a complete-half branch, its effective marker angles are $\theta_{s,j}=0$ for $0\leq j\leq h-2$.
This assigns the complete-half branch the all-zero marker word: its entire weight is selected by the branch bit, and its block-local profile is applied later under the branch control alone.

For a partial block $j$, the unary marker pattern has all marker bits preceding its boundary equal to one, and its boundary bit and all later marker bits equal to zero. A complete-half branch has all marker bits zero. If $h=1$, there is only one possible partial block position on each branch and no marker rotation is needed; the branch bit alone performs the selection. If $h=0$, both branches are complete halves and again only the branch bit is needed.

At the first marker position, if both branches are partial, apply the lower-branch rotation unconditionally and a branch-controlled correction to obtain the upper-branch angle. If exactly one branch is partial, apply its first marker rotation controlled on the branch bit; the complete branch has angle zero. At each subsequent marker position, the two branch-dependent angles are combined by the paired-angle identity below. Thus all selector gates have at most one control.

\begin{algorithm}[t]
	\caption{Weighted common-dyadic block selection}
	\label{alg:dyadic-selection}
	\begin{algorithmic}[1]
		\Require Common block exponents $\lambda_0>\cdots>\lambda_{h-1}$ and branch block weights $W_{s,j}$
		\State Prepare the branch qubit with angle $2\arccos\sqrt{W_0/Z}$
		\If{$h\geq2$}
		\State Compute the marker split angles for each partial branch; set all marker angles to zero on a complete-half branch
		\If{both branches are partial}
		\State Prepare the first marker using an unconditional lower-branch rotation and an upper-branch controlled correction
		\Else
		\State Prepare the first marker using the rotation controlled on the unique partial branch
		\EndIf
		\For{$j=1,\ldots,h-2$}
		\State Combine the two effective branch angles using the paired-angle identity
		\EndFor
		\EndIf
		\State After completing the selector, prepare each branch-specific local profile under its block controls
		\State Apply the effective coordinate masks and restore the fixed high prefix
	\end{algorithmic}
\end{algorithm}

\begin{lemma}[Paired-angle identity]\label{lem:paired-angle}
	Let $c,d,t$ be distinct qubits, with target $t$ initially in $|0\rangle$. For real angles $\theta_0,\theta_1$, put $\Delta=\theta_1-\theta_0$ and $\bar\theta=(\theta_0+\theta_1)/2$. Applying, in the listed order, $\cgate{d=1}{R_y(-\Delta/2)_t}$, $\cgate{c=1}{Z_t}$, and $\cgate{d=1}{R_y(\bar\theta)_t}$ implements $R_y(\theta_c)|0\rangle$ when $d=1$ and leaves $|0\rangle_t$ unchanged when $d=0$. The controls may be in coherent superposition.
\end{lemma}
\begin{proof}
	For $d=1,c=0$, the two rotations combine to
	$R_y(\bar\theta-\Delta/2)=R_y(\theta_0)$. For $d=1,c=1$, use
	$Z R_y(\alpha)|0\rangle=R_y(-\alpha)|0\rangle$ to obtain
	$R_y(\bar\theta+\Delta/2)|0\rangle=R_y(\theta_1)|0\rangle$. If $d=0$, the two controlled rotations do nothing and the middle $Z$ fixes $|0\rangle_t$.
\end{proof}

For a complete-half branch, one of the paired angles is zero, so the same identity implements the required branch-dependent marker rotation without a two-control gate. The selector is completed before any block-local profile gate is applied. Hence, at a paired-angle stage, the target marker has not previously been used as a profile qubit. Whenever the continuation control $d$ is zero, that target is still $|0\rangle_t$, and the controlled-$Z$ acts trivially. This is why no signed-angle correction is needed in the selector-first ordering. An interleaved ordering would require a separate phase analysis and is not used here.

\subsection{Finite geometric windows}

Consider the finite geometric target
\begin{equation}\label{eq:finite-geom-target}
	|\Psi_{a,M}(r)\rangle=\frac1{\sqrt{Z_M(\rho)}}
	\sum_{x=a}^{a+M-1}r^{x-a}|x\rangle,
	\qquad Z_M(\rho)=\sum_{j=0}^{M-1}\rho^{2j},
\end{equation}
where $\rho=|r|>0$. The phase of $r$ is supplied by the product phase layer
$\bigotimes_{t=0}^{n-1}\operatorname{diag}(1,e^{i2^t\arg r})$; its overall phase $e^{-ia\arg r}$ is immaterial.

For a generic two-branch interval, the lower side is reflected and has local ratio $\beta_0=\rho^{-1}$, while the upper side is unreflected and has ratio $\beta_1=\rho$. On a block of exponent $e$, the normalized local product state is
\begin{equation}\label{eq:geom-local-product}
	|g_e(\beta)\rangle=\frac1{\sqrt{\mathcal G_e(\beta)}}
	\sum_{z=0}^{2^e-1}\beta^z|z\rangle
	=\bigotimes_{t=0}^{e-1}
	\frac{|0\rangle+\beta^{2^t}|1\rangle}{\sqrt{1+\beta^{2^{t+1}}}},
\end{equation}
where
$\mathcal G_e(\beta)=\sum_{z=0}^{2^e-1}\beta^{2z}
=\prod_{t=0}^{e-1}(1+\beta^{2^{t+1}})$.
The rotation on local bit $t$ is $R_y(\phi_{\beta,t})$ with
$\phi_{\beta,t}=2\arctan(\beta^{2^t})$.

For an occupied partial block, let $e=\lambda_j$ and use $p=p_{L,j}$ or $p=p_{R,j}$ according to the branch. For a complete-half block use $e=k$ and $p=0$. The exact squared block weights are then given uniformly by
\begin{equation}\label{eq:geom-window-weights}
	\omega_{0,j}=\rho^{2(L-1-p_{L,j})}\mathcal G_{\lambda_j}(\rho^{-1}),
	\qquad
	\omega_{1,j}=\rho^{2(L+p_{R,j})}\mathcal G_{\lambda_j}(\rho),
\end{equation}
for partial blocks, and by the same expressions with $e=k,p=0$ for a complete-half block. Absent partial blocks have weight zero. Their branch sums satisfy $W_0+W_1=Z_M(\rho)$.

The coordinate mask precompensates the local product profile. With $\mu_{s,j}=\tau_s\bmod 2^e$, the rotation angle on local bit $t<e$ is $\phi_{\beta_s,t}$ if bit $t$ of $\mu_{s,j}$ is zero, and $\pi-\phi_{\beta_s,t}$ if that bit is one. This uses
$X R_y(\phi)|0\rangle=R_y(\pi-\phi)|0\rangle$.
On a lower block, the profile reflection and final XOR combine into the effective mask $X^{\sigma_L\bmod 2^e}$; on an upper block, the effective mask is $X^{\sigma_R\bmod 2^e}$. The local product rotations are applied only after the block selector is complete.

For an occupied partial block $j$, the rotation on bit $t<\lambda_j$ is controlled on the branch bit, all earlier marker bits being one, and the boundary/address bit at position $\lambda_j$ being zero. A complete dyadic half requires only the branch control. The lower-side reflection $X^{\otimes e}$ is applied under the same block control. The final branch XOR $X^{\tau_s}$ on the low $k$ parent bits restores the physical block base; on the local bits, the reflection and coordinate XOR combine to the effective mask just described. The complex phase layer is applied last.

\begin{theorem}[Exact finite-window geometric preparation]\label{thm:finite-geom}
	Every nonempty finite geometric window \eqref{eq:finite-geom-target} can be prepared exactly without additional ancillas. A singleton is a basis state. For $M>1$, the common dyadic partition, with the complete-half and one-sided specializations above, selects blocks with their exact squared weights, and the within-block states are the product profiles \eqref{eq:geom-local-product}.
\end{theorem}
\begin{proof}
	The singleton case is immediate. For $M>1$, the occupied dyadic blocks partition the interval, and \eqref{eq:geom-window-weights} gives the exact squared norm on each block. The branch and block selector therefore assigns block $(s,j)$ amplitude $\sqrt{\omega_{s,j}/Z_M(\rho)}$. Within the selected block, the product state \eqref{eq:geom-local-product}, together with the precompensated coordinate mask, gives the required local profile: on the lower side the exponent is $2^e-1-w$, while on the upper side it is $w$, where $w$ is the physical local offset from that branch's beginning. The scalar block weight then gives amplitude $\rho^{x-a}/\sqrt{Z_M(\rho)}$ at physical coordinate $x$. The selector-first ordering ensures that the controlled-$Z$ in each paired-angle step acts trivially whenever its continuation control is zero. The final phase layer supplies the phase of $r^{x-a}$ up to a global phase.
\end{proof}

\begin{example}[A two-branch geometric window]\label{ex:finite-geom}
	Take $n=4$, the interval $[1,11)$, and $r=1/2$. Its midpoint split has $L=7$, $R=3$, and $S=L\lor R=7=2^2+2^1+2^0$, so the common block exponents are $(2,1,0)$. The masks are $\sigma_L=0$ and $\sigma_R=4$, hence $\tau_0=7$ and $\tau_1=4$. In descending block order, the exact unnormalized weights are $(\omega_{0,0},\omega_{0,1},\omega_{0,2})=(85/4096,5/16,1)$ on the lower side and $(\omega_{1,0},\omega_{1,1},\omega_{1,2})=(0,5/65536,1/262144)$ on the upper side. The first upper block is absent. Their sum is
	$Z_{10}(1/2)=\frac43(1-2^{-20})$. After normalization, the lower-branch
	weights are approximately $0.015564$, $0.234375$, and $0.750001$;
	the upper-branch weights are $0$, $5.72205\times10^{-5}$, and
	$2.86102\times10^{-6}$. The paired-angle selector is prepared first,
	followed by branch- and block-controlled local product rotations, the
	lower reflection, and coordinate restoration.
\end{example}

\subsection{One-sided intervals and selector resources}\label{sec:dyadic-onesided}

When an interval is presented as a one-sided segment of a prescribed dyadic parent, the construction reduces to the usual one-sided binary decomposition. Write its length as
$M=\sum_{j=0}^{h-1}2^{e_j}$ with $e_0>\cdots>e_{h-1}=v$, and put $A_j=\sum_{i<j}2^{e_i}$. For a decreasing geometric prefix with ratio $\rho\leq1$, block $j$ has squared norm $\rho^{2A_j}\mathcal G_{e_j}(\rho)$. The block-selection angles are the scalar residual-mass splits from \eqref{eq:weighted-block-angle}, and each block is prepared by \eqref{eq:geom-local-product}. An aligned upper tail is obtained by reflecting the corresponding initial segment and replacing $\rho$ by $\rho^{-1}$ before reflection. A complete dyadic half is treated as one block; a complete dyadic parent is also one block and requires no branch or block selector. Singleton intervals reduce directly to basis states.

For the two-branch common selector, the branch rotation uses one gate. If $h\geq2$, the first marker position uses at most two gates, and each of the remaining $h-2$ positions uses three gates by the paired-angle identity. Thus the scalar selector uses at most $3h-3$ one-qubit or singly controlled one-qubit gates. This count applies whether both branches are partial or one branch is a complete half, because the complete branch is assigned zero effective marker angles. If $h=1$ or $h=0$, only the branch rotation is needed. If only one side of a one-sided construction is occupied, the branch rotation is omitted.

\subsection{Finite Fourier windows}

Now consider the real Fourier profile $f(x)=\sin(\phi+\omega x)$ on $[a,a+M)$, where $N=2^n$, $\nu\in\{0,\ldots,N-1\}$, and $\omega=2\pi\nu/N$. The target is $|\Phi_{a,M}(\omega,\phi)\rangle=Z_{a,M}^{-1/2}\sum_{x=a}^{a+M-1}\sin(\phi+\omega x)|x\rangle$, where $Z_{a,M}=\sum_{x=a}^{a+M-1}\sin^2(\phi+\omega x)$, provided $Z_{a,M}>0$. For $\nu=0$ the profile is constant, while for $\nu=N/2$ it is $(-1)^x\sin\phi$. Provided $\sin\phi\ne0$, these cases reduce to the uniform interval state, which is the finite geometric construction with $r=1$; for $\nu=N/2$, an additional $Z$ on the least significant qubit supplies the factor $(-1)^x$. We therefore assume $\nu\notin\{0,N/2\}$ and $Z_{a,M}>0$. Then $|1+e^{2i\omega}|<2$; since this factor occurs in every positive-exponent local product, every occupied local remaining-norm matrix of positive exponent is full rank.

The common partition and coordinate maps are those of \eqref{eq:dyadic-midpoint}--\eqref{eq:dyadic-coordinate}. On branch $s$, the masked local phase increment on bit $t$ is $\beta_{s,t}=(-1)^{(\tau_s)_t}\omega2^t$, and the phase at the start of block $(s,j)$ is $\psi_{s,j}=\phi+\omega(X_{s,j}+\mu_{s,j})$. Thus, for a local coordinate $z=\sum_{t=0}^{e-1}z_t2^t$, the block profile is
$\sin(\psi_{s,j}+\sum_{t<e}\beta_{s,t}z_t)$, where $e$ is the block exponent. Its exact squared norm is determined by the single complex product $\Pi_{s,e}=\prod_{t=0}^{e-1}(1+e^{2i\beta_{s,t}})$:
\begin{equation}\label{eq:fourier-window-weight}
	W_{s,j}=\frac12\left(2^e-\operatorname{Re}\left(e^{2i\psi_{s,j}}\Pi_{s,e}\right)\right).
\end{equation}
This follows from $\sin^2\alpha=(1-\cos2\alpha)/2$ and factorization of the sum over all local bit strings. Hence the block-selection probabilities can be computed without enumerating the individual amplitudes. The total normalization is $Z_{a,M}=\sum_{s,j}W_{s,j}$, with absent blocks contributing zero. For a complete-half block, the same formula is used with $e=k$, $X_{0,*}=C$ or $X_{1,*}=D$, and its corresponding mask $\mu_{s,*}$.

The local Fourier profile has a two-dimensional real representation. For $1\leq q\leq e$, put $m_q=2^q$,
$\Pi_{s,q}=\prod_{t=0}^{q-1}(1+e^{2i\beta_{s,t}})$, and
$\Delta_{s,q}=\sqrt{m_q^2-|\Pi_{s,q}|^2}$. Define
\begin{equation}\label{eq:finite-fourier-root}
	F_{s,q}=\frac{1}{2\sqrt{m_q+\Delta_{s,q}}}
	\begin{pmatrix}
		m_q+\operatorname{Re}\Pi_{s,q}+\Delta_{s,q}&\operatorname{Im}\Pi_{s,q}\\
		\operatorname{Im}\Pi_{s,q}&m_q-\operatorname{Re}\Pi_{s,q}+\Delta_{s,q}
	\end{pmatrix}.
\end{equation}
This is the positive square root of the local sine-profile Gram matrix
\[
H_{s,q}=\frac12\begin{pmatrix}
	m_q+\operatorname{Re}\Pi_{s,q}&\operatorname{Im}\Pi_{s,q}\\
	\operatorname{Im}\Pi_{s,q}&m_q-\operatorname{Re}\Pi_{s,q}
\end{pmatrix}.
\]
Its entries are the sums of the products of the two components of $R(\beta_{s,z})e_0$ over all $z\in\{0,1\}^q$, where $\beta_{s,z}=\sum_{t=0}^{q-1}z_t\beta_{s,t}$ and $e_0=(1,0)^T$. The complex product $\Pi_{s,q}$ is precisely the factorized sum $\sum_z e^{2i\beta_{s,z}}$. Splitting this sum according to the value of bit $t$, for $1\leq t<e$, gives the Gram-matrix recurrence
\begin{equation}\label{eq:finite-fourier-gram-recursion}
	H_{s,t+1}=H_{s,t}+R(\beta_{s,t})H_{s,t}R(\beta_{s,t})^T.
\end{equation}

Put $q_{s,j}=(\sin\psi_{s,j},\cos\psi_{s,j})$. Then
$W_{s,j}=\|q_{s,j}F_{s,e}\|^2$. For a block of nonzero weight, its normalized initial memory row is $\alpha_{s,j}=q_{s,j}F_{s,e}/\sqrt{W_{s,j}}$; it can be prepared on the most significant local qubit by
$R_y(\varphi_{s,j})$, with
$\varphi_{s,j}=2\atanTwo((\alpha_{s,j})_2,(\alpha_{s,j})_1)$.
The row notation $q_{s,j}$ is used here to distinguish this Fourier phase row from the block boundary column $\ell_{s,j}$ in \eqref{eq:block-boundary-balanced}.

For $1\leq t<e$, the norm-preserving maps for the transition that emits local bit $t$ are
\begin{equation}\label{eq:finite-fourier-transitions}
	C_{s,t}^{(0)}=F_{s,t+1}^{-1}F_{s,t},\qquad
	C_{s,t}^{(1)}=F_{s,t+1}^{-1}R(\beta_{s,t})F_{s,t}.
\end{equation}
By \eqref{eq:finite-fourier-gram-recursion}, they satisfy
$C_{s,t}^{(0)}C_{s,t}^{(0)T}+C_{s,t}^{(1)}C_{s,t}^{(1)T}=I_2$. The transitions are applied in decreasing order $t=e-1,e-2,\ldots,1$; the least significant bit $t=0$ is handled by the terminal pair. Each shared local gate $U_{s,t}$ is obtained from these maps by the local construction of Section~\ref{sec:local-circuit}: diagonalize $C_{s,t}^{(0)}C_{s,t}^{(0)T}$ to obtain the two controlled-$R_y$ angles, and use the polar factors for the branch-dependent memory updates. The gate depends on the branch and bit position, but not on the block phase or block length. The block phase enters only through the initial memory row.

The terminal pair for branch $s$ is $g_{s,0}=F_{s,1}^{-1}e_0$ and $g_{s,1}=F_{s,1}^{-1}R(\beta_{s,0})e_0$, with $G_s=[g_{s,0}\;g_{s,1}]$. The Gram identity makes $G_s$ orthogonal, and the final local bit is produced by $G_s^T$ on the least significant qubit. For a singleton block ($e=0$), the amplitude is the scalar $\sin\psi_{s,j}$; its sign is supplied by a phase on the selected block address. This includes a complete-half block when $k=0$.

Blocks have different lengths, so the shared Fourier transitions are applied with conditional cancellations. For each branch $s$ and bit position $t$, first apply $U_{s,t}$ under control of the branch bit. If the branch is a complete half, all its local transitions belong to that single block and are retained; no inverse cancellation is applied on that branch. If the branch is partial, let $j_t$ be the first index for which $\lambda_{j_t}\leq t$, if such an index exists. The blocks with index $j\geq j_t$ are precisely those too short to contain bit $t$. In the unary marker encoding, these blocks are selected by requiring all marker/address bits at exponents greater than $t$ to be one. Apply $U_{s,t}^{-1}$ under the branch control and this full conjunction to cancel the transition on precisely those shorter blocks. For a selected block of exponent $e\leq t$, every marker at exponent greater than $t$ precedes its boundary and is one. For a block with exponent $e>t$, its boundary bit at exponent $e$ is greater than $t$ and is zero, so the cancellation control is false. Process the positions in decreasing order so that the marker pattern is preserved. The inverse operations are well-defined because the local transitions are real orthogonal; their inverses are their transposes.

The terminal gate is applied under the branch control. On a partial branch it is cancelled only on a selected singleton block, using the inverse terminal gate controlled on the branch and all positive-exponent marker/address bits being one. Every nonsingleton partial block has a positive-exponent boundary bit equal to zero, so the inverse is not activated there. A complete-half branch retains its terminal gate when $k>0$; when $k=0$, it is a singleton and its scalar sign is handled by the block-address phase described above.

\begin{theorem}[Finite-window Fourier state preparation]\label{thm:finite-fourier}
	Let $[a,a+M)$ be a nonempty interval, let $N=2^n$, and let $\nu\in\{0,\ldots,N-1\}$ with $\omega=2\pi\nu/N$. Assume $\nu\notin\{0,N/2\}$ and that $f(x)=\sin(\phi+\omega x)$ is nonzero at at least one point of the interval. Then the normalized finite-window Fourier state can be prepared exactly without additional ancillas. For $M=1$ this is a basis state up to a global sign. For $M>1$, the block weights are given by \eqref{eq:fourier-window-weight}; each selected block is prepared from its initial memory row, the norm-preserving transitions, and the terminal gate, with the variable-length cancellation schedule described above.
\end{theorem}
\begin{proof}
	The singleton case is immediate. For $M>1$, the common partition gives a disjoint partition of the interval, and \eqref{eq:fourier-window-weight} gives the exact squared norm on each occupied block. The branch and block selector therefore assigns block $(s,j)$ total amplitude proportional to $\sqrt{W_{s,j}}$. Its normalized initial memory row and local norm-preserving circuit then prepare the exact profile on that block. The cancellation schedule leaves each partial block with exactly the transitions corresponding to its length, while a complete-half branch retains all transitions of its single block. The terminal cancellation removes the terminal operation only on singleton partial blocks. Finally, the branch masks map local coordinates to physical indices and the singleton phase corrections supply the required signs. Thus every basis state in the interval receives its required amplitude, and no basis state outside the interval is populated.
\end{proof}

\subsection{Resource interpretation of the shared architecture}

The construction separates scalar block selection from profile preparation. For $h\geq2$ common block positions in the two-branch selector, the branch rotation, first marker position, and remaining paired-angle stages use at most $3h-3$ one-qubit or singly controlled one-qubit gates. This bound includes the complete-half case because its effective marker angles are set to zero and the paired-angle identity is used at every later marker position. For $h\leq1$, only the branch rotation is needed. The local profile circuit then shares transitions across blocks; controlled inverse operations cancel transitions on partial blocks too short to contain the corresponding bit. The geometric lower-side reflection and coordinate mask combine into the effective mask described above, but its activation still depends on the selected block.

These statements describe the logical circuit and its selector cost, not a total native-gate count. The construction uses the output register without additional helper qubits, but the block-local controls and Fourier cancellation predicates may require multi-controlled operations. The accompanying statevector verifiers check the selector-first ordering for geometric windows and the variable-length cancellation schedule for Fourier windows, including one-sided and complete-half cases. They verify the coherent state-preparation action of the high-level schedules.

\section{Standard-gate synthesis and resource accounting}\label{sec:synthesis}

\subsection{Real local gates and the \texorpdfstring{$SO(4)$}{SO(4)} basis transformation}

The shifted-geometric and real Fourier constructions lead to real local update matrices. Their memory-basis changes can therefore be chosen real, and the completed two-qubit local gates can be taken in $SO(4)$. This real structure is useful: a general two-qubit unitary may require three CNOT gates, whereas every matrix in $SO(4)$ admits an exact implementation with at most two CNOT gates and one-qubit gates. We give an explicit construction and a procedure for recovering its one-qubit factors from the local matrix, so that the synthesis is reproducible from the norm-distribution data rather than delegated to a black-box unitary compiler.

We use the computational basis $\{\ket{00},\ket{01},\ket{10},\ket{11}\}$, with the memory qubit first and the next-output qubit second. Define
\begin{equation}\label{eq:so4-Omega}
	\Omega_{\mathrm{SO}}=\frac1{\sqrt2}
	\begin{pmatrix}
		1&i&0&0\\
		0&0&i&1\\
		0&0&i&-1\\
		1&-i&0&0
	\end{pmatrix}.
\end{equation}
This unitary is a change of basis in which the real special-orthogonal action separates into two independent one-qubit actions. The group-theoretic reason is the double-cover relation $\operatorname{Spin}(4)\cong SU(2)\times SU(2)$, with $SO(4)\cong (SU(2)\times SU(2))/\{(I,I),(-I,-I)\}$. In the present basis, this action is represented explicitly as follows: for every $U\in SO(4)$ there are $A,B\in SU(2)$ such that
\begin{equation}\label{eq:so4-factorization}
	\Omega_{\mathrm{SO}}U\Omega_{\mathrm{SO}}^\dagger=A\otimes B,
	\qquad
	U=\Omega_{\mathrm{SO}}^\dagger(A\otimes B)\Omega_{\mathrm{SO}}.
\end{equation}
Here $SO(4)$ is the group of real orthogonal $4\times4$ matrices with determinant one, and $SU(2)$ is the group of unitary $2\times2$ matrices with determinant one. Thus, after a fixed two-qubit basis change, synthesizing $U$ reduces to implementing two ordinary one-qubit gates.

The factors can be recovered from $K=\Omega_{\mathrm{SO}}U\Omega_{\mathrm{SO}}^\dagger$ by reshuffling its entries. For a $4\times4$ matrix $K$, define
\begin{equation}\label{eq:so4-reshuffling}
	\operatorname{Resh}(K)_{2i+k,\,2j+\ell}=K_{2i+j,\,2k+\ell},
	\qquad i,j,k,\ell\in\{0,1\}.
\end{equation}
With the convention $\operatorname{vec}(A)_{2i+k}=A_{ik}$, a tensor product $K=A\otimes B$ satisfies
\begin{equation}\label{eq:so4-rank-one}
	\operatorname{Resh}(K)=\operatorname{vec}(A)\operatorname{vec}(B)^T.
\end{equation}
Thus the reshuffled matrix has rank one. Conversely, a rank-one factorization of $\operatorname{Resh}(K)$ gives the two factors by reshaping its left and right vectors.

For numerical recovery, compute a singular-value decomposition $\operatorname{Resh}(K)=\sigma uv^\dagger$, where $u$ and $v$ are normalized singular vectors corresponding to the nonzero singular value. Then
\begin{equation}\label{eq:so4-svd-factors}
	\operatorname{vec}(A_0)=\sqrt{\sigma}\,u,
	\qquad
	\operatorname{vec}(B_0)=\sqrt{\sigma}\,\overline v
\end{equation}
gives $K=A_0\otimes B_0$. The complex conjugation in the second expression is essential because \eqref{eq:so4-rank-one} contains a transpose, not a conjugate transpose. The factors have the reciprocal scalar freedom $(A_0,B_0)\mapsto(cA_0,c^{-1}B_0)$. Since $K$ is unitary, $(A_0^\dagger A_0)\otimes(B_0^\dagger B_0)=I\otimes I$, so $A_0^\dagger A_0=\lambda I$ and $B_0^\dagger B_0=\lambda^{-1}I$ for some $\lambda>0$. The balanced SVD normalization in \eqref{eq:so4-svd-factors} gives $\|A_0\|_F=\|B_0\|_F$, hence $\lambda=1$; both factors are therefore unitary.

It remains to normalize their determinants. Since $\det(A_0\otimes B_0)=(\det A_0\det B_0)^2=\det K=1$, the product $\det A_0\det B_0$ is either $1$ or $-1$. This product is unchanged by the reciprocal scalar freedom and varies continuously with $U$. Because $SO(4)$ is connected and the product equals $1$ at the identity, it equals $1$ throughout $SO(4)$. We may therefore choose $c$ so that both $\det(cA_0)=1$ and $\det(c^{-1}B_0)=1$. The normalized factors belong to $SU(2)$. The reshuffling identity is exact; the SVD is a convenient numerical means of extracting its rank-one factors. In floating-point calculations, one should also check the rank-one residual and, when necessary, project the recovered factors onto $SU(2)$ using polar decomposition followed by determinant normalization.

There is a small completion issue in applying this result to the norm-distribution construction. That construction specifies the local gate only when the next-output input is $\ket0$. The controlled-rotation--SWAP--memory-update sequence gives a unitary extension to the full two-qubit space, but its determinant can be $-1$. In the input ordering $\{\ket{00},\ket{01},\ket{10},\ket{11}\}$, multiply this extension on the right by $\operatorname{diag}(1,-1,1,1)$ when its determinant is negative. This changes only the unused $\ket{01}$ input column, leaving the prescribed blank-output subspace $\operatorname{span}\{\ket{00},\ket{10}\}$ unchanged. The resulting representative has determinant one and hence lies in $SO(4)$. It implements exactly the same state-preparation action.

\subsection{Circuit realization and one-qubit factors}

The fixed basis transformation in \eqref{eq:so4-factorization} itself requires only one CNOT. Let $S=\operatorname{diag}(1,i)$. With the memory qubit first, the next-output qubit second, and the rightmost operation applied first, we have
\begin{equation}\label{eq:so4-basis-circuit}
	\Omega_{\mathrm{SO}}
	=\operatorname{CNOT}_{q_{\rm out}\rightarrow q_{\rm mem}}
	(I\otimes H)(S\otimes S).
\end{equation}
Thus $\Omega_{\mathrm{SO}}$ is implemented by applying $S$ to both qubits, then $H$ to the output qubit, and finally a CNOT controlled by the output qubit and targeting the memory qubit. The adjoint is implemented by reversing this sequence and taking the adjoint of each gate.

It remains to implement $A$ and $B$. Every $X\in SU(2)$ has a $ZYZ$ Euler decomposition
\begin{equation}\label{eq:so4-euler}
	X=R_z(\alpha)R_y(\beta)R_z(\gamma),
\end{equation}
where $R_z(\theta)=\operatorname{diag}(e^{-i\theta/2},e^{i\theta/2})$ and $R_y(\theta)=\begin{psmallmatrix}\cos(\theta/2)&-\sin(\theta/2)\\\sin(\theta/2)&\cos(\theta/2)\end{psmallmatrix}$. If the first column of $X$ is $(x_{00},x_{10})^T$, one choice of angles is
\begin{equation}\label{eq:so4-euler-angles}
	\beta=2\arccos|x_{00}|,\qquad
	\alpha=\arg(x_{10})-\arg(x_{00}),\qquad
	\gamma=-\arg(x_{10})-\arg(x_{00}).
\end{equation}
When $x_{10}=0$, take $\beta=0$, $\alpha=0$, and $\gamma=-2\arg(x_{00})$. When $x_{00}=0$, take $\beta=\pi$, $\alpha=2\arg(x_{10})$, and $\gamma=0$. These conventions cover the boundary cases; elsewhere the angles are subject to the usual Euler-angle nonuniqueness.

Apply \eqref{eq:so4-euler} to $A$ and $B$, with triples $(\alpha_A,\beta_A,\gamma_A)$ and $(\alpha_B,\beta_B,\gamma_B)$. The chronological implementation of $U=\Omega_{\mathrm{SO}}^\dagger(A\otimes B)\Omega_{\mathrm{SO}}$ is:
\begin{enumerate}
	\item Apply $S$ to both qubits, followed by $H$ on $q_{\rm out}$.
	\item Apply a CNOT with control $q_{\rm out}$ and target $q_{\rm mem}$.
	\item Apply $R_z(\gamma_A)$, $R_y(\beta_A)$, and $R_z(\alpha_A)$ to $q_{\rm mem}$, while applying the corresponding three rotations for $B$ to $q_{\rm out}$.
	\item Apply a second CNOT with the same control and target.
	\item Apply $H$ to $q_{\rm out}$ and $S^\dagger$ to both qubits.
\end{enumerate}
The middle layer contains only parallel one-qubit rotations. Before adjacent one-qubit gates are merged or decomposed into a chosen native gate set, the construction uses twelve one-qubit gates and two CNOT gates. The two CNOTs implement the basis changes, while the two $SU(2)$ factors are applied between them. This is a joint synthesis of the completed $SO(4)$ gate; it need not have the same cost as compiling the controlled rotation, wire exchange, and relative memory update as separate operations.

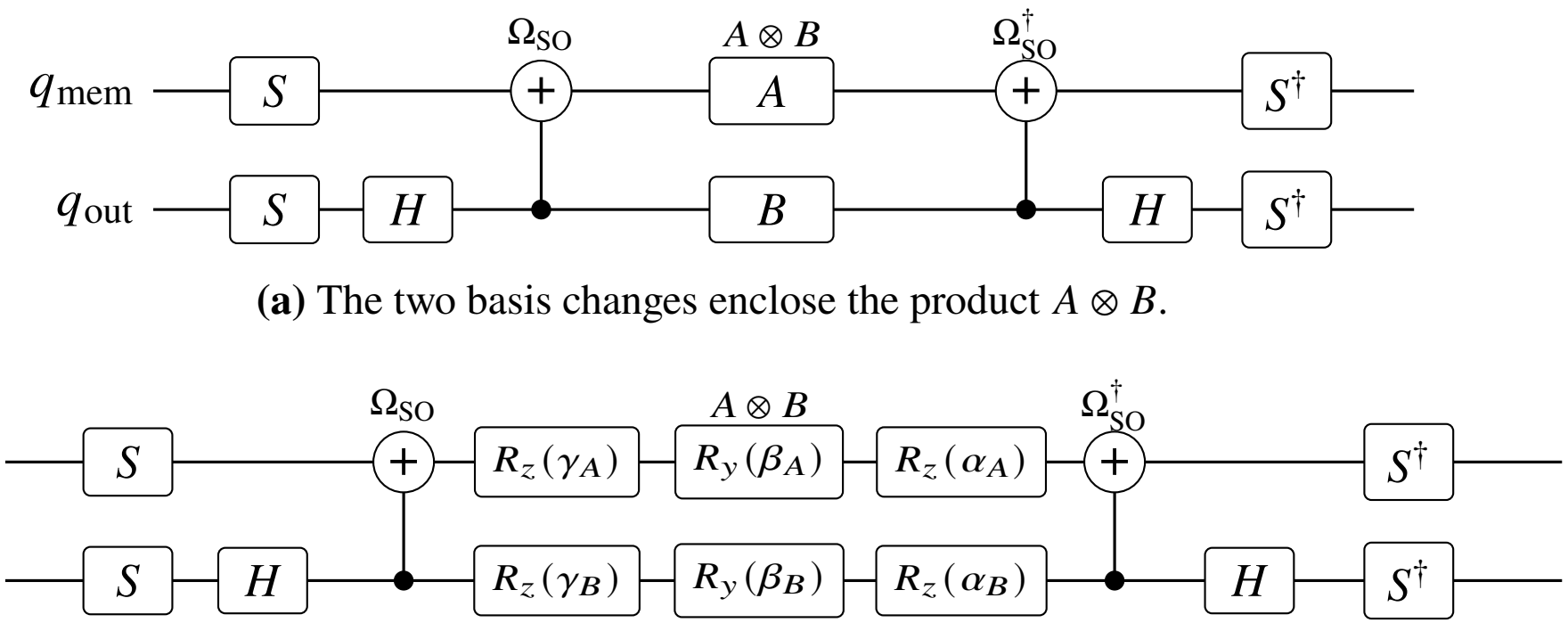
\begin{figure}[tbp]
	\centering
	\begin{subfigure}{0.98\textwidth}
		\centering
		\begin{tikzpicture}[
			x=0.90cm,y=0.88cm,
			wire/.style={draw,line width=0.6pt},
			gate/.style={draw,fill=white,rounded corners=1.5pt,minimum width=0.66cm,minimum height=0.44cm,font=\small},
			target/.style={draw,circle,fill=white,minimum size=4.5mm,inner sep=0pt}
			]
			\draw[wire] (0,0)--(10.4,0);
			\draw[wire] (0,-1)--(10.4,-1);
			\node[left] at (0,0) {$q_{\rm mem}$};
			\node[left] at (0,-1) {$q_{\rm out}$};
			\node[gate] at (1,0) {$S$};
			\node[gate] at (1,-1) {$S$};
			\node[gate] at (2.1,-1) {$H$};
			\draw[line width=0.6pt] (3.2,0)--(3.2,-1);
			\node[target] at (3.2,0) {$+$};
			\fill (3.2,-1) circle (2.1pt);
			\node[gate,minimum width=0.92cm] at (5.1,0) {$A$};
			\node[gate,minimum width=0.92cm] at (5.1,-1) {$B$};
			\draw[line width=0.6pt] (7.2,0)--(7.2,-1);
			\node[target] at (7.2,0) {$+$};
			\fill (7.2,-1) circle (2.1pt);
			\node[gate] at (8.2,-1) {$H$};
			\node[gate] at (9.35,0) {$S^\dagger$};
			\node[gate] at (9.35,-1) {$S^\dagger$};
			\node[font=\scriptsize] at (3.2,0.48) {$\Omega_{\rm SO}$};
			\node[font=\scriptsize] at (5.1,0.48) {$A\otimes B$};
			\node[font=\scriptsize] at (7.2,0.48) {$\Omega_{\rm SO}^\dagger$};
		\end{tikzpicture}
		\caption{The two basis changes enclose the product $A\otimes B$.}
	\end{subfigure}
	\par\medskip
	\begin{subfigure}{0.98\textwidth}
		\centering
		\begin{tikzpicture}[
			x=0.91cm,y=0.88cm,
			wire/.style={draw,line width=0.6pt},
			gate/.style={draw,fill=white,rounded corners=1.5pt,minimum width=1.18cm,minimum height=0.44cm,font=\scriptsize},
			smallgate/.style={draw,fill=white,rounded corners=1.5pt,minimum width=0.66cm,minimum height=0.44cm,font=\small},
			target/.style={draw,circle,fill=white,minimum size=4.5mm,inner sep=0pt}
			]
			\draw[wire] (0,0)--(12.7,0);
			\draw[wire] (0,-1)--(12.7,-1);
			\node[left] at (0,0) {$q_{\rm mem}$};
			\node[left] at (0,-1) {$q_{\rm out}$};
			\node[smallgate] at (1,0) {$S$};
			\node[smallgate] at (1,-1) {$S$};
			\node[smallgate] at (2.1,-1) {$H$};
			\draw[line width=0.6pt] (3.25,0)--(3.25,-1);
			\node[target] at (3.25,0) {$+$};
			\fill (3.25,-1) circle (2.1pt);
			\node[gate] at (4.50,0) {$R_z(\gamma_A)$};
			\node[gate] at (6.15,0) {$R_y(\beta_A)$};
			\node[gate] at (7.80,0) {$R_z(\alpha_A)$};
			\node[gate] at (4.50,-1) {$R_z(\gamma_B)$};
			\node[gate] at (6.15,-1) {$R_y(\beta_B)$};
			\node[gate] at (7.80,-1) {$R_z(\alpha_B)$};
			\draw[line width=0.6pt] (9.05,0)--(9.05,-1);
			\node[target] at (9.05,0) {$+$};
			\fill (9.05,-1) circle (2.1pt);
			\node[smallgate] at (10.15,-1) {$H$};
			\node[smallgate] at (11.45,0) {$S^\dagger$};
			\node[smallgate] at (11.45,-1) {$S^\dagger$};
			\node[font=\scriptsize] at (3.25,0.48) {$\Omega_{\rm SO}$};
			\node[font=\scriptsize] at (6.15,0.48) {$A\otimes B$};
			\node[font=\scriptsize] at (9.05,0.48) {$\Omega_{\rm SO}^\dagger$};
		\end{tikzpicture}
		\caption{The elementary circuit, with both $SU(2)$ factors expanded into parallel $ZYZ$ rotations.}
	\end{subfigure}
	\caption{Two-CNOT realization of a real local gate $U\in SO(4)$. Both CNOT gates have control $q_{\rm out}$ and target $q_{\rm mem}$. The circuit implements $U=\Omega_{\mathrm{SO}}^\dagger(A\otimes B)\Omega_{\mathrm{SO}}$.}
	\label{fig:so4-two-cnot}
\end{figure}

\subsection{When are two CNOT gates necessary?}

The preceding construction gives an upper bound of two CNOT gates for every completed local gate in $SO(4)$. For some gates, fewer suffice. The appropriate test is the operator-Schmidt rank. A two-qubit operator $U$ has operator-Schmidt rank $r$ if $r$ is the smallest integer for which $U=\sum_{j=1}^r A_j\otimes B_j$. This is exactly the matrix rank of $\operatorname{Resh}(U)$, and the singular values of that reshuffled matrix are its operator-Schmidt singular values. They are unchanged by multiplication on either side by product unitaries.

\begin{proposition}[Gatewise CNOT count for a real local gate]
	\label{prop:cnot-rank}
	Let $U\in SO(4)$, and let $s_1\geq s_2\geq s_3\geq s_4\geq0$ be the singular values of $\operatorname{Resh}(U)$. The minimum CNOT count when arbitrary one-qubit gates are allowed before and after the CNOT gates is
	\begin{equation}\label{eq:cnot-rank-count}
		N_{\rm CX}^{\min}(U)=
		\begin{cases}
			0,&\operatorname{rank}\operatorname{Resh}(U)=1,\\
			1,&\operatorname{rank}\operatorname{Resh}(U)=2\text{ and }s_1=s_2,\\
			2,&\text{otherwise}.
		\end{cases}
	\end{equation}
	Thus the construction in Figure~\ref{fig:so4-two-cnot} is optimal in the third case.
\end{proposition}
\begin{proof}
	A circuit with no CNOT gates is a product unitary, which is equivalent to operator-Schmidt rank one. A circuit with one CNOT has operator-Schmidt rank two, with two equal nonzero operator-Schmidt singular values. Since these singular values are invariant under local unitaries, rank two and $s_1=s_2$ are necessary for a one-CNOT implementation.
	
	For sufficiency, every two-qubit unitary of operator-Schmidt rank two is locally equivalent to a controlled unitary \cite{cohen2013all}. If its controlled target operation is $W\in U(2)$, write $W=e^{i\delta}V$ with $V\in SU(2)$; the factor $e^{i\delta}$ is absorbed into a phase gate on the control. We may therefore write the controlled unitary as
	\begin{equation}\label{eq:controlled-V-schmidt}
		C(V)=\ket0\!\bra0\otimes I+\ket1\!\bra1\otimes V,\qquad V\in SU(2).
	\end{equation}
	Its two nonzero operator-Schmidt singular values are $\sqrt{2+|\operatorname{tr}V|}$ and $\sqrt{2-|\operatorname{tr}V|}$. They are equal precisely when $\operatorname{tr}V=0$. A traceless matrix in $SU(2)$ has eigenvalues $i$ and $-i$, so it is unitarily conjugate to $iZ$. Its controlled version is therefore locally equivalent to controlled-$iZ$, which differs from controlled-$Z$ by a one-qubit phase on the control; controlled-$Z$ is locally equivalent to CNOT. Hence rank two together with $s_1=s_2$ is also sufficient for one CNOT.
	
	If neither the zero- nor the one-CNOT condition holds, the construction above still implements $U\in SO(4)$ with two CNOT gates, while the preceding arguments rule out fewer. This proves \eqref{eq:cnot-rank-count}.
\end{proof}

\begin{remark}[Why rank two alone is insufficient]
	A rank-two two-qubit unitary is locally equivalent to a controlled unitary, but a generic controlled rotation is not locally equivalent to CNOT. Equality of the two nonzero operator-Schmidt singular values is the additional condition that makes one CNOT sufficient. When they are unequal, two CNOT gates are necessary and sufficient.
\end{remark}

The proposition concerns the minimum CNOT count of one completed two-qubit gate. It does not give a lower bound on the total CNOT count of every circuit that prepares the same target state. The cut arguments in Sections~\ref{sec:geometric} and \ref{sec:fourier} establish a different statement: the optimal number of adjacent interactions on the prescribed linear qubit ordering. The two resource claims should not be conflated.

\subsection{The factorized architecture and joint synthesis}

The norm-distribution construction also gives a useful factorized description of the local gate: a two-angle memory-controlled $R_y$, a wire exchange, and a relative branch-controlled memory unitary, together with single-qubit memory-basis changes. In the rank-one geometric cases, the relative memory operation can be chosen as a reflection with eigenvalues $+1$ and $-1$, hence it is unitarily conjugate to $Z$. Its controlled implementation is therefore locally equivalent to controlled-$Z$ and thus to CNOT. It is tempting to count the controlled rotation and controlled reflection separately. That count need not be optimal, since the complete two-qubit unitary may admit cancellations between these factors. We therefore distinguish the factorized description, which exposes the norm split and memory update, from the joint $SO(4)$ synthesis, which compiles the completed local unitary as a whole. We make no CNOT reduction claim based solely on counting the factors separately.

\subsection{Numerical implementation and verification}

We numerically verified the full-register and finite-window
constructions using NumPy statevector simulations and Qiskit.
The tests covered the norm-distribution identities, local
transition matrices, block-weight formulas, and coherent
state-preparation procedures. In the tested cases, the
simulated states agreed with their corresponding theoretical
targets to high numerical accuracy.

The finite-window constructions were additionally checked
using independent coherent reference simulations with ideal
logical controlled operations. These tests establish the
correctness of the logical state-preparation procedures.

\begin{figure}[t]
	\centering
	\begin{subfigure}{\textwidth}
		\centering
		\includegraphics[width=\textwidth]
		{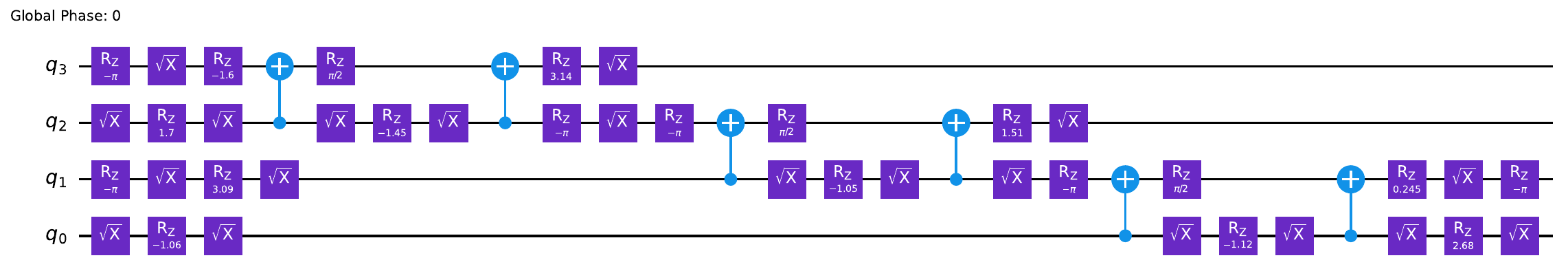}
		\caption{Shifted-geometric example
			($m=4$, $k=5$, $r=0.5$).}
	\end{subfigure}
	
	\par\medskip
	
	\begin{subfigure}{\textwidth}
		\centering
		\includegraphics[width=\textwidth]
		{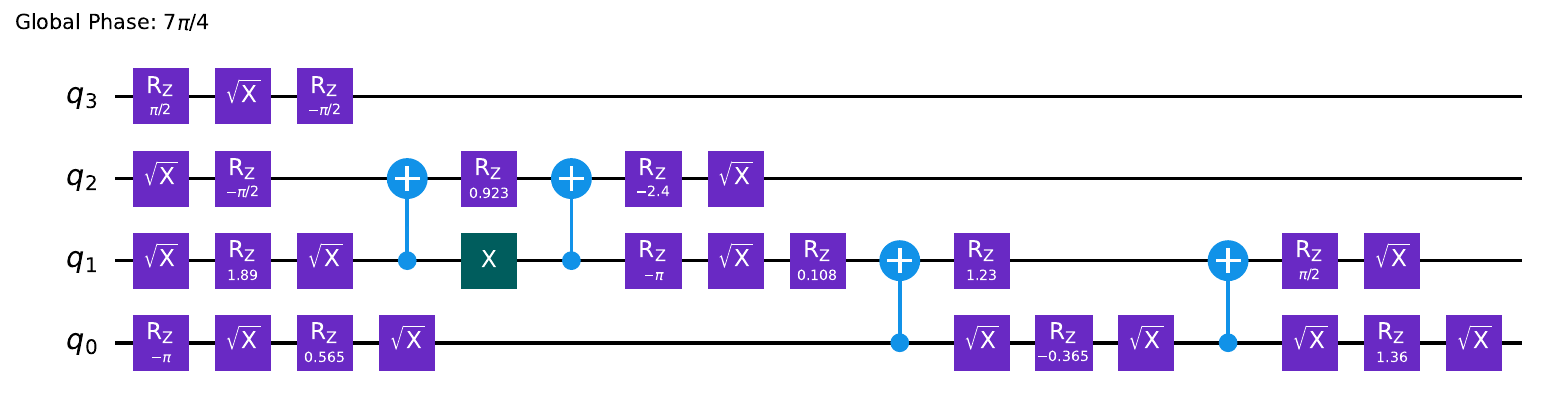}
		\caption{Single-mode Fourier example
			($m=4$, $\nu=1$, $\phi=0.37$).}
	\end{subfigure}
	
	\caption{Quantum circuits implementing the proposed
		shifted-geometric and single-mode Fourier
		state-preparation algorithms. The local two-qubit
		transformations are constructed using the explicit
		$SO(4)$-adapted magic-basis factorization developed
		in this work. The resulting circuits are transpiled
		using Qiskit 2.6.1 to the basis
		$\{rz,sx,x,cx\}$ at optimization level 3.
		The examples show the full-register implementations;
		the finite-window block-selection architecture is
		not represented.}
	\label{fig:our-circuits}
\end{figure}

For representative full-register examples, we implemented
the local two-qubit transformations using the explicit
$SO(4)$-adapted magic-basis factorization developed in this
work. The resulting circuits were transpiled using Qiskit
2.6.1 to the basis $\{rz,sx,x,cx\}$ at optimization level 3.
We considered four-qubit shifted-geometric and real
single-frequency Fourier states. The resulting circuits
are shown in Figure~\ref{fig:our-circuits}, and their
compiled resource requirements and numerical accuracy
are summarized in Table~\ref{tab:compiled-resources}.

The statevector simulations confirm that the transpiled
circuits reproduce their respective target states to
numerical precision, with phase-aligned Euclidean errors
below $2\times10^{-15}$. The shifted-geometric example
requires six CNOT gates, while the single-frequency
Fourier example requires four CNOT gates. These numerical
results provide an implementation-level verification of
the explicit local circuit constructions developed in
the preceding sections.

The reported gate counts and depths characterize these
specific four-qubit implementations and are not claimed
as general complexity bounds. The theoretical resource
bounds established earlier concern adjacent two-qubit
interactions in the prescribed register ordering, whereas
the present experiments illustrate their realization
using a standard gate basis.
\begin{table}[tbp]
	\centering
	\small
	\caption{Compiled resource requirements and
		statevector verification for representative
		four-qubit implementations of the proposed
		state-preparation constructions.}
	\label{tab:compiled-resources}
	
	\setlength{\tabcolsep}{5pt}
	\begin{tabular}{@{}lrrrrrr@{}}
		\toprule
		State family & Qubits & 1-qubit & CNOT
		& Depth & Size & Error\\
		\midrule
		Shifted geometric
		& 4 & 40 & 6 & 26 & 46
		& $1.42\times10^{-15}$\\
		
		Single-mode Fourier
		& 4 & 28 & 4 & 17 & 32
		& $7.07\times10^{-16}$\\
		\bottomrule
	\end{tabular}
	
	\vspace{2pt}
	\begin{minipage}{0.98\linewidth}
		\footnotesize
		Qiskit 2.6.1; target basis
		$\{rz,sx,x,cx\}$; optimization level 3;
		transpiler seed 12345. The error is the
		phase-aligned Euclidean statevector error
		$\min_{\theta\in\mathbb R}
		\|\psi_{\mathrm{transpiled}}
		-e^{i\theta}\psi_{\mathrm{target}}\|_2$.
	\end{minipage}
\end{table}

\section{Conclusion}\label{sec:conclusion}

The central contribution of this work is an exact sequential algorithm for quantum state preparation, built on a norm-distribution framework. Its main idea is
to exploit the low-dimensional structure of the residual
amplitude profiles. For the families studied
here, the residual states associated with different prefixes can
be represented using a two-dimensional coherent memory. This
memory is carried by the register itself, so no additional
ancilla qubit is required. At each step, the positive branch
effects determine how the squared norm is distributed between
the two possible next bits, while the polar factors retain the
coherent transformation of the unprepared suffix. A common
memory-basis change absorbs one branch factor and leaves a
single relative unitary as the branch-dependent correction.
Thus, norm distribution and coherent memory transport are
handled within one reusable local circuit.

The algorithm has two distinct ingredients: finding a suitable norm-preserving representation of the target amplitudes and converting that representation into a quantum circuit through successive local updates. The second step follows a common procedure once the representation is available, whereas the first depends on the algebraic structure of the target state. In this work, we have developed explicit representations for shifted geometric and real single-frequency Fourier profiles, exploiting their respective structures to derive efficient local operations and sharp adjacent-interaction bounds. The common dyadic partition further extends these constructions to arbitrary finite integer intervals. Thus, the results provide both a sequential algorithmic framework and detailed, resource-characterized realizations for these two families of structured quantum states.

For shifted-geometric states, the comparison structure reduces
the norm data to scalar formulas, and rank-one branch maps
permit reflection completions of the relative update. For real
single-frequency Fourier states, products of the remaining
phase increments determine the norm effects and the relative
planar rotation. The two families therefore provide distinct
examples of how structured residual amplitudes can be converted
into explicit local quantum operations.

A central complexity gain is obtained for the full-register
constructions. On a prescribed linear nearest-neighbour
ordering, the shifted-geometric state with offset $k$ requires
exactly $E_m(k)$ adjacent two-qubit interactions, where
$E_m(k)=0$ for $k\in\{0,2^m-1\}$ and
$E_m(k)=\max\{0,\lfloor\log_2(2^m-k-1)\rfloor-v_2(k)\}$
otherwise. For a nontrivial Fourier frequency
$\nu=2^s u$, with $u$ odd, removal of product factors leaves
an active register of width $d=m-s-1$, and the construction
requires exactly $d-1$ adjacent interactions. Both counts
are optimal, as follows from Schmidt-rank lower bounds across
the non-product cuts. Therefore, the number of entangling
interactions grows linearly with the active register width.
Moreover, the real local gates admit exact
synthesis using at most two CNOT gates each, giving linear
CNOT upper bounds of $2E_m(k)$ and $2(d-1)$, respectively.
The optimality established here concerns adjacent
two-qubit interactions; it does not assert that the
two-CNOT synthesis of every individual local gate is
itself optimal.

The framework also extends both families to arbitrary finite
integer intervals with nonzero restricted norm.
 A common
dyadic partition assigns the exact squared norm to each block,
after which the corresponding local profile is prepared
coherently. The scalar branch and block selector requires at
most $3h-3$ one-qubit or singly controlled one-qubit gates
when the common partition contains $h\geq2$ blocks. The
resulting finite-window circuits are exact and ancilla-free
at the logical-circuit level. Their state-preparation action
has been checked through coherent statevector simulations.
However, the block-dependent controls and the
variable-length controlled-inverse operations have not been
fully decomposed into a native gate basis. Thus, a complete
finite-window CNOT count and hardware-resource comparison
remain open.

More broadly, the norm-preserving condition is a coordinate
normalization:
every full-rank two-dimensional update representation can be
transformed to this form, as established in
Appendix~\ref{app:balanced-normalization}. The approach
therefore suggests extensions to sums of geometric modes,
multiple Fourier modes, complex exponential sums, and
profiles generated by small finite-state recurrences.
Such extensions may require memory dimensions greater than
two, together with higher-dimensional norm-distribution
operations. Understanding the resulting complexity and
identifying further structured families with optimal
interaction counts are natural directions for future work.

\section*{Declaration on the Use of Generative AI}

Generative AI tools were used to assist with language editing and improving the presentation of the manuscript, as well as in developing simulation code in support of the research. The author reviewed the AI-assisted material and takes full responsibility for the content of the manuscript, including the correctness of all mathematical results and reported computational findings.

	\section*{Appendix}
	\appendix
	
	\section{Norm-preserving coordinates and representation choice}\label{app:balanced-normalization}
	
	The main text uses updates that already preserve the total
	squared norm at each step. This appendix explains how such
	coordinates arise from a more general two-dimensional update
	representation and records the freedom to change the memory
	basis. A three-qubit example illustrates how the choice of
	terminal columns and memory coordinates can simplify
	the resulting update matrices.
	
	\subsection{Normalization of a general update representation}
	
	Start with arbitrary matrices $B_j^{(b)}\in\C^{2\times2}$, an initial row $\lambda$, and a terminal column $\rho$, with amplitudes
	\[
	a(\mathbf b)=\lambda B_1^{(b_1)}\cdots B_n^{(b_n)}\rho.
	\]
	For $0\leq j\leq n$, define the positive semidefinite matrices $H_j$ recursively by
	\begin{equation}\label{eq:app-H}
		H_n=\rho\rho^\dagger,\qquad
		H_{j-1}=\sum_{b=0}^1 B_j^{(b)}H_jB_j^{(b)\dagger},
		\quad j=n,n-1,\ldots,1.
	\end{equation}
	The total squared norm of the unnormalized state is $Z=\sum_{\mathbf b}|a(\mathbf b)|^2=\lambda H_0\lambda^\dagger$. If $v$ is a row vector entering cut $j$, then $vH_jv^\dagger$ is the total squared norm of all suffix amplitudes generated from $v$. In particular,
	\[
	vH_jv^\dagger=
	\sum_{b_{j+1},\ldots,b_n\in\{0,1\}}
	\bigl|vB_{j+1}^{(b_{j+1})}\cdots B_n^{(b_n)}\rho\bigr|^2.
	\]
	Thus $H_j$ records the remaining squared norm as a quadratic form on the memory space.
	
	\begin{theorem}[Normalization to norm-preserving coordinates]\label{thm:app-balanced-normalization}
		Assume that $H_0,\ldots,H_{n-1}$ are positive definite and that $Z=\lambda H_0\lambda^\dagger>0$. Choose invertible factors $F_j$ such that $H_j=F_jF_j^\dagger$, for $0\leq j<n$, and set
		\begin{align}
			A_j^{(b)}&=F_{j-1}^{-1}B_j^{(b)}F_j,
			&&1\leq j<n,\label{eq:app-balanced-mid}\\
			g_b&=F_{n-1}^{-1}B_n^{(b)}\rho,
			&&b=0,1,\label{eq:app-balanced-terminal}\\
			\widetilde\lambda&=\frac{\lambda F_0}{\sqrt Z}.
			&&\label{eq:app-balanced-boundary}
		\end{align}
		Then
		\[
		\frac{a(\mathbf b)}{\sqrt Z}
		=\widetilde\lambda A_1^{(b_1)}\cdots A_{n-1}^{(b_{n-1})}g_{b_n},
		\]
		and the transformed data satisfy
		\begin{align*}
			\widetilde\lambda\widetilde\lambda^\dagger=1, \quad
			\sum_{b=0}^1A_j^{(b)}A_j^{(b)\dagger}=I_2
			\text{ for } 1\leq j<n,\quad \text{and} \quad 
			\sum_{b=0}^1g_bg_b^\dagger=I_2.
		\end{align*}
		Thus, with $\widetilde\lambda$ as the boundary row, the normalized amplitudes have the norm-preserving form used in the main text.
	\end{theorem}
	
	\begin{proof}
		For $1\leq j<n$, the recursion \eqref{eq:app-H} gives
		\begin{align*}
			\sum_b A_j^{(b)}A_j^{(b)\dagger}
			&=F_{j-1}^{-1}
			\left(\sum_bB_j^{(b)}F_jF_j^\dagger B_j^{(b)\dagger}\right)
			F_{j-1}^{-\dagger}
			=F_{j-1}^{-1}H_{j-1}F_{j-1}^{-\dagger}=I_2.
		\end{align*}
		At the terminal step,
		\[
		\sum_bg_bg_b^\dagger
		=F_{n-1}^{-1}H_{n-1}F_{n-1}^{-\dagger}=I_2,
		\]
		and $\widetilde\lambda\widetilde\lambda^\dagger
		=\lambda H_0\lambda^\dagger/Z=1$. Finally, the factors $F_j$ cancel successively in the product defining the transformed amplitude, leaving $a(\mathbf b)/\sqrt Z$.
	\end{proof}
	
	The terminal object is a pair of columns because $H_n=\rho\rho^\dagger$ has rank one. The theorem assumes positive definiteness only at the intermediate cuts. If an intermediate $H_j$ is singular, the effective memory at that cut can instead be restricted to its support, whose dimension is $\operatorname{rank}H_j$. If this rank varies with $j$, the memory dimension may vary as well, or an explicit embedding may be required. The one-qubit-memory constructions in the main text apply when the intermediate support has dimension two, or when the profile has already been reduced to a two-dimensional active memory.
	
	\subsection{Freedom in the memory basis}\label{app:memory-basis-freedom}
	
	The norm-preserving representation is not unique. Let $Q_0,\ldots,Q_{n-1}$ be arbitrary unitary $2\times2$ matrices and define
	\begin{align*}
		\lambda'&=\lambda Q_0,\\
		A_j'^{(b)}&=Q_{j-1}^\dagger A_j^{(b)}Q_j,
		\qquad 1\leq j<n,\\
		g_b'&=Q_{n-1}^\dagger g_b.
	\end{align*}
	The factors telescope in the amplitude product, so the physical state is unchanged. The norm-preserving identities are also unchanged; for example,
	\[
	\sum_bA_j'^{(b)}A_j'^{(b)\dagger}
	=Q_{j-1}^\dagger
	\left(\sum_bA_j^{(b)}A_j^{(b)\dagger}\right)Q_{j-1}
	=I_2.
	\]
	The memory basis may therefore be selected independently at each cut.
	
	At a given step, the incoming basis can be chosen to diagonalize $E_0=A_0A_0^\dagger$. If $C_b=D_bW_b$ are the left polar decompositions of the resulting branch maps, a common outgoing basis change $T$ replaces $W_b$ by $W_bT$ and leaves $\mathcal R=W_1W_0^\dagger$ invariant. Taking $T=W_0^\dagger$ makes the zero-branch polar factor the identity and leaves $\mathcal R$ on the one-branch map. The only branch-dependent memory-ket operation is then $\mathcal R^T$, controlled on output one. This outgoing basis choice becomes the incoming basis choice at the next step. The eigenvalues of $E_0$ are basis invariant, whereas its eigenvectors and the individual polar factors depend on the chosen coordinates.
	
	When a branch map has rank one, its unitary polar completion is free on the one-dimensional null space. In a real two-dimensional memory, that freedom permits either orientation of the missing row and can be used to make the relative update a reflection in the geometric cases. For full-rank real Fourier updates, the polar factors are unique, so the relative planar rotation is fixed by the maps. One can also choose a convenient terminal pair and sparse raw updates before applying Theorem~\ref{thm:app-balanced-normalization}; the example below illustrates this design freedom.
	
	\subsection{A sparse three-qubit example}\label{app:sparse-example}
	
	Consider
	\[
	\ket\Psi=\frac{\ket{000}+\ket{001}+2\ket{111}}{\sqrt6}.
	\]
	The choice of terminal columns affects the complexity of the corresponding updates. Taking $g_0=e_0$ and $g_1=e_1$ (the two columns of the identity, with $e_1=Xe_0$), and writing $D_1=\diag(0,1)$, gives the norm-preserving representation
	\begin{align*}
		\lambda&=(1,0),\qquad g_0=e_0,\quad g_1=e_1,\\
		A_2^{(0)}&=\frac1{\sqrt2}
		\begin{pmatrix}1&1\\0&0\end{pmatrix},
		&A_2^{(1)}&=D_1,\\
		A_1^{(0)}&=\begin{pmatrix}1/\sqrt3&0\\0&1/\sqrt2\end{pmatrix},
		&A_1^{(1)}&=\begin{pmatrix}0&\sqrt{2/3}\\1/\sqrt2&0\end{pmatrix}.
	\end{align*}
	Direct multiplication gives amplitudes $1/\sqrt6$, $1/\sqrt6$, and $2/\sqrt6$ on $000$, $001$, and $111$, respectively, and zero elsewhere. The terminal pair satisfies $g_0g_0^T+g_1g_1^T=I_2$; the second and first update pairs satisfy $\sum_bA_2^{(b)}A_2^{(b)T}=I_2$ and $\sum_bA_1^{(b)}A_1^{(b)T}=I_2$, respectively. Hence all intermediate remaining-norm matrices are the identity in these coordinates. The example shows that a complicated raw factorization need not be intrinsic to the state: a suitable terminal completion and memory basis can expose sparse updates and simple branch effects.

	\section{Geometric comparison formulas and stable evaluation}\label{app:geom}
	
	For the geometric comparison, let $q$ bits remain and let $d$ be the residual threshold. The upper-tail masses $S_q(d)$ and $T_q(d)$ are defined in Section~\ref{sec:geometric}; the initial-segment convention interchanges their roles according to whether the accepted prefix or its complement is being tracked. The raw comparison updates, in the memory order (equal/active, already decided), can be chosen as follows. Set $D_1=\diag(0,1)$ and $M_*=\begin{psmallmatrix}0&1\\0&1\end{psmallmatrix}$. If the current comparison bit is $\kappa$ and $\gamma=\tau^{2^{q-1}}$, then
	\begin{center}\small
		\begin{tabular}{@{}c c c c@{}}
			\toprule
			Support form & $\kappa$ & $B^{(0)}$ & $B^{(1)}$\\
			\midrule
			Upper tail & 0 & $I_2$ & $\gamma M_*$\\
			Upper tail & 1 & $D_1$ & $\gamma I_2$\\
			Initial segment & 0 & $I_2$ & $\gamma D_1$\\
			Initial segment & 1 & $M_*$ & $\gamma I_2$\\
			\bottomrule
		\end{tabular}
	\end{center}
	The initial row is $(1,0)$; the terminal comparison column is $(1,1)^T$ for an upper tail and $(0,1)^T$ for an initial segment. An induction over the processed comparison bits shows that these updates reproduce the support indicator and the geometric weight. Applying the normalization of Theorem~\ref{thm:app-balanced-normalization} gives the explicit norm-preserving updates used in Section~\ref{sec:geometric}.
	
	For stable numerical evaluation, put $z=2\log\tau$. When $\tau\ne1$, evaluate
	\[
	T_q(d)=\frac{\operatorname{expm1}(dz)}{\operatorname{expm1}(z)},\qquad
	S_q(d)=\tau^{2d}\frac{\operatorname{expm1}((2^q-d)z)}{\operatorname{expm1}(z)},
	\]
	using the direct integer formulas at $\tau=1$. The parameter should be formed as $\tau=\exp(-|2^\nu\log\sigma|)$ instead of first evaluating $\sigma^{2^\nu}$. This avoids overflow for large $\sigma$ and reduces cancellation when $\tau$ is close to one. Very small positive tail sums can still underflow in finite precision; this is a limitation of floating-point evaluation, not of the exact formulas.
	
	\section{Further details for Fourier products}\label{app:fourier}
	
	The product formula \eqref{eq:fourier-P} follows by summing over the suffix bits independently. If $\Omega=\sum_{\ell=j+1}^d\theta_\ell b_\ell$, then
	\[
	\sum_{b_{j+1},\ldots,b_d\in\{0,1\}}\e^{2i\Omega}
	=\prod_{\ell=j+1}^d(1+\e^{2i\theta_\ell})=P_j.
	\]
	Using $\cos^2\Omega=(1+\cos2\Omega)/2$, $\sin^2\Omega=(1-\cos2\Omega)/2$, and $\sin\Omega\cos\Omega=\sin2\Omega/2$ gives the real symmetric remaining-norm matrix whose positive square root is $F_j$ in \eqref{eq:fourier-F}. Its determinant is $(m_j^2-|P_j|^2)/4$. For $j<d$, the oddness of $u$ implies $|P_j|<m_j$, so this matrix is positive definite and the square root is nonsingular. At the first active stage, $P_0=0$ and $F_0=2^{(d-1)/2}I_2$.
	
	The eigenvalue formula \eqref{eq:fourier-eigenvalues} follows from the trace and determinant of the first branch effect. The two eigenvalues sum to one, giving the complementary-angle identity in \eqref{eq:fourier-angles}. To compute the relative memory angle, the orthogonal polar factor of a real $2\times2$ matrix $M$ with positive determinant is $R(\chi(M))$, where
	\[
	\chi(M)=\atanTwo(M_{21}-M_{12},M_{11}+M_{22}).
	\]
	Applying this formula to $C_j^{(0)}$ and $C_j^{(1)}$ gives the relative angle in \eqref{eq:fourier-relative-angle}. Because the amplitudes are represented by row vectors, the controlled operation on memory kets is the transpose of the row-amplitude update and therefore has the opposite angle.
	
	\section{Common-partition identities and block-selection schedule}\label{app:dyadic}
	
	The common dyadic partition in Section~\ref{sec:dyadic} is inherited from the interval construction of \cite{ShuklaVedulaInterval2026}. The blocks $B_j=[u_j,u_j+2^{\lambda_j})$ partition the coordinate intervals on both sides, with the same exponents $\lambda_j$. For geometric profiles, the lower-side reflection and final coordinate mask combine into the effective mask $\sigma_L\bmod 2^e$, while the upper-side effective mask is $\sigma_R\bmod 2^e$. For Fourier profiles, the signed phase increments account directly for the XOR coordinate map. The coordinate masks in \eqref{eq:sigma-m}--\eqref{eq:dyadic-coordinate} map each local block bijectively to its physical aligned interval. The block-selection register encodes block $j$ by setting all earlier selection bits to one and the boundary bit and all later selection bits to zero.
	
	For a local transition $U_{s,t}$ acting on bits $q_t,q_{t-1}$, the variable-length schedule first applies $U_{s,t}$ under control of the branch bit and then applies $U_{s,t}^{-1}$ under control of that branch bit and the condition that every selection bit above $t$ equals one. If the selected block has exponent $e\leq t$, all selection bits above $t$ are earlier selection bits and equal one, so the inverse cancels the transition. If $e>t$, the selected boundary bit above $t$ is zero, so the inverse is inactive and the transition remains. Processing $t$ in decreasing order preserves this distinction. The final one-bit transition is cancelled on singleton blocks in the same way. This establishes the logical schedule; it does not decompose its multiple controls into CNOT gates.
	
%	\bibliographystyle{unsrt}
%	\bibliography{references}

\end{document}